\documentstyle[11pt,epsf,psfig]{article}

\newskip\humongous \humongous=0pt plus 1000pt minus 100pt
\def\caja{\mathsurround=0pt}
\def\eqalign#1{\,\vcenter{\openup1\jot \caja
       \ialign{\strut \hfil$\displaystyle{##}$&$
        \displaystyle{{}##}$\hfil\crcr#1\crcr}}\,}
\newif\ifdtup
\def\panorama{\global\dtuptrue \openup1\jot \caja
        \everycr{\noalign{\ifdtup \global\dtupfalse
        \vskip-\lineskiplimit \vskip\normallineskiplimit
        \else \penalty\interdisplaylinepenalty \fi}}}
\def\eqalignno#1{\panorama \tabskip=\humongous
        \halign to\displaywidth{\hfil$\displaystyle{##}$
        \tabskip=0pt&$\displaystyle{{}##}$\hfil
        \tabskip=\humongous&\llap{$##$}\tabskip=0pt
        \crcr#1\crcr}}

\newcounter{eqnumber}[section]
\renewcommand{\theeqnumber}{\thesection.\arabic{eqnumber}}
\def\equn{
\refstepcounter{eqnumber}
\eqno({\rm \theeqnumber})
}
\def\equnno{\refstepcounter{eqnumber}
({\rm \theeqnumber})}

\def\eqn#1{eq.~(\ref{#1})}
\def\Eqn#1{Eq.~(\ref{#1})}
\def\eqns#1#2{eqs.~(\ref{#1}) and~(\ref{#2})}

\def\fig#1{fig.~{\ref{#1}}}
\def\figs#1#2{figs.~{\ref{#1}} and {\ref{#2}}}

\def\sec#1{section~{\ref{#1}}}
\def\Sec#1{Section~{\ref{#1}}}
\def\secs#1#2{sections~{\ref{#1}} and {\ref{#2}}}
\def\app#1{appendix~\ref{#1}}
\def\App#1{Appendix~\ref{#1}}
\def\apps#1#2{appendices~{\ref{#1}} and {\ref{#2}}}
\def\Apps#1#2{Appendices~{\ref{#1}} and {\ref{#2}}}
\def\tab#1{table~\ref{#1}}
\def\Tab#1{Table~\ref{#1}}
\def\tr{\mathop{\rm tr}\nolimits}
\def\trplus{\mathop{\rm tr}\nolimits_+}
\def\trminus{\mathop{\rm tr}\nolimits_-}

\newbox\charbox
\newbox\slabox
\def\s#1{{      
        \setbox\charbox=\hbox{$#1$}
        \setbox\slabox=\hbox{$/$}
        \dimen\charbox=\ht\slabox
        \advance\dimen\charbox by -\dp\slabox
        \advance\dimen\charbox by -\ht\charbox
        \advance\dimen\charbox by \dp\charbox
        \divide\dimen\charbox by 2
        \raise-\dimen\charbox\hbox to \wd\charbox{\hss/\hss}
        \llap{$#1$}
}}

\def\spa#1.#2{\left\langle#1\,#2\right\rangle}
\def\spb#1.#2{\left[#1\,#2\right]}
\def\lor#1.#2{\left(#1\,#2\right)}

\catcode`@=11  

\def\Tr{\, {\rm Tr}}
\def\P{{\rm P}}
\def\NP{{\rm NP}}

\def\eps{\epsilon}

\def\I{{\cal I}}

\def\pol{\eps}
\def\zb{\bar z}

\def\gT{{\tilde g}}
\def\hT{{\tilde h}}
\def\vT{{\tilde v}}
\def\e{\epsilon}

\def\half{{1\over 2}}
\def\threehalf{{3\over 2}}
\def\la{\langle}
\def\ra{\rangle}
\def\oneloop{{1 \mbox{-} \rm loop}}
\def\twoloop{{2 \mbox{-} \rm loop}}
\def\Lloop{{L \mbox{-} \rm loop}}
\def\sumYM{\sum_{S_1, S_2 \in \{N=4\} }}

\def\S{{\hbox{\tiny$S$}}}

\def\vp{\vphantom{\Big|}}
\def\lsl{\not{\hbox{\kern-2.3pt $\ell$}}}
\def\ksl{\not{\hbox{\kern-2.3pt $k$}}}

\begin{document}

\begin{titlepage}

\begin{flushright}

hep-th/9802162 \hfill SLAC--PUB--7751\\
UCLA/98/TEP/03\\
SWAT-98-183\\
February, 1998\\
\end{flushright}

\vskip 2.cm

\begin{center}
\begin{Large}
{\bf On the Relationship between Yang-Mills Theory and Gravity}

{\bf and its Implication for Ultraviolet Divergences}
\end{Large}

\vskip 2.cm

{\large Z. Bern$^{\star,1}$, L. Dixon$^{\dagger,2}$, 
D.C. Dunbar$^{\sharp,3}$, 
M. Perelstein$^{\dagger,2}$ and J.S. Rozowsky$^{\star,1}$}

\vskip 0.5cm

$^\star${\it Department of Physics,
University of California at Los Angeles,
Los Angeles,  CA 90095-1547}

\vskip .3cm

$^\dagger${\it Stanford Linear Accelerator Center,
Stanford University, Stanford, CA 94309}

\vskip .3 cm

$^\sharp${\it Department of Physics, University
    of Wales Swansea, Swansea, SA2 8PP, UK }

\vskip .3cm
\end{center}

\begin{abstract}
String theory implies that field theories containing gravity are in a
certain sense `products' of gauge theories.  We make this product
structure explicit up to two loops for the relatively simple case of
$N=8$ supergravity four-point amplitudes, demonstrating that they are
`squares' of $N=4$ super-Yang-Mills amplitudes.  This is accomplished
by obtaining an explicit expression for the $D$-dimensional two-loop
contribution to the four-particle $S$-matrix for $N=8$ supergravity,
which we compare to the corresponding $N=4$ Yang-Mills result.  From
these expressions we also obtain the two-loop ultraviolet divergences
in dimensions $D=7$ through $D=11$.  The analysis relies on
the unitarity cuts of the two theories, many of which can be recycled
from a one-loop computation.  The two-particle cuts, which may be
iterated to all loop orders, suggest that squaring relations between
the two theories exist at any loop order.  The loop-momentum
power-counting implied by our two-particle cut analysis indicates that
in four dimensions the first four-point divergence in $N=8$
supergravity should appear at five loops, contrary to the earlier
expectation, based on superspace arguments, of a three-loop
counterterm.
\end{abstract}

\vfill
\noindent\hrule width 3.6in\hfil\break
${}^{1}$Research supported in part by the US Department of Energy
under grant DE-FG03-91ER40662 
\hfil\break
${}^{2}$Research supported by the US Department of
Energy under grant DE-AC03-76SF00515.\hfil\break
${}^3$Research supported in part by the Leverhulme Foundation.\hfil\break
\end{titlepage}

\baselineskip 16pt


\section{Introduction}

Gravity and gauge theories both contain a local symmetry, mediate
long-range forces at the classical level, and have a number of other
well-known similarities.  Indeed, one might naively suspect that a
spin two graviton can in some way be interpreted as a product of spin
one gluons.  There are, however, some important differences.  In
particular, the Feynman rules for gauge field perturbation theory
include only three- and four-point vertices, while those for gravity
can have arbitrarily many external legs.  A related fact is that gauge
theories are renormalizable in four dimensions, while gravity is not.
Given the rather disparate forms of the Yang-Mills and
Einstein-Hilbert Lagrangians, it is not obvious how to make a
relationship between graviton and gluon scattering precise, starting
from the Lagrangians.

String theory, however, suggests that the $S$-matrix elements
satisfy a relationship of the form 
$$
\hbox{gravity} \sim \hbox{(gauge theory)} \times \hbox{(gauge theory)} \,.
\equn\label{GravityYMRelation}
$$
This relationship follows from the string representation of amplitudes
as integrals over world-sheet variables --- the integrands for closed
string amplitudes can be factorized into products of left-moving and
right-moving modes, each of which is nearly identical to an
open-string integrand,
$$
\hbox{closed string} \sim \hbox{(left-mover open string)} \times  
                          \hbox{(right-mover open string)} \,.
\equn\label{StringRelation}
$$
In order to show that the string relation~(\ref{StringRelation})
implies the field theory relation (\ref{GravityYMRelation}) one must
inspect the string amplitudes in the field theory (or infinite
tension) limit.  In this limit a closed string should reduce to a
theory of gravity and an open string to a gauge theory.  

In this paper we make the heuristic relation (\ref{GravityYMRelation})
precise for four-point amplitudes up to two loops in $N=8$
supergravity and $N=4$ Yang-Mills field theory by investigating their
unitarity cuts~\cite{Cutting}.  (The number of supersymmetries refers to their
four-dimensional values; in any dimension we define $N=4$ Yang-Mills
theory to be the dimensional reduction of ten-dimensional $N=1$
Yang-Mills theory, and $N=8$ supergravity to be the dimensional
reduction of eleven-dimensional $N=1$ supergravity.)  These theories
have a high degree of supersymmetry which considerably simplifies
explicit calculations, making it relatively easy to make relation
(\ref{GravityYMRelation}) precise and to verify it.

At tree level, Kawai, Lewellen and Tye (KLT) \cite{KLT} have used the
string relation (\ref{StringRelation}) to provide explicit formulae
for closed string amplitudes as sums of products of open string
amplitudes.  In the field-theory limit, the higher excitations of the 
string decouple and the KLT relations directly relate gravity and 
gauge theory amplitudes.  
These results were exploited by Berends, Giele and Kuijf~\cite{BGK}
to obtain an infinite set of maximally helicity violating (MHV)
gravity amplitudes using the known~\cite{ParkeTaylor} MHV
Yang-Mills amplitudes.

At one loop, the close relation between four-point open and closed
superstring amplitudes is also well known~\cite{GSW}.  This relation
leads in the field-theory limit~\cite{GSB} to a relation between the
one-loop four-point amplitudes for $N=4$ super-Yang-Mills theory and $N=8$
supergravity.  More generally, one-loop gravity amplitudes have been
obtained from string theory~\cite{BDS,DN} using rules~\cite{Long} for
systematically extracting the field-theory limit of one-loop string
amplitudes.  In many ways, the rules for gravity are double copies of
the gauge theory rules.  In particular, in ref.~\cite{BDS} the
Feynman parameter integrands appearing in a non-supersymmetric
four-graviton amplitude are squares of the integrands appearing in the
corresponding four-gluon amplitude.

Beyond one loop, the general structure of multi-loop string theory
amplitudes~\cite{MultiLoopStrings} leads one to suspect that relation
(\ref{GravityYMRelation}) will continue to hold in some form.  Since
it is non-trivial to take the field-theory limit of a multi-loop
string amplitude, one would like an alternative approach to
investigate \eqn{GravityYMRelation} beyond one loop.  Our approach
will be to evaluate the unitarity cuts of multi-loop amplitudes. The
methodology for performing computations using the analytic properties
of amplitudes \cite{AllPlus,SusyFour,SusyOne,DNb,Rozowsky} has been
reviewed in ref.~\cite{Review} and applied to multi-loop $N=4$
supersymmetric amplitudes in ref.~\cite{BRY}.  At one loop this is a
proven technology, having been used in the calculation of analytic
expressions for the QCD one-loop matrix elements for $Z \rightarrow 4$
partons~\cite{Zjets} and in the construction of infinite
sequences~\cite{AllPlus,SusyFour,SusyOne} of one-loop MHV amplitudes.
This technique allows for a {\it complete} reconstruction of the
amplitudes from the cuts, provided that all cuts are known in
dimensional regularization for arbitrary dimension.  Because on-shell
expressions are used throughout such calculations, gauge invariance,
Lorentz covariance and unitarity are manifest.  These techniques, of
course, do not depend on taking the field-theory limit of any string
theory.

Here we use the KLT relations to generate gravity tree amplitudes from
gauge theory amplitudes, which we then use as input into cutting
rules.  In this way one can compute (super) gravity loop amplitudes
without any reference to a Lagrangian or to Feynman rules.  Indeed, we
will obtain the complete two-loop amplitude for $N=8$ supergravity, in
terms of scalar integral functions, without having evaluated a single
Feynman diagram.  Furthermore the structure of the results suggests a
simple relationship between four-point $N=8$ supergravity and $N=4$
Yang-Mills amplitudes.  The multi-loop structure of the latter has
already been outlined~\cite{BRY} for the planar contributions.  We
also compute the two-loop ultraviolet divergences of $N=4$ Yang-Mills
theory in $D=7,9$ from the amplitude in ref.~\cite{BRY}, and compare
them to prior results of Marcus and Sagnotti~\cite{MarcusSagnotti};
the agreement provides a nice additional check on our general
techniques.

We investigate the multi-loop relationship between four-point $N=8$
supergravity and $N=4$ Yang-Mills amplitudes by computing the
$D$-dimensional two-particle cuts at an arbitrary order in the loop
expansion.  When evaluating the cuts, much of the algebra performed in
evaluating the cuts of $N=4$ amplitudes can be recycled to obtain the
$N=8$ supergravity cuts.  Moreover, the two-particle cut algebra iterates
to {\it all} loop orders.  Using the one-loop $N=8$ supergravity
four-point amplitude~\cite{GSB,DN} as a starting point, it takes little
effort to obtain certain `entirely two-particle constructible' terms in
the $L$-loop amplitude.

The ultraviolet divergences of theories of gravity have been under 
investigation for quite some time.  In four dimensions, pure gravity 
was shown to be finite on-shell at one loop by 't Hooft and 
Veltman~\cite{tHooftVeltmanAnnPoin,tHooftGrav},
but the addition of scalars~\cite{tHooftVeltmanAnnPoin,tHooftGrav},
fermions or photons~\cite{DeserEtal} renders it non-renormalizable at
that order.   At two loops, a potential counterterm for pure gravity of the 
form $R^3 \equiv R_{\mu\nu}^{\lambda\rho} 
R_{\lambda\rho}^{\sigma\tau} R_{\sigma\tau}^{\mu\nu}$
was identified in 
refs.~\cite{PureGravityInfinityPredictionK,PureGravityInfinityPredictionNW}.
An explicit computation by Goroff and Sagnotti~\cite{PureGravityInfinityGS}, 
and later by van de Ven~\cite{PureGravityInfinityV}, 
verified that the coefficient of this counterterm was indeed nonzero.  

On the other hand, in any supergravity theory, 
supersymmetry Ward identities (SWI)~\cite{SWI} forbid all possible
one-loop~\cite{OneLoopSUGRA} and two-loop~\cite{Grisaru} 
counterterms.  For example, the $R^3$ operator, when added to the 
Einstein Lagrangian, produces a non-vanishing four-graviton scattering 
amplitude for the helicity configuration $(-{}+{}+{}+)$, where all gravitons
are considered outgoing~\cite{PureGravityInfinityPredictionNW}.
But this configuration is forbidden by the SWI, hence $R^3$ cannot
belong to a supersymmetric multiplet of counterterms~\cite{Grisaru}.

At three loops, the square of the Bel-Robinson
tensor~\cite{BelRobinson}, which we denote by $R^4$, has been
identified as a potential counterterm in supergravity (or more
accurately, as a member of a supermultiplet of potential
counterterms)~\cite{DKS}.  This operator does not suffer from the
obvious problem that $R^3$ did, in that its four-graviton matrix
elements populate only the $(-{}-{}+{}+)$ helicity configuration which
is allowed by the ($N=1$) supersymmetry Ward identities.  In fact, the
$R^4$ operator has been shown to belong to a full $N=8$ supermultiplet
at the linearized level~\cite{KalloshNeight}.  (A manifestly
$SU(8)$-invariant form for the supermultiplet has also been
given~\cite{HSTNeight}.) Furthermore, this operator
appears in the first set of corrections to the $N=8$ supergravity
Lagrangian, in the inverse string-tension expansion of the effective
field theory for the type II superstring~\cite{GrossWitten}.
Therefore we know it has a completion into an $N=8$ supersymmetric
multiplet of operators, even at the non-linear level.  However, no
explicit counterterm computation has been performed in any
supergravity theory beyond one loop (until now), leaving it an open
question whether supergravities actually do diverge at three (or more)
loops.

The analysis of which counterterms can be generated can often
be strengthened when the theory is quantized in a manifestly
supersymmetric fashion, using superspace techniques.  In particular, 
ref.~\cite{HoweStelle} used an off-shell covariant $N=2$
superspace formalism to perform a power-counting analysis of
divergences in $N=4$ Yang-Mills theory, and ref.~\cite{HoweStelleTownsend} 
similarly used an $N=4$ superspace formalism to study $N=8$ supergravity.
However, it is not possible to covariantly quantize either of
the maximally extended theories, $N=4$ or $N=8$, while maintaining
all of the supersymmetries.
For example, in the $N=4$ Yang-Mills analysis of ref.~\cite{HoweStelle}, 
the complete $N=4$ spectrum falls into an $N=2$
gauge multiplet plus an $N=2$ matter multiplet.  The `superspace
arguments' consist of applying power-counting rules to manifestly
$N=2$ supersymmetric counterterms made out of the $N=2$ gauge
multiplet.  An important point is that these rules might not fully take 
into account all the constraints of $N=4$ supersymmetry, because only
the $N=2$ is manifest~\cite{StellePrivate};
similar remarks apply to the superspace analysis of $N=8$ supergravity.

Here we shall examine the ultraviolet divergences of $N=8$
supergravity as a function of dimension and number of loops, exploiting its
relationship to $N=4$ Yang-Mills theory.  From the exact results for
the two-loop $N=8$ amplitudes, we can extract the precise two-loop
divergences in $D=7,9,11$.  (The manifest $D$-independence of the
sewing algebra allows us to extend the calculation to $D=11$, even
though there is no corresponding $D=11$ super-Yang-Mills theory.)  For
$D<7$ there is no divergence.  This behavior is less divergent than
expected based on superspace
arguments~\cite{HoweStelleTownsend,HoweStelle}.  Moreover, we have
investigated the supersymmetry cancellations of the $L$-loop
four-point amplitudes by inspecting the general two-particle cut, plus
{\it all} cuts where the amplitudes on the left- and right-hand sides
of the cuts are maximally helicity-violating (MHV).  Our investigation
detects no divergences in $D=4$ at three or four loops, contrary to
expectations from the same types of superspace arguments.  Assuming
that the additional contributions to the cuts do not alter this power
count, we conclude that the potential $D=4$ three-loop counterterm
vanishes.

At five loops, our investigation of the cuts for the four-point $N=8$
amplitude indeed indicates a non-vanishing counterterm, of the generic form
$\partial^4 R^4$.  This suggests that $N=8$ supergravity is a
non-renormalizable theory, with a four-point counterterm arising at
five loops.  Thus our cut calculations represent the first hard
evidence that a four-dimensional supergravity theory is
non-renormalizable, {\it albeit
at a higher loop order than had been expected}.  Since superspace
power counting amounts to putting a bound on allowed divergences, our
results are compatible with the discussion of ref.~\cite{HoweStelle}.
Our results are inconsistent, however, with some earlier work
\cite{GrisaruSiegel,SuperSpace} based on the speculated existence of
an unconstrained covariant off-shell superspace for $N=8$
supergravity, which in $D=4$ would imply finiteness up to seven loops.
The non-existence of such a superspace has already been
noted~\cite{HoweStelle}.

This paper is organized as follows.  In
section~\ref{StringRelationSection} we review known tree and one-loop
relationships between gravity and Yang-Mills theory.  Following this,
in section~\ref{YMPowerCountingSection} we examine the multi-loop
$N=4$ Yang-Mills amplitudes previously discussed in ref.~\cite{BRY} as
a precursor to the $N=8$ supergravity calculations.  We will also
extract the $D=7,9$ two-loop divergences from the amplitudes
calculated in ref.~\cite{BRY} and compare them to a previous
diagrammatic calculation of Marcus and Sagnotti~\cite{MarcusSagnotti}.
This provides a non-trivial check on our methods.  Following this, in
section~\ref{SuperGravitySection} we present our result for the
two-loop four-point amplitude and relationships between $N=8$
supergravity and $N=4$ Yang-Mills amplitudes.  We then examine the
implied $D$-dimensional divergence structure for $N=8$ supergravity.
\Sec{CutConstructionSection} contains the evaluation of the
$D$-dimensional two-particle cuts to all loop orders, as well as the
three-particle cuts for the two-loop amplitudes.  In
\sec{MultiParticleCutsSection} we illustrate the supersymmetry
cancellations occurring in multi-particle MHV cuts at any loop order.
\Sec{ConclusionsSection} contains our conclusions.

Various appendices are also included.
\Apps{TwoParticleCutsAppendix}{ThreeParticleCutsAppendix} contain the
calculations of two- and three-particle cuts via helicity techniques.
The ultraviolet behavior of the two-loop scalar integrals appearing in
the amplitudes is given in \app{UVExtractionAppendix}.
\App{YMColorAppendix} describes rearrangements of the color algebra
necessary for comparing our results for the two-loop ultraviolet 
divergences of $N=4$ Yang-Mills theory in $D>4$ with those of Marcus 
and Sagnotti.  Finally, some useful on-shell supersymmetry Ward identities
are collected in \app{SusyIdentityAppendix}.


\section{Known Relationships Between Gravity and Yang-Mills Theory}
\label{StringRelationSection}

In this section, we review the known relations between the 
$S$-matrices of gravity and gauge theories.  As yet, these relationships
have only been investigated in any detail at the tree and one-loop
levels.

\subsection{Relations Between Interaction Vertices}

In field theory, if one starts from the Einstein-Hilbert Lagrangian,
$$
{\cal L}_{\rm gravity} = -{2\over\kappa^2} \sqrt{g}\, R \, , 
\equn
$$
and the Yang-Mills Lagrangian, 
$$
{\cal L}_{\rm YM} = - {1\over 4 g^2} F^a_{\mu\nu} F^{a\, \mu\nu} \, , 
\equn
$$
it is not clear how to relate amplitudes in these two theories. In
particular, pure gravity contains an infinite number of interaction vertices
and is not renormalizable in 
$D=4$~\cite{PureGravityInfinityPredictionK,PureGravityInfinityPredictionNW,
PureGravityInfinityGS,PureGravityInfinityV}. 
In contrast, gauge theories, of course, contain only up to
four-point vertices and are renormalizable in $D=4$.

Although a complete off-shell realization of the relation
(\ref{GravityYMRelation}) is not yet known, it is possible
to choose field variables which make the relation manifest for
three-point vertices.  Specifically, one may express the three-graviton 
vertex as
$$
\eqalign{
G_3{}_{\mu\alpha, \nu\beta, \rho\gamma}(k, p , q) & = -{i \over 32} \kappa
\Bigl[
V_3{}_{\mu\nu\rho}(k, p, q) \times V_{3}{}_{\alpha \beta \gamma}(k, p, q)
 + \{\mu \leftrightarrow \alpha\} , \{\nu\leftrightarrow\beta \},
     \{\rho\leftrightarrow\gamma\} \Bigr] \,, \cr }
\equn\label{GravityYMThreeVertex}
$$
where $G_3$ is the three-graviton vertex, with coupling constant
$\kappa^2 = 32\pi G_N = 32\pi/M_{\rm Planck}^2$, and $V_3$ is the Yang-Mills
three-gluon vertex (stripped of color and coupling constant factors).
In order to satisfy this relation, starting from the Lagrangians, one
must choose appropriate gauges, or more generally field
variables.

Conventional choices of gauge, such as Feynman gauge for Yang-Mills
theory and harmonic (de Donder) gauge for gravity, preclude any simple
relations between the vertices.  For example, the de Donder gauge
three-vertex is
$$
G_{3\mu\alpha,\nu\beta,\rho\gamma}(k_1,k_2,k_3) 
\sim   k_1\cdot
k_2\eta_{\mu\alpha}\eta_{\nu\beta}\eta_{\rho\gamma} +\cdots \, , 
\equn\label{deDonder}
$$
where the dots represent the many remaining terms.  The complete
vertex is given in ref.~\cite{DeWitt,tHooftVeltmanAnnPoin}, 
but for the purpose of rewriting
the vertex in the factorized form~(\ref{GravityYMThreeVertex}) the
term given in eq.~(\ref{deDonder}) is already problematic ---
it contains an $\eta_{\mu\alpha}$ trace which does not appear in the 
desired relation~(\ref{GravityYMThreeVertex}).

When both gravity and gauge theory are quantized in an appropriate 
background-field gauge~\cite{Background}, 
relationship (\ref{GravityYMThreeVertex}) does indeed
hold~\cite{BDS}. (The observation that the field-theory limit of
one-loop string theory amplitudes~\cite{Long} closely resembles the
form obtained when using background-field gauges was made in
ref.~\cite{BernDunbar}.)  The Feynman-gauge background-field Yang-Mills
three-vertex is
$$
V_3{}_{\mu\nu\rho}(k, p , q) = \eta_{\nu\rho} (p - q)_\mu
- 2\eta_{\mu\rho} k_\nu +  2\eta_{\mu\nu}k_\rho \,,
\equn\label{Gauge3Vertex}
$$
with $k$ the momentum of the background-field line and $p$
and $q$ the momenta of the internal lines.  With the equivalent
Feynman-gauge background-field choice in gravity, the three-vertex $G_3$
is of the desired form~(\ref{GravityYMThreeVertex}), with $V_3$ given by
eq.~(\ref{Gauge3Vertex}).

In principle, this process of adjusting the field variables of
both gauge and gravity theories to make the relationship of gravity
and gauge theories more apparent can be continued, but the process
becomes increasingly tedious.  One would need to rearrange an 
infinite set of gravity vertices in terms of Yang-Mills vertices which
have been combined via the cancellation of propagators.  Instead of
examining off-shell vertices, we will proceed by examining the
relationships between gauge-invariant $S$-matrix elements.

\subsection{Kawai-Lewellen-Tye Tree-Level Relations}

At tree-level, KLT have given a complete description of the
relationship between closed string amplitudes and open string
amplitudes.  This relationship arises because any closed string vertex
operator is a product of open string vertex operators,
$$
V^{\rm closed} = V_{\rm left}^{\rm open}\, 
\overline{V}_{\rm right}^{\rm  open} \,.
\equn\label{ClosedVertex}
$$
The left and right string oscillators appearing in $V_{\rm left}$ and
$\overline{V}_{\rm right}$ are distinct, but the zero mode momentum is
shared.  This property of the string vertex operators is then
reflected in the amplitudes.

For example, the open string amplitude for gluons is
$$
A_n \sim \int {dx_1 \cdots d x_n \over   {\cal V}_{abc} }
\prod_{1  \le i < j \le n}  |x_ i - x_j|^{k_i \cdot k_j}
\exp\biggl[ \sum_{i <j} \Bigl({\pol_i \cdot \pol_j \over (x_i - x_j)^2}
                       + {k_i \cdot \pol_j - k_j \cdot \pol_i
                          \over (x_i - x_j)} \Bigr) \biggr] \,
                        \biggr|_{\rm multi-linear} \,,
\equn\label{OpenKN}
$$
where
$$
{\cal V}_{abc} = {dx_a\, dx_b\, dx_c \over |(x_a - x_b)
     (x_b - x_c) (x_c-x_a)|} \, , 
\equn
$$
and $x_a,x_b,x_c$ are any three of the $x_i$.  In \eqn{OpenKN}
we have suppressed the inverse string tension $\alpha'$, and 
the `multi-linear' denotes that after expanding the exponential one only
keeps terms linear in each polarization vector $\pol_i$.

The corresponding $n$-graviton tree amplitude amplitude in string theory
is
$$
\eqalign{
M_n & \sim \int {d^2z_1 \cdots d^2 z_n \over  \Delta_{abc} }
\prod_{1  \le i < j \le n}  (z_ i - z_j)^{ k_i \cdot k_j}
\exp\biggl[ \sum_{i <j} \Bigl({\pol_i \cdot \pol_j \over (z_i - z_j)^2}
                       + {k_i \cdot \pol_j - k_j \cdot \pol_i
                          \over (z_i - z_j)} \Bigr) \biggr] \cr
& \hskip 1 cm \times
\prod_{1  \le i < j \le n}  (\zb_ i - \zb_j)^{ k_i \cdot k_j}
\exp\biggl[ \sum_{i <j} \Bigl({\overline{\pol}_i \cdot 
\overline{\pol}_j \over (\zb_i - \zb_j)^2}
+ {k_i \cdot \overline{\pol}_j - k_j \cdot \overline{\pol}_i
                          \over (\zb_i - \zb_j)} \Bigr) \biggr] 
          \Bigr|_{\rm multi-linear} \,, 
\cr}
\equn\label{ClosedKN}
$$
where
$$
\Delta_{abc} = {d^2z_a \, d^2 z_b \, d^2 z_c \over
            |z_a - z_b|^2 |z_b - z_c|^2 |z_c-z_a|^2 } \,,
\equn
$$
$z_a,z_b,z_c$ are any three of the $z_i$, and `multi-linear'
means linear in each $\pol_i$ and each $\overline{\pol}_i$.
In this expression we have taken the graviton 
polarization vector to be a product of gluon
polarization vectors
$$
\pol^{\mu\nu}_i = \pol^\mu_i \, \overline{\pol}^\nu_i \,.
\equn
$$
(We distinguish between $\pol$ and $\bar{\pol}$ merely as a convenience.)
It is not difficult to verify that in a helicity basis~\cite{SpinorHelicity}
this expression satisfies all required properties of a graviton 
polarization vector, including its tracelessness.

The closed string integrand in \eqn{ClosedKN} is a product of two open
string integrands. This factorization into products of holomorphic and
anti-holomorphic integrands is a generic feature of any closed string
tree-level amplitude and does not depend on whether the external particles
under consideration are fermions or bosons, as long as the closed string
states are tensor products of open string states, so that the 
vertex operator relation~(\ref{ClosedVertex}) can be applied.

Using these string expressions, and carrying out various
contour-integral deformations, KLT obtained relationships between the
tree-level closed and open string amplitudes after all $z_i$ and $x_i$
integrations have been performed.  In ref.~\cite{BGK} the KLT
relations were used to obtain explicit formulae for field-theory
$n$-point MHV graviton amplitudes at tree level.  After taking the 
field-theory limit, the KLT relations for four- and five-point amplitudes are, 
$$
M_4^{\rm tree} (1,2,3,4) =  
     - i s_{12} A_4^{\rm tree} (1,2,3,4) \, A_4^{\rm tree}(1,2,4,3)\,, 
\equn\label{GravYMFour}
$$
$$
\eqalign{
M_5^{\rm tree}
(1,2,3,4,5) & = i s_{12} s_{34}  A_5^{\rm tree}(1,2,3,4,5)
                                     A_5^{\rm tree}(2,1,4,3,5)  \cr
& \hskip 2 cm 
             + i s_{13}s_{24} A_5^{\rm tree}(1,3,2,4,5) \, 
                           A_5^{\rm tree}(3,1,4,2,5) \,, \cr}
\equn\label{GravYMFive}
$$
where the $M_n$'s are the amplitudes in a gravity theory stripped of 
couplings,
the $A_n$'s are the color-ordered  amplitudes in a gauge
theory and $s_{ij}\equiv (k_i+k_j)^2$.  We have suppressed all
$\pol_j$ polarizations and $k_j$ momenta, but have kept the `$j$'
labels to distinguish the external legs.  These 
combine to give the full amplitude via,
$$
\eqalign{
{\cal M}_n^{\rm tree}(1,2,\ldots n) &= 
\left({  \kappa \over 2} \right)^{(n-2)} 
M_n^{\rm tree}(1,2,\ldots n)\,,
\cr
{\cal A}_n^{\rm tree}(1,2,\ldots n) &=  g^{(n-2)} \sum_{\sigma \in S_n/Z_n}
{\rm Tr}\left( T^{a_{\sigma(1)}} 
T^{a_{\sigma(2)} }\cdots  T^{a_{\sigma(n)}} \right)
 A_n^{\rm tree}(\sigma(1), \sigma(2),\ldots, \sigma(n)) \,,
\cr}
\equn
$$
where $S_n/Z_n$ is the set of all permutations, but with cyclic
rotations removed. The $T^{a_i}$ are fundamental representation 
matrices for the Yang-Mills gauge group $SU(N_c)$, normalized so that
$\Tr(T^aT^b) = \delta^{ab}$.  
(For more detail on the tree and one-loop color ordering of
gauge theory amplitudes see refs.~\cite{ManganoReview,Color}.)  
Our phase conventions differ from those of ref.~\cite{BGK} in that we have
introduced an `$i$' into the gravity tree amplitudes; this is to
maintain consistency with Minkowski-space Feynman rules.  In the
sewing procedure, which we use in later sections, these overall phases are
unimportant, since one can anyway fix phases at the end of the
calculation.

It is sometimes convenient to write the four-point Yang-Mills amplitude 
in terms of structure constants as 
$$
{\cal A}_4^{\rm tree}(1,2,3,4) =  g^2 \biggl[
 \tilde{f}^{a_2a_3c}\tilde{f}^{ca_4a_1} A_4^{\rm tree}(1,2,3,4)
+  \tilde{f}^{a_1a_3c}\tilde{f}^{ca_4a_2} A_4^{\rm tree}(2,1,3,4) \biggr]\,,
\equn\label{FourPointColor}
$$
where 
$$
\tilde{f}^{abc} = i\sqrt{2} f^{abc} = \Tr\bigl( [T^a,T^b] T^c \bigr)\,,
\equn\label{tildefabc}
$$
and $U(1)$ decoupling and reflection identities \cite{ManganoReview} have 
been used.

The relations~(\ref{GravYMFour}) and (\ref{GravYMFive}) hold for any
external states of a closed string and not just gravitons.  
We can thus obtain the amplitudes containing, for example, gravitinos:
$$
M_4^{\rm tree} (1_h,2_h,3_{\tilde h},4_{\tilde h}) =  
     -i s_{12} A_4^{\rm tree} (1_g,2_g,3_g,4_g) \, 
A_4^{\rm tree}(1_g,2_g,4_{\tilde g},3_{\tilde g} )\,, 
\equn\label{GravitinoFour}
$$
In this equation we have labeled the particle types of the $N=8$
multiplet, graviton ($h$), spin-$3/2$ gravitino (${\tilde h}$), vector
($v$), spin-$1/2$ fermion (${\tilde v}$) and scalar ($s$).  The labels
of the $N=4$ Yang-Mills multiplet are: gluon ($g$), gluino ($\tilde
g$) and scalar ($s$).  In general, a closed string state in four dimensions
with helicity $\lambda$ is composed of a `left' state of helicity $\lambda_L$
and a `right' state of helicity $\lambda_R$, where
$\lambda=\lambda_L+\lambda_R$.  We can then obtain the gravity
amplitudes for states of helicity $\{ \lambda_i \}$ in terms of gauge theory
amplitudes with states of helicity $\{ \lambda_{L,i} \}$ and 
$\{ \lambda_{R,i} \}$ just as in \eqn{GravitinoFour}.  
In an arbitrary dimension $D$, similar relations
hold, where the $N=8$ states are constructed as tensor products of $N=4$
states, in terms of representations of the massless little group $SO(D-2)$.

As a concrete example, consider the four gluon amplitude given by
$$
A^{\rm tree}_4 (1,2,3,4) 
= {- 4i  \over s t}\,  (t_8)_{\mu_1\nu_1\mu_2\nu_2\mu_3\nu_3\mu_4\nu_4} 
k_1^{\mu_1} k_2^{\mu_2} k_3^{\mu_3} k_4^{\mu_4}
\pol_1^{\nu_1}\pol_2^{\nu_2}\pol_3^{\nu_3}\pol_4^{\nu_4}
\equiv  {-4 i \, K  \over s t} \,,
\equn\label{TreeTensor}
$$
where $s= s_{12}$, $t= s_{14}$ and $u= s_{13}$ are
the usual Mandelstam variables and
where the tensor $t_8$ is defined in eq. (9.A.18) of ref.~\cite{GSW},
except that the term containing the eight-dimensional Levi-Civita
tensor should be dropped.  The kinematic factor $K$ is totally
symmetric under interchange of external legs.  Applying the relation
(\ref{GravYMFour}) yields the four-graviton amplitude
$$
M^{\rm tree}_4 (1,2,3,4) 
={ 16 i\,  K^2 \over stu} \,.
\equn
$$
A relation we will use later is 
$$
s t u M_4^{\rm tree}(1,2,3,4) \, = - i \left(st \, [A_4^{\rm tree}(1,2,3,4)]
\right)^2\,.
\equn\label{OneLoopRelation}
$$

These expressions may be made more concrete in four dimensions.
The only non-vanishing four-gluon helicity amplitudes at tree-level
(and at any loop order in supersymmetric Yang-Mills theory~\cite{SWI}),
are those with two positive and two negative gluon helicities.
All of these configurations are trivially related in $N=4$ Yang-Mills theory.
The one independent tree amplitude is
$$
A_4^{\rm tree}(1^-, 2^-, 3^+, 4^+) 
= i {\spa1.2^4 \over \spa1.2\spa2.3\spa3.4\spa4.1}\,,
\equn
$$
using the spinor helicity formalism~\cite{SpinorHelicity} to represent
the amplitudes.  (The reader may wish to consult review articles for
details \cite{ManganoReview}.)  
The $\pm$ superscripts label the helicity of the external gluon.  In
general, we will drop the $g$ subscript from a gluon leg in $A_n$ and
the $h$ subscript from a graviton leg in $M_n$. 
We use the notation $\la k_i^- |
k_j^+\ra = \spa{i}.j$ and $\la k_i^+| k_j^-\ra=\spb{i}.j$, where
$|k_i^{\pm}\ra$ are massless Weyl spinors, labeled with the sign of
the helicity and normalized by 
$\spa{i}.j \spb{j}.i = s_{ij} = 2k_i\cdot k_j$.
In this formalism the one nonvanishing four-graviton tree amplitude is
$$
M_4^{\rm tree}(1^-, 2^-, 3^+, 4^+)  =  - i s_{12} {\spa1.2^8 \over 
        \spa1.2^2\spa2.3\spa2.4\spa3.4^2 \spa3.1 \spa4.1}  \,.
\equn\label{CorrectFourGraviton}
$$

We can rewrite these amplitudes in a form which clarifies the relationship 
between gauge and gravity theories.   Firstly we may rearrange the 
Yang-Mills result,
$$
\eqalign{
& A_4^{\rm tree}(1^-, 2^-, 3^+, 4^+) =  -{i \over t}\, 
\times 
   \left[ s {\spa1.2\over \spb1.2} \, {\spb3.4 \over \spa3.4} \right]
\,, \cr 
& A_4^{\rm tree}( 1^-, 2^-, 4^+, 3^+) =  - {i \over u}\, 
\times
 \left[  s {\spa1.2\over \spb1.2} \, {\spb3.4 \over \spa3.4}  \right]
\,. \cr}
\equn\label{YMPoleForm}
$$
We have chosen to write the amplitudes 
as a pole times a non-singular term.

\begin{figure}[ht]
\centerline{\epsfxsize 2.8 truein \epsfbox{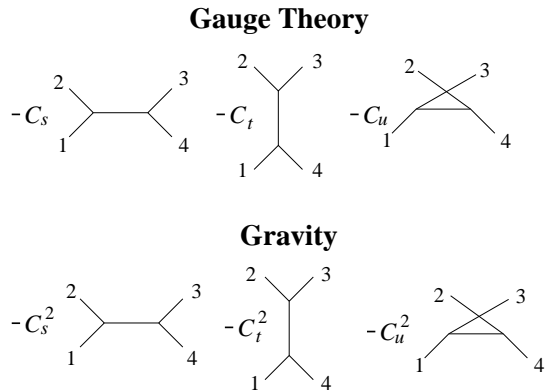}}
\vskip -.2 cm
\caption[]{
\label{TreeRelationFigure}
\small Tree-level gauge theory four-gluon amplitudes and gravity
four-graviton amplitudes expressed in terms of scalar $\phi^3$
diagrams.  The coefficients of the graviton amplitudes are squares of
the coefficients for four-gluon amplitudes.}
\end{figure}

In gravity we expect all algebraic factors associated with vertices to
be squares of corresponding gauge theory factors. However, the
propagators of gravity are the same propagators as in gauge theory.
This suggests that we can obtain the corresponding pure gravity
amplitude by squaring all kinematic factors in \eqn{YMPoleForm} except
for the poles and summing over permutations.  As a test of this
expectation note that the gravity amplitude
(\ref{CorrectFourGraviton}) can be rewritten in the form
$$
M_4^{\rm tree}(1^-, 2^-, 3^+, 4^+) =   -i
 \biggl( {1\over t} + {1\over u} \biggr) \biggl[s \, 
   {\spa1.2\over \spb1.2} \, {\spb3.4 \over \spa3.4} \biggr]^2 \, 
 \,,
\equn\label{TreeSquared}
$$
where the coefficient of each pole
is the square of the coefficients appearing in
\eqn{YMPoleForm}.  This relationship may be expressed in terms of
coefficients of $\phi^3$ scalar diagrams as we have done in
\fig{TreeRelationFigure}. (The value of the $\phi^3$ scalar diagram 
in the upper left of \fig{TreeRelationFigure} is $i/s$.)
From \eqns{YMPoleForm}{TreeSquared} we can read off 
the coefficients of
the $\phi^3$ scalar diagrams for $A_4^{\rm tree}(1^-, 2^-, 3^+, 4^+)$ 
and $M_4^{\rm tree}(1^-, 2^-, 3^+, 4^+)$:
$$
C_s = 0 \,, \hskip 2 cm C_t = C_u = s \, 
   {\spa1.2\over \spb1.2} \, {\spb3.4 \over \spa3.4} \,.
\equn
$$

Similar relations hold for supergravity amplitudes where some
of the gravitons are replaced by other states, for example the 
amplitudes 
$M_4^{\rm tree}(1^{-\lambda_L - \lambda_R}, 2^-, 
3^+, 4^{+\lambda_L+\lambda_R})$
where the gauge theory amplitudes are 
$A_4^{\rm tree}(1^{-\lambda}, 2^-, 3^+, 4^{+\lambda})$ and where 
$$
C_s = 0, \hskip 2 cm C_t = C_u =  
\Bigl({\spa4.2 \over \spa1.2} \Bigr)^{2(1-\lambda)}\times  \,
\left[ s\, 
 {\spa1.2\over \spb1.2} \, {\spb3.4 \over \spa3.4} \right]
\,.
\equn
$$
Here $M_4^{\rm tree}(1^{-\lambda_L - \lambda_R}, 2^-, 3^+, 4^{+\lambda_L+\lambda_R})$ is a
product of a `left' gauge theory factor times a `right' gauge theory
factor depending on the decomposition of each state in the gravity
theory in terms of gauge theory states.

In the remaining sections of this paper, we shall argue that, at least
for $N=8$ supergravity, similar relations hold to all loop orders.

%

\subsection{Relationships between Loop Amplitudes}

At one loop it is more difficult to take the field-theory limit of a
string.  Nevertheless, a technology exists for systematically
extracting such limits \cite{Long}.  This approach was followed in
refs.~\cite{BDS,DN} to compute four-graviton amplitudes
in a variety of supersymmetric and non-supersymmetric field theories.  
A feature of these calculations is that at the very beginning the loop
integrands for gravity are products of two Yang-Mills integrands.

This feature is clear when the one-loop string amplitudes are
expressed in terms of a loop momentum (i.e. bosonic zero mode)
integral.  An open string scattering amplitude for vector states is of
the form,
$$
\eqalign{
 & \int  {d^D p \over (2\pi)^D } \;  
{\exp\bigl[
\pol_i \cdot p_i  \bigr] \times ({\rm  oscillator\ contributions}) 
     \bigr|_{\rm multi-linear}
\over  p_1^2 p_2^2 \cdots p_n^2} 
 \,, \cr}
\equn\label{OpenStringMomIntegrals}
$$
where
$$
p_i = p - k_1 - k_2 - \cdots - k_{i-1} = p + k_i + \cdots + k_{n} \,.
\equn
$$
The $k_i$ are the momenta of the external vectors and $p$ is the loop
momentum.  The oscillator contributions depend on the external momenta,
but are independent of $p$.  In a supersymmetric theory there are
cancellations between the contributions from the different particle
types circulating in the loop. The supersymmetric cancellations occur
in the oscillator terms.  (For example, the cancellation of leading
powers of loop momentum follows from the vanishing of the string
partition function.)  The closed string graviton scattering amplitude
is of the form
$$
 \int  {d^D p \over (2\pi)^D } \;
{\exp\bigl[
\pol_i \cdot p_i  + \bar\pol_i\cdot p_i\bigr] 
\times ({\rm left\ oscillator\ contrib.}) 
\times ({\rm right\  oscillator\ contrib.}) \bigr|_{\rm multi-linear}
\over  p_1^2 p_2^2 \cdots p_n^2}
\,.
\equn\label{StringMomIntegrals}
$$
This generic structure of any one-loop closed string amplitude follows
from the factorization of a closed string vertex operator into a
product of left- and right-mover vertex operators given in
\eqn{ClosedVertex}. This structure leads us to look for similar
relations for loop amplitudes as the tree-level ones in
\fig{TreeRelationFigure}.

\begin{figure}[ht]
\centerline{\epsfxsize 1.8 truein \epsfbox{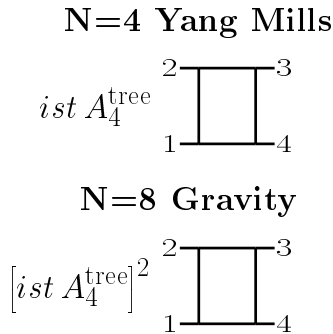}}
\vskip -.2 cm
\caption[]{
\label{OneLoopRelationFigure}
\small One-loop gauge theory and gravity four-point contributions
expressed in terms of scalar diagrams.  The coefficient of a given
scalar diagram in the $N=8$ four-graviton amplitude is the square of the
corresponding coefficient in the $N=4$ four-gluon amplitude.}
\end{figure}

For the cases of $N=4$ Yang-Mills theory and $N=8$ supergravity the
one-loop amplitudes are known~\cite{GSB}, so we can easily identify
any relations.  The one-loop $N=4$ amplitude may be expressed
as~\cite{GSB,Review}
$$
\eqalign{
{\cal A}_4^{N=4,{\oneloop} }
(1,2,3,4) & = i \, g^4 \, s_{12} s_{23} A_4^{\rm tree}(1,2,3,4)
    \Bigl( C_{1234}\, \I_4^{\oneloop}(s_{12},s_{23})  
    + C_{3124}\, \I_4^{\oneloop}(s_{12},s_{13})    \cr
& \hskip 4 cm 
    + C_{2314}\, \I_4^{\oneloop}(s_{23},s_{13})   \Bigr)\,, \cr}
\equn\label{OneLoopYMResult}
$$
where $C_{1234}$ is the color factor obtained by dressing each 
diagram in \fig{OneLoopRelationFigure} with a structure constant
$\tilde{f}^{abc}$, and each bond between vertices with a
$\delta^{ab}$.  The integral functions are defined as
$$
\I_4^{\oneloop}(s_{12},s_{23}) = \int {d^D p \over (2\pi)^D
} \; 
{1\over p^2 (p-k_1)^2 (p-k_1-k_2)^2 (p+k_4)^2 } \, .
\equn\label{OneLoopBox}
$$
An explicit representation of this integral in terms of hypergeometric
functions may be found in, for example, ref.~\cite{OneLoopInt}. The
amplitude~(\ref{OneLoopYMResult}) effectively was calculated in the
dimensional reduction scheme \cite{Siegel}, which preserves
supersymmetry.

Similarly, the one-loop $N=8$ four-graviton amplitude is,
$$
\eqalign{
{\cal M}_4^{N=8,{\oneloop} }(1, 2, 3, 4)
& =  -i \Bigl( {\kappa \over 2}\Bigr)^4 
s_{12} s_{23} s_{13} M_4^{\rm tree}(1,2,3,4)
 \Bigl(  \I_4^{\oneloop}(s_{12},s_{23})   
           + \I_4^{\oneloop}(s_{12},s_{13})  \cr
& \hskip 4 cm
           + \I_4^{\oneloop}(s_{23},s_{13})  \Bigr) \,. \cr}
\equn\label{OneLoopGravResult}
$$
These amplitudes were first obtained by Green, Schwarz and 
Brink~\cite{GSB} in the field-theory limit of superstring theory. 

From~\eqn{OneLoopRelation}, the coefficient of the scalar box integral
in the $N=8$ gravity amplitude (\ref{OneLoopGravResult}) is precisely
the square of the coefficient appearing in the gauge theory amplitude
(\ref{OneLoopYMResult}) (ignoring couplings and color factors), as
depicted in \fig{OneLoopRelationFigure}.  
These equations hold for amplitudes
with {\it any} external particles in the supermultiplet.  Furthermore,
the results of ref.~\cite{GSB} are valid in any dimension.

%

Although no systematic study has as yet been performed, at one loop
string theory does suggest that similar relations should hold in
gravity theories other than $N=8$ supergravity~\cite{BDS}.  Moreover,
at any loop order closed string amplitudes continue to be expressible
in a form where the integrands are products of `left' and `right'
factors.  This leads one to suspect that in the field-theory limit
there are relationships of the form (\ref{GravityYMRelation}) to all
orders of the perturbative expansion.  However, since the field-theory
limit of a multi-loop string amplitude is not known in a convenient
form, we will use cutting rules to study the relations between $N=4$
Yang-Mills theory and $N=8$ supergravity.


\section{$N=4$ Yang-Mills Amplitudes}
\label{YMPowerCountingSection}

In ref.~\cite{BRY} the two-loop four-point amplitudes for $N=4$
Yang-Mills theory were computed in terms of scalar integral functions
via cutting methods.  Furthermore, from an inspection of the
two-particle cuts, a conjecture for the planar parts of the four-point
amplitude was presented to all loop orders.  In this section we
examine the results of ref.~\cite{BRY} in preparation for the
analogous construction for $N=8$ supergravity.  We will, in addition,
extract the two-loop counterterms in various dimensions implied by the
$N=4$ amplitudes.  These counterterms were previously obtained by
Marcus and Sagnotti~\cite{MarcusSagnotti} via an explicit Feynman
diagram calculation, using a specialized computer program.  Comparison
to their calculation provides a non-trivial two-loop check on our
methods.  The results of ref.~\cite{BRY} are more general,
being the complete amplitudes and not just divergences.  We also
comment that a comparison of the two calculations illustrates the
computational efficiency of the cutting techniques: the 
calculation of the complete amplitudes in terms of scalar integrals can 
easily be performed without computer assistance.  Moreover, the
technicalities associated with overlapping divergences are
alleviated.  We will apply the same cutting techniques to obtain new
results for $N=8$ supergravity.

\begin{figure}[ht]
\centerline{
\epsfxsize 3. truein \epsfbox{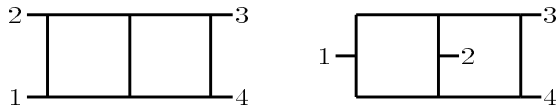}}
\vskip -.2 cm
\caption[]{
\label{PlanarNonPlanarFigure}
\small
The planar and non-planar scalar integrals. }
\end{figure}

The two-loop amplitude is given by
$$
\eqalign{ 
& {\cal A}_4^{\twoloop}(1,2,3,4) 
=  - g^6 s_{12} s_{23} \, 
              A_4^{\rm tree}(1,2,3,4) \Bigl(
    C^{\P}_{1234} \, s_{12} \, \I_4^{\twoloop,\P}(s_{12}, s_{23}) 
  + C^{\P}_{3421} \, s_{12} \, \I_4^{\twoloop,\P}(s_{12}, s_{24}) 
\cr
\null & \hskip 4.0 truecm  
  + C^{\NP}_{1234} \, s_{12}\, \I_4^{\twoloop , \NP}(s_{12},s_{23}) 
  + C^{\NP}_{3421} \, s_{12}\, \I_4^{\twoloop , \NP}(s_{12},s_{24})
+  {\rm cyclic} \Bigr)\,, }
\equn\label{TwoLoopYM}
$$
where `$+$~cyclic' instructs one to add the two cyclic permutations of
(2,3,4) and
$$
\eqalign{
 \I_4^{\twoloop,  \P}(s_{12},s_{23}) &= \int
 {d^{D}p\over (2\pi)^{D}} \;
 {d^{D}q\over (2\pi)^{D}} \;
 {1\over p^2 \, (p - k_1)^2 \,(p - k_1 - k_2)^2 \,(p + q)^2 q^2 \,
        (q-k_4)^2 \, (q - k_3 - k_4)^2 } \,, \cr
\I_4^{\twoloop , \NP}(s_{12},s_{23}) & = \int {d^{D} p \over (2\pi)^{D}} \, 
            {d^{D} q \over (2\pi)^{D}} \
{1\over p^2\, (p-k_2)^2 \,(p+q)^2 \,(p+q+k_1)^2\,
  q^2 \, (q-k_3)^2 \, (q-k_3-k_4)^2} \, ,  \cr}
\equn\label{TwoLoopScalarInts}
$$
are the planar and non-planar scalar integrals
(\ref{TwoLoopScalarInts}) depicted in \fig{PlanarNonPlanarFigure}.
The group theory factors $C^{\P}_{1234}$ and $C^{\NP}_{1234}$ are
obtained by dressing the diagrams in \fig{PlanarNonPlanarFigure} with
$\tilde{f}^{abc}$ factors at each vertex.  (In
ref.~\cite{BRY} the result was presented in a color decomposed form,
but to facilitate a comparison to the results of Marcus and Sagnotti
we choose not to do so here.)  The massless scalar integral functions
$\I_4^{\twoloop, \P}$ and $\I_4^{\twoloop , \NP}$ are not, as yet,
known in terms of elementary functions~\cite{TwoLoopIntegrals};
nevertheless, they are definite functions which can be manipulated and
whose divergences can be extracted.

%

As discussed in ref.~\cite{BRY}, for the case of $N=4$ Yang-Mills theory at
two loops all the ambiguities of constructing amplitudes from cuts were
resolved and eq.~(\ref{TwoLoopYM}) contains all terms to all
orders in the dimensional regulating parameter $\eps$. This allows us
to continue the amplitude to arbitrary dimension.

\subsection{Two Loop Ultraviolet Infinities and Counterterms}

Although $N=4$ Yang-Mills theory is ultraviolet finite in four dimensions,
for $D>4$ the theory is non-renormalizable.  We can use \eqn{TwoLoopYM} to
extract the two-loop $N=4$ counterterms in dimensions $D>4$.  
Before proceeding to two loops, we recall that the one-loop amplitude,
\eqn{OneLoopYMResult}, 
first diverges at $D=8$, and this divergence is proportional
to $st A^{\rm tree}$.  The corresponding gluonic counterterm is fixed
by supersymmetry to be
$$
\eqalign{
t_8 F^4 &\equiv
t_8^{\mu_1\nu_1\mu_2\nu_2\mu_3\nu_3\mu_4\nu_4}
F^a_{\mu_1\nu_1}F^b_{\mu_2\nu_2}F^c_{\mu_3\nu_3}F^d_{\mu_4\nu_4} C_{abcd}\cr
&= 4! \, \left( F^a_{\alpha\beta} F^{b\,\beta\gamma}
F^c_{\gamma\delta} F^{d\,\delta\alpha} 
- {1\over4} F^a_{\alpha\beta}
F^{b\,\alpha\beta} F^c_{\gamma\delta} F^{d\,\gamma\delta} \right)
 C_{abcd} \,, \cr}
\equn\label{Ffourterms}
$$
where $C_{abcd}$ is a group theory factor 
(which we shall mostly suppress for clarity).   In four dimensions we 
can rewrite this term as
$$
t_8 F^4 = {\textstyle{3\over2}} ( F-\tilde F)^2 ( F+\tilde F)^2 
= {\textstyle{3\over2}} \Bigl( F_{\alpha\beta}
-\tilde F_{\alpha\beta} \Bigr) \Bigl( F^{\alpha\beta} -\tilde
F^{\alpha\beta} \Bigr) \Bigl( F_{\gamma\delta} +\tilde
F_{\gamma\delta} \Bigr) \Bigl( F^{\gamma\delta} +\tilde
F^{\gamma\delta} \Bigr) \,,
\equn
$$
where $\tilde F$ is the dual of $F$. 
The full counterterm also includes the scalar and fermionic operators 
obtained by the $N=4$ completion of the $F^4$ terms~\cite{MarcusSagnotti}.
The two-loop counterterms will be specified in terms of derivatives acting 
on $t_8 F^4$.

We may extract the coefficient of the counterterm from the ultraviolet
divergences in our amplitude.  In general, to extract a counterterm
from a two-loop amplitude one must take into account sub-divergences
and one-loop counterterms.  Indeed, the divergences in
ref.~\cite{MarcusSagnotti} received contributions from a large number
of two-loop graphs with diverse topologies, many of which contained
sub-divergences (i.e. $1/\e^2$ poles) which required subtraction
before the $1/\e$ poles could be extracted.  The cancellation of
$1/\e$ poles in the $D=6$ theory occurred only after summing all
diagrams and taking into account one-loop off-shell counterterms.

In our case, however, \eqn{TwoLoopYM} is manifestly finite in $D=6$,
since both planar and non-planar double-box integrals first diverge in
$D=7$.  The manifest finiteness in $D=6$ is not an accident and is due
to the lack of off-shell sub-divergences when using on-shell cutting
rules. It is also simple to extract the $D=7$ and $D=9$ counterterms,
which are the ones evaluated in ref.~\cite{MarcusSagnotti}, by
evaluating the ultraviolet poles of the scalar integrals
(\ref{TwoLoopScalarInts}).  The evaluation of these poles
is performed in \app{UVExtractionAppendix}.

The counterterm in $D=7$ is of the form  $\partial^2 F^4$, as can
be seen from dimensional analysis.  The form is again unique (at the
on-shell linearized level) 
$$
t_8^{\mu_1\nu_1\mu_2\nu_2\mu_3\nu_3\mu_4\nu_4}
\partial_\alpha F^a_{\mu_1\nu_1}\partial^\alpha
F^b_{\mu_2\nu_2}F^c_{\mu_3\nu_3}F^d_{\mu_4\nu_4} \tilde C_{abcd} \,,
\equn\label{Fdfourterms}
$$
where $\tilde C_{abcd}$ is another group theory factor.
Again, in $D=4$ such a tensor takes on a simple schematic form,
$$
 ( F-\tilde F)^2 \partial^2 ( F+\tilde F)^2  \,.
\equn\label{GoodCounterTerm}
$$
In the notation of Marcus and Sagnotti \cite{MarcusSagnotti}
the counterterm is presented as
$$
T_D \Bigl( F_{\alpha\beta} F^{\beta\gamma} F_{\gamma\delta} 
F^{\delta \alpha}-{1\over4}F_{\alpha\beta} F^{\alpha\beta} F_{\gamma\delta}
F^{\gamma\delta} + \cdots \Bigr) \,.
\equn
$$
From the expression (\ref{TwoLoopYM}) and \app{UVExtractionAppendix}
we obtain
$$
\eqalign{
T_7&  = - {g^6 \, \pi \over (4\pi)^7 \, 2\e} \biggl[ s \biggl( {1 \over 10} 
(C^{\P}_{1234} + C^{\P}_{1243})+{2 \over 15} \, C^{\NP}_{1234} \biggr) 
+ {\rm cyclic} \biggr], \cr 
T_9& =  -{g^6 \, \pi \, s \over (4\pi)^9 \, 4\e}
\biggl[  {1\over99792} (-45s^2+18st+2t^2) \, C^{\P}_{1234}
\ +\  {1\over99792} (-45s^2+18su+2u^2)\, C^{\P}_{1243} \cr 
& \hskip 2 cm  -\ {2\over83160} (75s^2+2tu) \, C^{\NP}_{1234}
\biggr]\ +\ \hbox{cyclic}, \cr
}
\equn\label{SUSYdiv}
$$
corresponding to the $D=7$ and $D=9$ counterterms.  

To compare \eqn{SUSYdiv} to the results of ref.~\cite{MarcusSagnotti} we
must rearrange the group theory factors to coincide with their basis.  The
necessary rearrangements are presented in \app{YMColorAppendix}. 
Comparing \eqn{SevenNineResults} with eqs.~(4.5) and (4.6) of 
ref.~\cite{MarcusSagnotti} we find that all 
the relative factors agree (up to a typographical error in eq.~(4.6) 
in which the tree group theory factors accompanying the $s^3$ and $t^3$
factors were exchanged).  After accounting for a different normalization
of the operator $t_8 F^4$, as deduced from the one-loop case, the overall
factor for $T_9$ also agrees, while our result for $T_7$ is larger than
that in ref.~\cite{MarcusSagnotti} by a factor of $3/2$.
Nevertheless, the agreement of the relative
factors is rather non-trivial and provides a strong check that 
the amplitude in \eqn{TwoLoopYM} is correct.

\subsection{Higher Loop Structure}
\label{HigherLoopStructureYMSubSection}

\begin{figure}[ht]
\centerline{\epsfxsize 1.0 truein \epsfbox{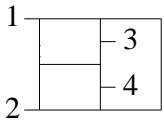} }
\vskip -.2 cm
\caption[]{
\label{NonTwoParticleGraph}
\small This three-loop non-planar graph is not `entirely two-particle
constructible'.  In fact, it has no two-particle cuts at all.}
\end{figure}

As shown in ref.~\cite{BRY} the two-particle cut sewing equation is
the same at any loop order, allowing one to iterate the sewing algebra
to all loop orders. As we discuss in \sec{TwoParticleSubSection}, the
two-particle cuts were performed to all orders in the dimensional
regularization parameter $\e$, and are therefore valid in any
dimension.  However, since this construction is based only on
two-particle cuts it is only reliable for integral functions which can
be built using such cuts.  We call a function which is successively
two-particle reducible into a set of four-point trees
`entirely two-particle constructible'.  Such contributions can be 
both planar and non-planar.  (Planar topologies give the leading
Yang-Mills contributions for a large number of colors.)
All two-loop contributions, and the three-loop contributions shown in
\fig{ThreeLoopExampleFigure}, are entirely two-particle constructible.
An example of a three-loop non-planar graph which is not entirely 
two-particle constructible is given in \fig{NonTwoParticleGraph}.

%

\begin{figure}[ht]
\centerline{\epsfxsize 3.7 truein \epsfbox{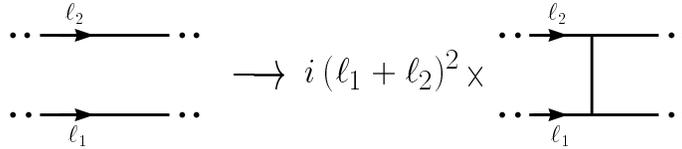} }
\vskip -.2 cm
\caption[]{
\label{AddLineFigure}
\small Starting from an $L$-loop planar integral function appearing in
an $N=4$ Yang-Mills amplitude we may add an extra line, using this
rule.  The two-lines on the left represent two lines buried in some
$L$-loop integral.}
\end{figure}

The two-particle cut sewing equation leads to a loop-momentum factor
insertion rule for planar contributions \cite{BRY}, as
shown in \fig{AddLineFigure}.  The pattern is that one
takes each $L$-loop graph in the $L$-loop amplitude and generates all
the possible $(L+1)$-loop graphs by inserting a new leg between each
possible pair of internal legs. Diagrams where triangle or bubble
subgraphs are created should not be included.  The new loop momentum
is integrated over, after including an additional factor of 
$i (\ell_1+\ell_2)^2$ in the numerator, where $\ell_1$ and $\ell_2$ are the
momenta flowing through each of the legs to which the new line is
joined. (This rule is depicted in \fig{AddLineFigure}).  This procedure
does not create any four-point vertices. Each distinct $(L+1)$-loop
contribution should be counted once, even though they can be generated
in multiple ways.  (Contributions which have identical diagrammatic
topologies but different numerator factors should be counted as
distinct.)  The $(L+1)$-loop amplitude is then the sum of all distinct
$(L+1)$-loop graphs.  This insertion rule has only been proven for the
entirely two-particle constructible contributions.

%

For both the planar and non-planar entirely two-particle constructible
contributions to the Yang-Mills amplitude, the color factors associated
with a given diagram are given, as in the two-loop case, by associating a 
$\tilde{f}^{abc}$ with each vertex and a $\delta^{ab}$ with each bond
of the $(L+1)$-loop graph.  On the other hand, we have investigated the 
three-particle cuts at three loops and have
found non-planar contributions (such as those associated with 
\fig{NonTwoParticleGraph}) which are not given by the rule of
\fig{AddLineFigure}.  Nevertheless, all contributions at three loops
have the same power count as the terms obtained from \fig{AddLineFigure}.


\subsection{Higher-Loop Divergences in $N=4$ Yang-Mills Theory}
\label{YMPowerCountingSubSection}

The superspace arguments \cite{HoweStelle} mentioned in the introduction
lead to predictions for the higher-loop ultraviolet divergences of
$N=4$ Yang-Mills theory in $D>4$.
We can compare these predictions to the divergences implied by our conjectured
form for the $N=4$ four-point amplitudes.  The power counting which we
will perform assumes that the terms we have are representative of the
amplitude.  The multi-particle cut analysis of
\sec{MultiParticleCutsSection} provides further evidence that this is
indeed the case.

First consider the planar three-loop case.  If we apply the insertion 
rule of \fig{AddLineFigure} to the two-loop amplitudes,
we can obtain up to two powers of loop momentum
in the numerators of the three-loop integrands.  
To find the dimension where the
amplitude first becomes divergent we focus on the diagrams with two
powers of loop momenta in the numerators, since they are the most
divergent ones. For these diagrams the superficial degree of
divergence is obtained by ignoring distinctions between the momenta of
the various loops and dropping all external momenta, thus reducing the
integral to
$$
\int (d^D p)^3  {(p^2)  \over (p^2)^{10}}  \,.
\equn
$$
This integral is ultraviolet divergent for $D\ge 6$.

%
\begin{figure}[ht]
\centerline{\epsfxsize 3.8 truein \epsfbox{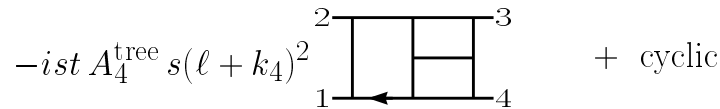} }
\caption[]{
\label{ThreeLoopLeadingYMFigure}
\small The leading-color diagrams that diverge in $D=6$. The arrow indicates
the line with momentum $\ell$.}
\end{figure}

This analysis easily generalizes to all loop orders.  From
\fig{AddLineFigure} for each additional loop the maximum number of powers
of loop momentum in the numerator increases by two.  Thus, for $L>1$ loops 
we expect that the most divergent integrals have $2L - 4$
powers of loop momenta in the numerator. These integrals will 
reduce to
$$
\int (d^D p)^L  {(p^2)^{(L-2)}  \over (p^2)^{3L +1}}  \,.
\equn\label{YMsimpInt}
$$
(The $L=1$ case is special and must be treated separately.)  
These integrals are finite for
$$
D < {6\over L} + 4 \,,  \hskip 2 cm (L > 1) \,.
\equn\label{N4Finiteness}
$$
This degree of divergence is eight powers less than the maximum
for non-supersymmetric Yang-Mills theory.

This $N=4$ power count has differences with the conventional one based on
superspace arguments~\cite{HoweStelle}.  Specifically, for dimensions
$D=5,6$ and $7$ the amplitudes first diverge at $L = 6, 3$ and 2 loops.
The corresponding superspace arguments indicate that the first divergence
may occur at $L=4,3$ and $2$, respectively.  Since the superspace
arguments of ref.~\cite{HoweStelle} only place a bound on finiteness, our
results at four and five loops are not inconsistent.  However, the
ultraviolet behavior of the amplitudes seems to indicate that the extra
symmetries in $N=4$ Yang-Mills theory, which are not taken into account by the
off-shell $N=2$ superspace arguments, are important to understanding 
its divergences in $D>4$.  Curiously, the finiteness condition
(\ref{N4Finiteness}) agrees with the power count based on the
assumption of the existence of an unconstrained off-shell covariant $N=4$
superspace formalism \cite{GrisaruSiegel,SuperSpace}.  
This agreement is probably accidental, because it is known that such a
formalism does not exist; for example, the two-loop $D=7$
counterterm has the wrong group-theory structure (although the right 
dimension) to be written as an $N=4$ superspace 
integral~\cite{MarcusSagnotti}.

Combining the $N=4$ finiteness condition (\ref{N4Finiteness}) with
those for $N=1,2$ \cite{SuperSpace} (for which off-shell superspaces for
the full supersymmetry exist) we find that an $L>1$ loop
amplitude is finite for
$$
D < {2 (N-1)\over L} + 4 \,, \hskip 2 cm  N=1,2,4\,.
\equn\label{YMFiniteness}
$$

\begin{table}
\vskip .5 cm
\hskip 1.9 truecm
\hbox{
\def\tend{\cr \noalign{\hrule}}
\def\t#1{\tilde{#1}}
\def\tw{\theta_W}
\vbox{\offinterlineskip
{
\hrule
\halign{
        &\vrule#
        &\strut\quad\hfil #\hfil\quad\vrule
        &\quad\hfil\strut#\hfil\quad\vrule
        & \quad\hfil\strut # \hfil \vrule
        \cr
height13pt  &{\bf Dimension $D$}  &{\bf Loop }  & {\bf Counterterm} \tend
height12pt  & 8
& 1 & $F^4$ \tend
height12pt  & $7$
& 2 & $\partial^2 F^4$ \tend
height12pt  & $6$
& 3 & $\partial^2 F^4$ \tend
height12pt   & $5 $
& 6  &$\partial^2 F^4$ \tend
}
}
}
}
\nobreak
\caption[]{
\label{CountertermsTableYM}
\small For $N=4$ Yang-Mills theory in $D$ dimensions, the number of
loops at which the {\it first} ultraviolet divergence occurs for
four-point amplitudes, and the generic form of the associated
counterterm.  In each case the degree of divergence is logarithmic,
but the specific color factors will differ.

\smallskip}
\end{table}

\section{$N=8$ Supergravity Amplitudes}
\label{SuperGravitySection}

In this section we present the exact result for the $N=8$ 
two-loop four-point amplitude, in terms of scalar integral functions.
Moreover, we present a conjecture for a precise relationship
between (parts of) the $N=4$ Yang-Mills and $N=8$ supergravity 
four-point amplitudes to all loop orders, extending the tree and one-loop 
relationships summarized in \figs{TreeRelationFigure}{OneLoopRelationFigure}.

Using the two-loop $N=4$ Yang-Mills amplitudes discussed in the previous
section, we can immediately obtain candidate expressions for the corresponding
$N=8$ amplitudes simply by dropping the color factors and squaring the
coefficients and numerators of every scalar integral.  In
\sec{TwoParticleSubSection} we will verify that this procedure is
valid to all loop orders for the entirely two-particle constructible terms.
For the two-loop case, in \sec{ThreeParticleSubSection} we also
evaluate the three-particle cuts, allowing for a complete
reconstruction of this amplitude, proving that the squaring procedure
gives the correct two-loop results.  (Direct evaluations of the two- and
three-particle cuts using the four-dimensional helicity basis may be found in
\apps{TwoParticleCutsAppendix}{ThreeParticleCutsAppendix}.)

\subsection{Two-Loop $N=8$ Supergravity Amplitudes}
\label{TwoLoopGravityAmplSubSection}

We now obtain the two-loop $N=8$ four-graviton
amplitudes by squaring the corresponding coefficients appearing in the
$N=4$ four-point amplitudes 
(\ref{TwoLoopYM}), after stripping away the color factors.  
The $N=8$ amplitudes are thus expected to be,
$$
\eqalign{ 
{\cal M}_4^{\twoloop}(1,2,3,4) & = 
 -i \Bigl({\kappa \over 2} \Bigr)^6 
 [s_{12} s_{23} \,  A_4^{\rm tree}(1,2,3,4)]^2
 \Bigl(s_{12}^2 \, \I_4^{\twoloop,\P}(s_{12}, s_{23}) 
+ s_{12}^2 \, \I_4^{\twoloop,\P}(s_{12}, s_{24})  \cr
& \hskip 3 cm  
+ s_{12}^2 \, \I_4^{\twoloop,\NP}(s_{12}, s_{23})
+ s_{12}^2 \, \I_4^{\twoloop,\NP}(s_{12}, s_{24}) 
 \hskip 0.5  cm 
+ \hbox{cyclic} \Bigr) \,. \cr}
\equn\label{TwoLoopGrSquare}
$$
Here $A_4^{\rm tree}$ is the $N=4$ Yang-Mills four-gluon tree
amplitude, the integrals are defined in \eqn{TwoLoopScalarInts} 
(see \fig{PlanarNonPlanarFigure}) and 
`$+$~cyclic' instructs one to add the two cyclic permutations of
(2,3,4), just as in \eqn{TwoLoopYM}.  Using \eqn{OneLoopRelation}
we may re-express the overall coefficient in terms of the gravity 
tree amplitude to obtain the final form for the amplitude, 
$$
\eqalign{ 
{\cal M}_4^{\twoloop}(1,2,3,4) & = 
 \Bigl({\kappa \over 2} \Bigr)^6 \, s_{12} s_{23} s_{13} \, 
 M_4^{\rm tree}(1,2,3,4)
 \Bigl(s_{12}^2 \, \I_4^{\twoloop,\P}(s_{12}, s_{23}) 
+ s_{12}^2 \, \I_4^{\twoloop,\P}(s_{12}, s_{24})  \cr
& \hskip 3 cm  
+ s_{12}^2 \, \I_4^{\twoloop,\NP}(s_{12}, s_{23})
+ s_{12}^2 \, \I_4^{\twoloop,\NP}(s_{12}, s_{24}) 
 \hskip 0.5  cm 
+ \hbox{cyclic} \Bigr) \,. \cr}
\equn\label{TwoLoopGr}
$$
 
%
\begin{figure}[ht]
\centerline{\epsfxsize 2.8 truein \epsfbox{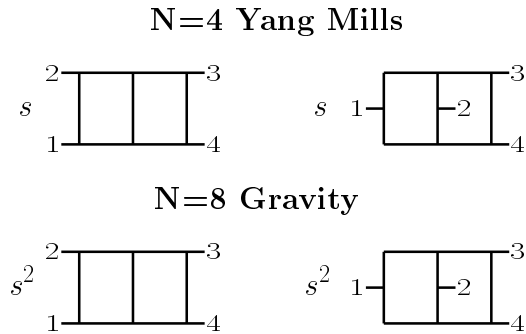}}
\caption[]{
\label{TwoLoopRelationFigure}
\small The expected relationship between two-loop contributions to
$N=8$ four-graviton amplitudes and $N=4$ four-gluon amplitudes: the
graviton coefficients are squares of the gluon coefficients.  The
$N=4$ and $N=8$ contributions depicted here are to be multiplied
respectively by factors of $-g^6 s t A_4^{\rm tree}$ (dropping the
group theory factor) and $-i(\kappa/2)^6 [st A_4^{\rm tree}]^2$.}
\end{figure}

The two-loop ultraviolet divergences for $N=8$ supergravity in $D=7$,
9 and 11, after using the results of \app{UVExtractionAppendix} 
and summing over the different double-box integrals which appear, are
$$
\eqalign{
{\cal M}_4^{\twoloop,\ D=7-2\e}\vert_{\rm pole} 
 &= 
 {1\over2\e\ (4\pi)^7} {\pi\over3} (s^2+t^2+u^2) \, 
\times 
\left( {\kappa \over 2}\right)^6 
\times stu M_4^{\rm tree} 
, \cr 
{\cal M}_4^{\twoloop,\ D=9-2\e}\vert_{\rm pole} 
 &=  {1\over4\e\ (4\pi)^9} {-13\pi\over9072} (s^2+t^2+u^2)^2 \, 
\times 
\left( {\kappa \over 2}\right)^6 \times stu M_4^{\rm tree} \, , \cr 
{\cal M}_4^{\twoloop,\ D=11-2\e}\vert_{\rm pole} 
 &=  {1\over48\e\ (4\pi)^{11}} {\pi\over5791500} 
  \Bigl( 438 (s^6+t^6+u^6) - 53 s^2 t^2 u^2 \Bigr) \,
\times 
\left( {\kappa \over 2}\right)^6 \times stu M_4^{\rm tree} \, . \cr 
\cr}\equn\label{gravtwolooppoles}
$$
There are no sub-divergences because one-loop divergences are absent in
odd dimensions when using dimensional regularization.  

In both $D=8$ and $D=10$, the box integrals encountered in the
one-loop $N=4$ Yang-Mills amplitude~(\ref{OneLoopYMResult}) and $N=8$
supergravity amplitude~(\ref{OneLoopGravResult}) do have ultraviolet
poles.  Curiously, for the $N=8$ case in $D=10$ the coefficient of the
pole cancels in dimensional regularization.  (The pole in the box
integral $\I_4^{\oneloop}(s_{12},s_{23})$ is proportional to
$s_{12}+s_{23}$, and the sum over boxes in \eqn{OneLoopGravResult}
cancels using $s_{12}+s_{23}+s_{13}=0$.)  This cancellation between
quadratically divergent integrals may well be an artifact of
dimensional regularization.  In any case, we can investigate whether
the cancellation persists to two loops.  To do so, we have calculated
the ultraviolet divergences of subtracted planar and non-planar
double-box integrals in $D=10$.  For completeness we have also
calculated the divergence in $D=8$.  A subtraction is required because
the individual integrals have one-loop sub-divergences.  The results
for the integrals are presented in \app{UVExtractionAppendix}.  The
sum over double-box integrals in \eqn{OneLoopGravResult} then yields
$$
\eqalign{
{\cal M}_4^{\twoloop,\ D=8-2\e}\vert_{\rm pole} 
 &=  {1\over2 \ (4\pi)^{8}} \Bigl( - {1\over 24 \, \e^2} 
                  + {1\over 144\e} \Bigr) \bigl(s^3 + t^3 + u^3 \bigr)
 \times 
\left( {\kappa \over 2}\right)^6 \times stu M_4^{\rm tree} \,, \cr 
{\cal M}_4^{\twoloop,\ D=10-2\e}\vert_{\rm pole} 
 &=  {1\over12\e\ (4\pi)^{10}} { - 13 \over 25920 } \, stu \, 
  \bigl( s^2+t^2+u^2 \bigr) \times 
\left( {\kappa \over 2}\right)^6 \times stu M_4^{\rm tree} \,, \cr 
\cr}\equn\label{gravtwolooppolestend}
$$
so there is indeed a two-loop divergence in $D=10$.

In the five cases, for four graviton external states, the linearized
counterterms take the form of derivatives acting on
$$
t_8 t_8 R^4 \equiv
t_8^{\mu_1\mu_2\cdots \mu_8}\, 
t_8^{\nu_1\nu_2\cdots \nu_8} \, 
R_{\mu_1\mu_2 \nu_1 \nu_2} \, 
R_{\mu_3\mu_4 \nu_3 \nu_4} \,
R_{\mu_5\mu_6 \nu_5 \nu_6} \,
R_{\mu_7\mu_8 \nu_7 \nu_8}  \,,
\equn\label{Rfourterms}
$$
plus the appropriate $N=8$ completion.  As mentioned in the
introduction, the operator (\ref{Rfourterms}) appears in the
tree-level superstring effective action.  It also appears as the
one-loop counterterm for $N=8$ supergravity in $D=8$.  Finally, it is
thought to appear in the M-theory one-loop effective action
\cite{StringR4}.

\subsection{Higher-loop conjecture}

We conjecture that to all orders in the perturbative expansion the
four-point $N=8$ supergravity amplitude may be found by
squaring the coefficients and numerator factors of all the loop integrals 
that appear in the $N=4$ Yang-Mills amplitude at the same order,
after stripping away the color factors.
In \sec{CutConstructionSection} we show that the exact two-particle cuts
and the two-loop three-particle cuts confirm this picture, making the
conjecture precise at two loops. 

\begin{figure}[ht]
\centerline{\epsfxsize 3 truein \epsfbox{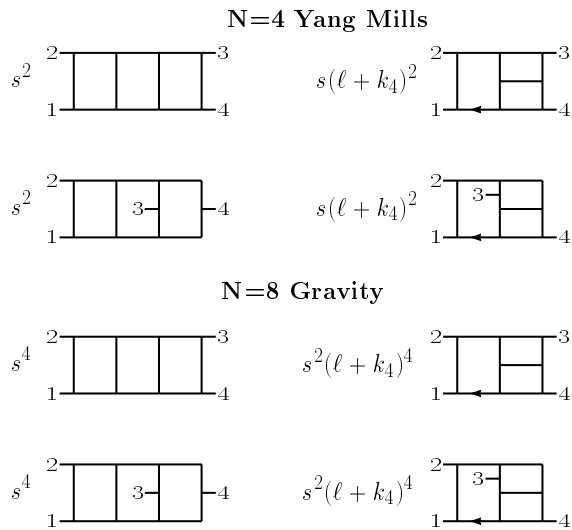} }
\vskip -.1 cm
\caption[]{
\label{ThreeLoopExampleFigure}
\small
Some sample three-loop integrals and their coefficients for $N=4$ 
Yang-Mills theory and for $N=8$ supergravity.
The coefficients for $N=8$ supergravity are just the squares of those for 
$N=4$ Yang-Mills theory. 
The $N=4$ and $N=8$ contributions depicted here 
are to be multiplied respectively by overall factors of 
$-i g^8 s t A_4^{\rm tree}$ and $(\kappa/2)^8 [st A_4^{\rm tree}]^2$.}
\end{figure}

In order to specify the precise form of the conjecture at $L$ loops
one would need to investigate cuts with up to $(L+1)$ intermediate
particles.
Nevertheless, some of the integral coefficients and numerators 
can be obtained from the known two-particle cuts.  
For example, \fig{ThreeLoopExampleFigure}
contains a few sample three-loop integrals which are entirely 
two-particle constructible, and their
associated coefficients for the case of $N=4$ Yang-Mills
theory~\cite{BRY}; the $N=8$ supergravity coefficients are given by 
squaring the super-Yang-Mills coefficients.  This provides an
explicit example of three-loop relationships between contributions
to the Yang-Mills and gravity amplitudes for the cases of $N=4$ and
$N=8$ supersymmetry.  The supergravity coefficients can again be 
expressed in terms of the tree-level gravity amplitudes 
using \eqn{OneLoopRelation}.  It would be interesting to determine 
whether squaring relations do indeed continue
to hold for all remaining three-loop diagram topologies.

%


\subsection{Higher-Loop Divergences in $N=8$ Supergravity} 
\label{GravityPowerCountingSubSection}

In this subsection we discuss the ultraviolet behavior of $N=8$
supergravity arising from our all-loop conjecture for the form of the
four-point amplitude.  This `squaring' conjecture gives twice as
many powers of loop momentum in the numerator of the integrand as for
Yang-Mills theory.  The power-counting equation that describes the leading
divergent behavior for $N=4$ Yang-Mills theory, 
\eqn{YMsimpInt}, becomes for $N=8$ supergravity at $L$ loops,
$$
\int (d^D p)^L {(p^2)^{2(L-2)} \over (p^2)^{3L +1}} \,. \equn
$$
This integral will be finite when
$$
D < {10 \over L} + 2 \,,  \hskip 2 cm (L > 1) \,.
\equn\label{N8Finiteness}
$$

The results of this analysis are summarized in
\tab{CountertermsTableGR}.  In particular, in $D=4$ no three-loop
divergence appears --- contrary to expectations from a superspace
analysis~\cite{HoweStelleTownsend,HoweStelle} 
--- and the first $R^4$-type counterterm occurs at five loops.  
The divergence will have the same kinematical structure as the $D=7$
divergence in \eqn{gravtwolooppoles}, but with a different
non-vanishing numerical coefficient.

\begin{table}[ht]
\vskip .5 cm
\hskip 1.9 truecm
\hbox{
\def\tend{\cr \noalign{\hrule}}
\def\t#1{\tilde{#1}}
\def\tw{\theta_W}
\vbox{\offinterlineskip
{
\hrule
\halign{
        &\vrule#
        &\strut\quad\hfil #\hfil\quad\vrule
        &\quad\hfil\strut#\hfil\quad\vrule
        &\quad\hfil\strut#\hfil\quad\vrule
        & \quad\hfil\strut # \hfil \vrule
        \cr
height13pt  &{\bf Dimension }  &{\bf Loop }   &{\bf Degree of Divergence}    & {\bf Counterterm} \tend
height12pt  & 8 
& 1 & logarithmic &  $R^4$ \tend
height12pt  & $7$
& 2 &  logarithmic & $\partial^4 R^4$ \tend
height12pt  & $6$
& 3 & quadratic & $\partial^6 R^4$ \tend
height12pt  & $5$
& 4 & quadratic & $\partial^6 R^4$ \tend
height12pt   & $4 $
& 5  &  logarithmic & $\partial^4 R^4$ \tend
}
}
}
}
\nobreak
\caption[]{
\label{CountertermsTableGR} 
\small The relationship between dimensionality and the number of loops 
at which the {\it first} ultraviolet divergence should occur in the 
$N=8$ supergravity four-point amplitude.  The form of the associated 
counterterm assumes the use of dimensional regularization.
\smallskip}
\end{table}

The one- and two-loop entries in \tab{CountertermsTableGR} are based
on complete calculations of the amplitudes.  Beyond two loops we do
not have complete calculations, but in \sec{TwoParticleSubSection} we
will show that the divergence structure given in \eqn{N8Finiteness}
is consistent with the two-particle cuts to all loop orders, and
in \sec{MultiParticleCutsSection} we will demonstrate related
ultraviolet cancellations in the $m$-particle MHV cuts.  Continuing
along these lines, a complete proof of the ultraviolet behavior of the
$L$-loop amplitude would require an analysis of all contributions to
the cuts.

\section{Cut Constructions}
\label{CutConstructionSection}

In this section we justify the form of the $N=8$ four-point amplitude
presented in \sec{SuperGravitySection} by iterating the exact
$D$-dimensional two-particle cuts to {\it all} loop orders.  We will
also compute the $D$-dimensional three-particle cuts at two loops,
demonstrating that \eqn{TwoLoopGr} gives the complete two-loop $N=8$
amplitude.  First we briefly review the cut construction method
\cite{SusyFour,SusyOne,Review,BRY,Rozowsky} for constructing complete
amplitudes.  This approach leads to relatively compact expressions
because the calculation is organized in terms of gauge-invariant
quantities at intermediate steps.

In this method the amplitude is reconstructed from its analytic
properties.  In general, to reconstruct an $L$-loop amplitude one must
calculate all cuts, which can have up to $(L+1)$ intermediate states.  
However, the various cuts are related to each other, so one can often write
down complete expressions for the amplitudes based on a
calculation of a small subset of cuts.  (When combining the cuts into
a single function, care must be exercised not to over-count a
particular term.)  Once one has a robust ansatz for the form of the
amplitude, the remaining cuts become much easier to calculate, since
one has a definite final form to compare with.  As one calculates
additional cuts, one obtains cross-checks on the terms inferred from the 
earlier cuts; the consistency of the different cuts provides a rather
powerful check that one is calculating correctly.

We find it convenient to perform the cut construction using
components instead of superfields.  The potential advantage of a
superfield formalism would be that one would simultaneously include
contributions from all particles in a supersymmetry multiplet.
However, for the cases we investigate, the supersymmetry Ward identities,
discussed in \app{SusyIdentityAppendix}, are sufficiently powerful
that once the contribution from one component is known the others
immediately follow. In a sense these identities are equivalent to
using an on-shell superspace formulation.  A component formulation is
also more natural for extensions to non-supersymmetric theories.

\subsection{Review of Cut Construction Method}

\begin{figure}[ht]
\centerline{
{\epsfxsize 2.3 truein \epsfbox{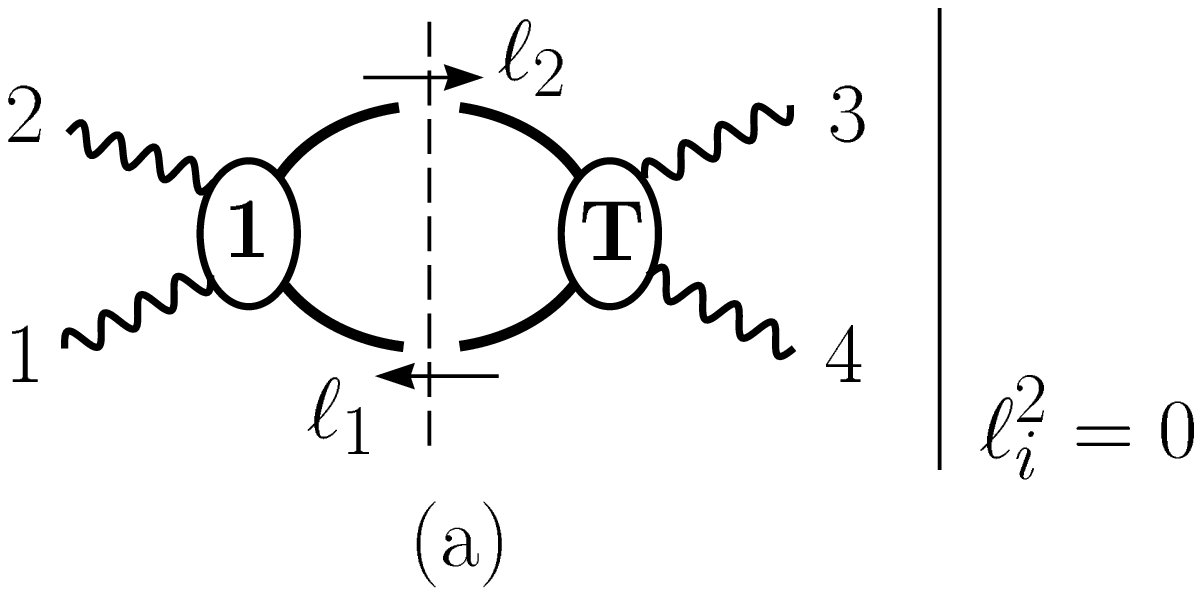}
\hskip 1.5 cm
\epsfxsize 2.3 truein \epsfbox{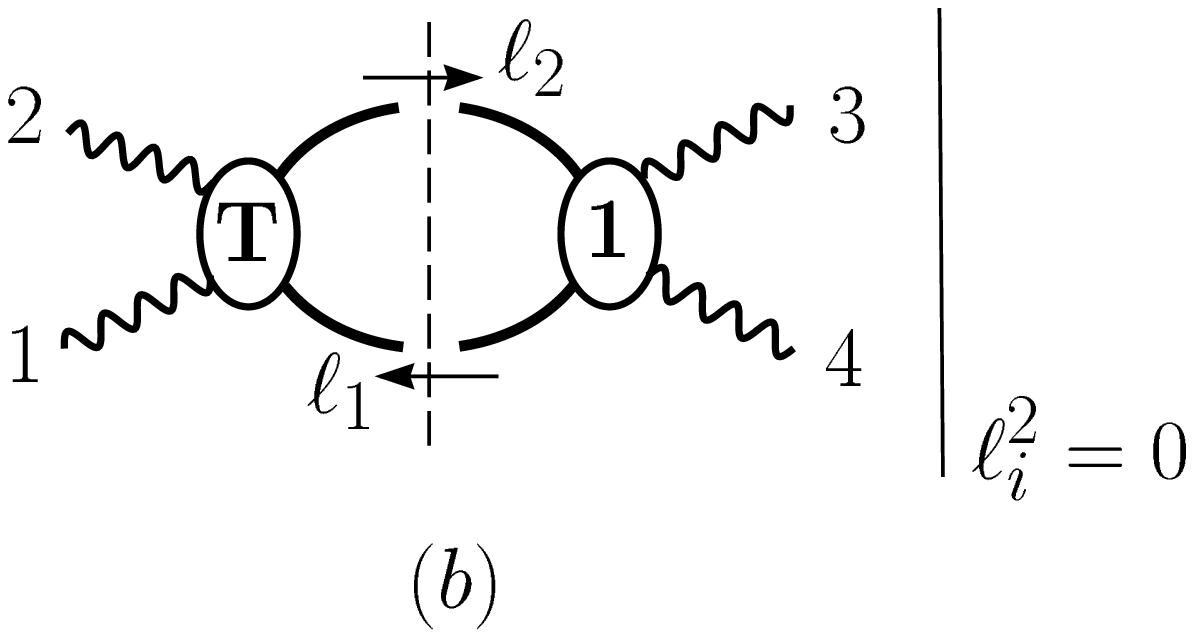}} }
\vskip -.5 cm
\caption[]{
\label{TwoParticleFigure}
\small
The two-particle $s$-channel cut has two contributions: one with the 
four-point one-loop amplitude `1' to the left and the tree amplitude `T' to 
the right (a) and the other with the reverse assignment (b).  }
\end{figure}

As an example, consider the cut construction method for a two-loop
amplitude ${\cal M}_4^\twoloop(1,2,3,4)$.  At two loops one must consider
both two- and three-particle cuts.  
In each channel there can be multiple contributing cuts.  
For example, in the $s$ channel there are
two two-particle cuts, as depicted in \fig{TwoParticleFigure}.  The
first of these has the explicit representation
$$
\eqalign{
M_{4}^{\twoloop}&(1, 2, 3, 4)
\Bigr|_{\rm cut (a)}  = 
\left. \int\! {d^{D}\ell_1\over (2\pi)^{D}} \, \sum_{\S_1, \S_2} \,
  {i\over \ell_1^2 } \,
         M_4^{\rm tree} (-\ell_1^{\S_1} ,1,2,\ell_2^{\S_2}) 
\,{i\over \ell_2^2} \,  M_{4}^{\oneloop} 
    (-\ell_2^{\S_2}, 3,4,\ell_1^{\S_1}) 
\right|_{\ell_1^2 = \ell_2^2 =0} \, ,\cr}  
\equn
\label{CutProductDef}
$$
where $\ell_1$ and $\ell_2$ are the momenta of the cut legs and the
sum runs over all particle states $S_1$ and $S_2$ which may propagate
across the two cut lines.  Following the discussion in
ref.~\cite{Review}, it is useful to have replaced the phase-space
integrals with cuts of unrestricted loop momentum integrals, even
though we use the on-shell conditions on the amplitudes appearing in
the integrand.  In this way, one may simultaneously construct the
imaginary and associated real parts of the cuts, avoiding the need for
dispersion relations.

%

\begin{figure}[ht]
\centerline{
\epsfxsize 2.5 truein \epsfbox{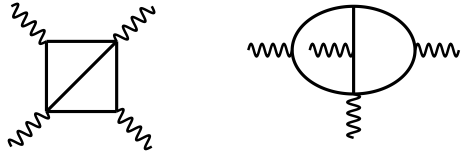} }
\vskip -.5 cm
\caption[]{
\label{NoTwoParticleCutFigure}
\small Examples of loop integrals with no two-particle cuts.}
\end{figure}

For a two-loop amplitude one must also calculate three-particle cuts.
If one does not calculate these cuts one could, in principle, miss
functions that have no two-particle cuts.  Examples of such functions
are the integral functions shown in \fig{NoTwoParticleCutFigure}.


\begin{figure}[ht]
\centerline{\epsfxsize 2.5 truein \epsfbox{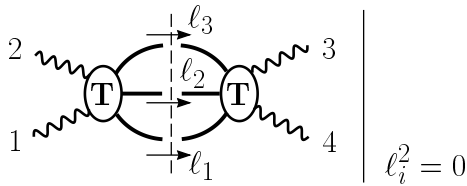}}
\vskip -.3 cm
\caption[]{
\label{ThreeParticleFigure}
\small
The three-particle $s$-channel cut for a two-loop amplitude.}
\end{figure}

The three-particle $s$-channel cut is
$$
\eqalign{
&M_{4}^{\twoloop}(1, 2, 3, 4)
\Bigr|_{3\rm \hbox{-}cut} \cr
& \hskip0.1cm 
= \int
{d^{D}\ell_1\over (2\pi)^{D}} \, 
{d^{D}\ell_2\over (2\pi)^{D}} \; 
 \sum_{S_1, S_2, S_3} 
M_5^{\rm tree} (1,2,\ell_3^{\S_3},\ell_2^{\S_2}, \ell_1^{\S_1})   
\,{i\over \ell_1^2} \, {i\over \ell_2^2 } \,\,{i\over \ell_3^2} \, 
M_5^{\rm tree} (3,4,-\ell_1^{\S_1}, -\ell_2^{\S_2}, -\ell_3^{\S_3}) 
\biggr|_{\ell_i^2  = 0} \, ,\cr}  
\equn
\label{ThreeCutProductDef}
$$
which is depicted in
\fig{ThreeParticleFigure}.  The other channels are, of course, similar.

%

The complete two-loop amplitude may be constructed by finding a single
function which has the correct two- and three-particle cuts in $D$
dimensions. As discussed in
refs.~\cite{TwoLoopUnitarity,Massive,Review}, by computing all cuts to
all orders in the dimensional regularization parameter $\eps =
(4-D)/2$, one may perform a complete reconstruction of a massless loop
amplitude in any dimension.  This follows from dimensional analysis,
since every term in an $L$-loop amplitude must carry $L$ prefactors of the
form $(-s_{ij})^{-\eps}$, which have cuts away from integer
dimensions.  As with Feynman diagrams, the result is unique for a
given dimensional regularization scheme.  As mentioned in the
introduction, we define the $N=4$ Yang-Mills amplitudes to be the
dimensional reduction of ten-dimensional $N=1$ Yang-Mills amplitudes,
and $N=8$ supergravity amplitudes to be the dimensional reduction of
eleven-dimensional $N=1$ supergravity amplitudes; these definitions
should also include the non-integer dimensions implied by dimensional
regularization.

This same technique may be applied at any loop order; at $L$ loops one
would need to compute cuts with up to $(L+1)$ intermediate particles.  
In general, when performing cut calculations, it is convenient to ignore
normalizations until the end of the calculation.  There are sufficiently
many cross-checks between the various cuts so that relative normalizations
can usually be fixed.

For reasons of technical simplicity, we sometimes evaluate the cuts using
four-dimensional helicities and momenta.  In particular,
\apps{TwoParticleCutsAppendix}{ThreeParticleCutsAppendix} contain such
evaluations of the two- and three-particle cuts for the two-loop $N=8$
amplitudes.  In principle, the use of helicity amplitudes with
four-dimensional momenta can drop terms necessary for analytically
continuing to $D$ dimensions.  However, it turns out that for the $N=8$
two-loop four-point amplitudes there are no such dropped terms and one
obtains the exact $D$-dimensional expressions.  For the case of $N=4$
Yang-Mills theory an argument has been outlined in ref.~\cite{BRY}.  For the
$N=8$ case it is more convenient to calculate directly in arbitrary
dimensions, by recycling the previously-obtained $N=4$ Yang-Mills cuts.

\subsection{Exact Two-Particle Cuts}
\label{TwoParticleSubSection}

We now compute the two-particle cuts of four-point $N=8$ amplitudes
iteratively to all loop orders in $D$ dimensions.
For four-point $N=4$ Yang-Mills amplitudes it was noted in
ref.~\cite{BRY} that the algebraic steps needed to obtain the
two-particle cuts at one loop recycle to all loop orders.  Here we
show that the same situation holds for four-point $N=8$ supergravity
amplitudes.

The key relation for evaluating the $N=4$ two-particle cuts exactly 
to all loop orders is,
$$
\eqalign{
\sumYM A_4^{\rm tree}(-\ell_1^{\S_1}, & 1, 2, \ell_2^{\S_2}) \times
  A_4^{\rm tree}(-\ell_2^{\S_2}, 3, 4, \ell_1^{\S_1}) 
= - ist A_4^{\rm tree}(1, 2, 3, 4) 
   {1\over (\ell_1 - k_1)^2 } 
   {1\over (\ell_2 - k_3)^2 } \,, 
\cr}
\equn\label{BasicYMCutting}
$$
where all momenta are on shell and the sum is over all states in the
$N=4$ multiplet.  In the remainder of this section we will suppress
the particle state labels on the $\ell_i$ for simplicity.  This
equation is true in all dimensions without any integrations over the
$\ell_i$.  One convenient way to verify \eqn{BasicYMCutting} for the
case of external gluons is by using background field Feynman gauge and
second order fermions~\cite{Background,BernDunbar,SecondOrder}.  In
this way, all powers of $\ell_i$ in the numerator cancel
algebraically. (In this particular scheme \eqn{BasicYMCutting} is true
even for off-shell $\ell_i$.)  The uniqueness of the supersymmetric completion
of the operator given in \eqn{Ffourterms} assures that one obtains
an $A_4^{\rm tree}$ no matter what the external states are.

Using the KLT relation (\ref{GravYMFour}) we can use \eqn{BasicYMCutting}
to obtain the equivalent relation for $N=8$ supergravity, 
$$
\eqalign{
\sum_{N=8 \rm\ states} 
& M_4^{\rm tree}(-\ell_1,  1, 2, \ell_2) \times
  M_4^{\rm tree}(-\ell_2, 3, 4, \ell_1) \cr 
& =  
- s^2 \biggl(\sum_{N=4\rm\ states} 
  A_4^{\rm tree}(-\ell_1,  1, 2, \ell_2) \times
                A_4^{\rm tree}(-\ell_2, 3, 4, \ell_1) \biggr)
     \cr
\null & \hskip 5.0 truecm
\times     
\biggl(\sum_{N=4\rm\ states}  A_4^{\rm tree}(\ell_2,  1, 2, -\ell_1) \times
                A_4^{\rm tree}(\ell_1, 3, 4, -\ell_2) \biggr) \cr
& =s^2 (st)^2 \bigl[A_4^{\rm tree}(1, 2, 3, 4)\bigr]^2
   { 1\over (\ell_1 - k_1)^2  (\ell_2 - k_3)^2 
 (\ell_2 + k_1)^2  (\ell_1 + k_3)^2 }  \cr
& =  i\, s^2\, stu\, M_4^{\rm tree} (1, 2, 3, 4)
     { 1\over (\ell_1 - k_1)^2  (\ell_2 - k_3)^2 
   (\ell_1 - k_2)^2 (\ell_2 - k_4)^2 } \,. \cr}
\equn\label{N8TPC}
$$
In \eqn{N8TPC} we have used the decomposition of a state in the 
$N=8$ multiplet into a `left' and a `right' $N=4$ state,
and the fact that summing over the $N=8$ multiplet is equivalent
to summing over the left and right $N=4$ multiplets independently. 
We then perform a partial-fraction decomposition of the denominators
(using the on-shell conditions),
$$
\eqalign{
& -{s \over  (\ell_1 - k_1)^2 (\ell_1 - k_2)^2} 
 = {1\over (\ell_1 - k_1)^2} + {1\over (\ell_1 - k_2)^2} \,, \cr
& -{s \over  (\ell_2 - k_3)^2 (\ell_2 - k_4)^2} 
 = {1\over (\ell_2 - k_3)^2} + {1\over (\ell_2 -  k_4)^2} \,, \cr}
\equn
$$
to obtain the $N=8$ basic two-particle on-shell sewing relation, 
$$
\eqalign{
\sum_{N=8 \rm\ states} M_4^{\rm tree}& (-\ell_1, 1, 2, \ell_2) \times
  M_4^{\rm tree}(-\ell_2, 3, 4, \ell_1) \cr \
& = i\, stu M_4^{\rm tree}(1, 2, 3, 4) 
 \biggl[{1\over (\ell_1 - k_1)^2 } + {1\over (\ell_1 - k_2)^2} \biggr]
\biggl[{1\over (\ell_2 - k_3)^2 } + {1\over (\ell_2 - k_4)^2} \biggr]\,, \cr}
\equn\label{BasicGravityCutting}
$$
where the $\ell_i$ are on-shell.  The $t$- and
$u$-channel formulas are given by relabelings of the legs in 
\eqn{BasicGravityCutting}. From this sewing relation it is trivial to obtain
the $N=8$ one-loop amplitude~(\ref{OneLoopGravResult}).

A key feature of the sewing relation (\ref{BasicGravityCutting}) is
that when one sews two tree amplitudes and sums over all intermediate
$N=8$ states, one gets back a tree amplitude multiplied by scalar
functions.  This means that we may recycle the sewing
relation~(\ref{BasicGravityCutting}) to obtain two-particle cuts of
higher-loop amplitudes.  The only difference is that now we must
keep track of the scalar function prefactors.

Consider, for example, the two-particle $s$-cut with a tree amplitude
on the left and a one-loop amplitude on the right,
$$
M_4^{\twoloop}(1, 2, 3, 4) \Bigl|_{s\rm \hbox{-}cut} 
=  \int {d^D \ell_1 \over (2\pi)^D} 
\sum_{N=8\ {\rm states}}
\; {i \over \ell_1^2} M_4^{\rm tree}(-\ell_1, 1, 2, \ell_2) \,
   {i \over \ell_2^2} \, M_4^{\oneloop}(-\ell_2, 3, 4, \ell_1) 
\biggr|_{\ell_1^2 = \ell_2^2 =0} \,.
\equn
$$
Inserting \eqn{OneLoopGravResult} for $M_4^{\oneloop}$ and applying 
the sewing relation~(\ref{BasicGravityCutting}), we have
$$
\eqalign{
M_4^{\twoloop} & (1, 2, 3, 4) \Bigl|_{s\rm \hbox{-}cut} 
 =   - s t u M_4^{\rm tree}
\int {d^D \ell_1 \over (2\pi)^D}  
  s (\ell_2 - k_3)^2  (\ell_2 - k_4)^2  \cr
& \hskip 1 cm \times
 \Bigl[{1\over (\ell_1 - k_1)^2} + {1\over (\ell_1 - k_2)^2} \Bigr] \, 
{i\over  \ell_1^2 } \, 
  \Bigl[ {s \over (\ell_2 - k_3)^2 (\ell_2 -  k_4)^2} \Bigr] \,
  {i\over  \ell_2^2 }  \cr
& \hskip 1 cm \times
\bigl[ \I_4^{\oneloop}(s, (\ell_2 - k_3)^2) 
     + \I_4^{\oneloop}((\ell_2 - k_3)^2, (\ell_2 - k_4)^2)
     + \I_4^{\oneloop}((\ell_2 - k_4)^2, s) \bigr]
  \biggr|_{\ell_1^2 = \ell_2^2 =0}\,, \cr}
\equn
$$
where we have combined the two propagators on a common denominator
in the second brackets in order to make a cancellation manifest.
The unwanted propagators cancel and our final result is remarkably simple,
$$
\eqalign{
M_4^{\twoloop}(1, 2, 3, 4) 
\Bigl|_{s\rm \hbox{-}cut}  
&=
 s t u  M_4^{\rm tree} s^2  \Bigl( 
\I_4^{\twoloop,\P}(s,t) + \I_4^{\twoloop,\P}(s,u)
\cr \null & \hskip 3.0 truecm 
+ \I_4^{\twoloop,\NP}(s,t) + \I_4^{\twoloop,\NP}(s,u)
\Bigr)\Bigl|_{s\rm \hbox{-}cut} \,,
\cr}
\equn
$$
where the scalar integrals $\I_4^{\twoloop,\P}$ and
$\I_4^{\twoloop,\NP}$ are defined in \eqn{TwoLoopScalarInts}.  The
analysis of the two-particle $t$- and $u$-channel cuts is identical.

It is simple to find a single function with the correct two-particle cuts 
in all three channels; the answer is given in \eqn{TwoLoopGr}.  
We emphasize that the results for the two-particle
cuts are valid in any dimension and do not rely on any
four-dimensional properties.  (A more direct evaluation of the
two-particle cuts using four-dimensional helicity states may be found in
\app{TwoParticleCutsAppendix}.)

Since the right-hand side of sewing relation (\ref{BasicGravityCutting}) is
proportional to the tree amplitude, one may iterate this procedure to
obtain the `entirely two-particle constructible' contributions,
defined in \sec{HigherLoopStructureYMSubSection}, at any desired loop order.

\subsection{Two-loop Three-Particle Cuts}
\label{ThreeParticleSubSection}

We now evaluate the $D$-dimensional two-loop three-particle cuts for
the $N=8$ four-point amplitude, by recycling the analogous calculation
for $N=4$ Yang-Mills theory.  (In \app{ThreeParticleCutsAppendix}
we also evaluate the $N=8$ three-particle cuts directly, using the
helicity formalism, which yields the same result.)

Consider first the three-particle $s$-channel cut
(\ref{ThreeCutProductDef}).  Just as for the two-particle sewing, the
sum over three-particle states for $N=8$ supergravity may be
expressed as a double sum over $N=4$ states.  Then we may apply the
five-point KLT formula (\ref{GravYMFive}), which expresses the $N=8$
supergravity tree amplitudes appearing in \eqn{ThreeCutProductDef} in
terms of the corresponding $N=4$ Yang-Mills tree amplitudes.

The tree amplitude on the left side of the cut is
$$
\eqalign{
M_5^{\rm tree}(\ell_1, 1, 2, \ell_3, \ell_2) 
& = i\, (\ell_1 + k_1)^2 (\ell_3 + k_2)^2 
A_5^{\rm tree}(\ell_1, 1, 2, \ell_3, \ell_2)\,
A_5^{\rm tree}(1, \ell_1, \ell_3, 2, \ell_2) 
+ \{ 1 \leftrightarrow 2 \} \,, \cr}
\equn\label{ThreePartRightCutTree}
$$
where we have chosen a convenient representation in terms of the gauge
theory amplitudes.  (The $ \{ 1 \leftrightarrow 2 \}$ interchange acts
only on $k_1$ and $k_2$, and {\it not} on $\ell_1$ and $\ell_2$.)
Similarly, for the tree amplitude on the right side of the cut, 
$$
\eqalign{
M_5^{\rm tree}(-\ell_3, 3, 4, -\ell_1, -\ell_2) 
 &= i \, (\ell_3 - k_3)^2 (\ell_1 - k_4)^2 
A_5^{\rm tree}(-\ell_3, 3, 4, -\ell_1, -\ell_2) \, 
A_5^{\rm tree}(3, -\ell_3, -\ell_1, 4, -\ell_2) \cr
& \hskip 4 cm 
+ \{3 \leftrightarrow 4\} \,. \cr}
\equn\label{ThreePartLeftCutTree}
$$
Thus we have, 
$$
\eqalign{
\sum_{N=8\ {\rm states}}
M_5^{\rm tree} & (1,  2, \ell_3, \ell_2, \ell_1)
M_5^{\rm tree} (3, 4, -\ell_1, -\ell_2, -\ell_3) \cr
& = 
-(\ell_1 + k_1)^2 (\ell_3 + k_2)^2 (\ell_3 - k_3)^2 (\ell_1 - k_4)^2 \cr
& \hskip 1 cm \times
\biggl[ \sum_{N=4\ {\rm states}} 
A_5^{\rm tree}(\ell_1, 1, 2, \ell_3, \ell_2)
A_5^{\rm tree}(-\ell_3, 3, 4, -\ell_1, -\ell_2) \biggr] \cr 
& \hskip 1 cm \times
\biggl[\sum_{N=4\ {\rm states}} 
A_5^{\rm tree}(1, \ell_1, \ell_3, 2, \ell_2)
A_5^{\rm tree}(3, -\ell_3, -\ell_1, 4, -\ell_2) \biggr] \cr
& \hskip 4 cm 
 + \{1 \leftrightarrow 2 \} + \{3 \leftrightarrow 4\}
 + \{ 1 \leftrightarrow 2 ,\,  3 \leftrightarrow 4 \}
\,,\cr}
\equn\label{ThreeCutSewingRelation}
$$
where all momenta are to be taken on-shell.
The sum over $N=4$ states in each set of brackets can be simplified 
using the results for the two-loop $N=4$ Yang-Mills amplitudes 
(\ref{TwoLoopYM}).  Taking the three-particle cuts of the planar 
contributions (see figure~3 of ref.~\cite{BRY}) yields
the on-shell phase-space integral of 
$$
\eqalign{
 \sum_{N=4\ {\rm states}} &
A_5^{\rm tree}(\ell_1, 1, 2, \ell_3, \ell_2)
A_5^{\rm tree}(-\ell_3, 3, 4, -\ell_1, -\ell_2) 
 = -i\, st  \, A_4^{\rm tree}(1,2,3,4) \, \cr
& \times 
\Bigl[
  {s \over 
  (\ell_1 - k_4)^2 (\ell_3 + k_2)^2 (\ell_1 + \ell_2)^2 (\ell_2 + \ell_3)^2}
+ {s \over 
  (\ell_3 - k_3)^2 (\ell_1 + k_1)^2 (\ell_1 + \ell_2)^2 (\ell_2 + \ell_3)^2}\cr
& \hskip 2 cm 
+ {t \over 
    (\ell_3 - k_3)^2 (\ell_1 - k_4)^2 (\ell_3 + k_2)^2 (\ell_1 + k_1)^2} 
   \Bigr]   \,.
\cr}
\equn\label{N4ThreeCut}
$$
This equation actually holds even before carrying out the loop-momentum 
(or phase-space) integral.
In the calculations used to derive the results of
ref.~\cite{BRY} \eqn{N4ThreeCut} was obtained at the level of the 
integrands, with all states carrying four-dimensional momenta and helicities, 
but then it was argued that the light-cone superspace
power-counting of Mandelstam~\cite{Mandelstam} ruled out 
corrections coming from the $(-2\eps)$-dimensional components of the loop
momenta.  Since this argument is based on superspace cancellations it
applies to the integrands before integration over loop momenta, and works
for $D$-dimensional external states as well.  (A similar argument is also 
applied in \app{CutExactnessSubsection}.)

The second sum over $N=4$ states is similar, but more complicated,
involving three different cuts of planar double-boxes and ten of
non-planar ones,
$$
\eqalign{
\sum_{N=4\ {\rm states}} & 
A_5^{\rm tree}(1, \ell_1, \ell_3, 2, \ell_2)
A_5^{\rm tree}(3, -\ell_3, -\ell_1, 4, -\ell_2)  =
- i \, s t A_4^{\rm tree}(1,2,3,4)  \cr 
&  \times
\Bigl[ 
- {s \over (\ell_1+\ell_3)^2  (\ell_3+k_2)^2 (\ell_1+k_1)^2 (\ell_2-k_3)^2}
+{t \over (\ell_1+k_1)^2 (\ell_2+k_2)^2 (\ell_2-k_3)^2(\ell_1-k_4)^2}\cr
& \hskip .3 cm 
+{t \over (\ell_1+k_1)^2 (\ell_3+k_2)^2 (\ell_1-k_4)^2 (\ell_2-k_3)^2} 
-{s \over (\ell_1+\ell_3)^2 (\ell_2+k_2)^2 (\ell_3-k_3)^2 (\ell_1-k_4)^2}\cr
& \hskip .3 cm 
+{t \over (\ell_1+k_1)^2 (\ell_2+k_2)^2 (\ell_1-k_4)^2 (\ell_3-k_3)^2}
+{t \over (\ell_1+k_1)^2 (\ell_3+k_2)^2 (\ell_1-k_4)^2 (\ell_3-k_3)^2} \cr
& \hskip .3 cm 
%
-{u \over (\ell_2+k_1)^2 (\ell_3+k_2)^2 (\ell_1-k_4)^2 (\ell_2-k_3)^2}
-{s \over (\ell_1+\ell_3)^2 (\ell_2+k_1)^2 (\ell_3-k_3)^2 (\ell_1-k_4)^2}\cr
& \hskip .3 cm 
+{t \over (\ell_2+k_1)^2 (\ell_3+k_2)^2 (\ell_3-k_3)^2 (\ell_1-k_4)^2}
%
-{s \over (\ell_1+\ell_3)^2 (\ell_1+k_1)^2 (\ell_3+k_2)^2 (\ell_2-k_4)^2}\cr
& \hskip .3 cm 
-{u \over (\ell_1+k_1)^2 (\ell_2+k_2)^2 (\ell_3-k_3)^2 (\ell_2-k_4)^2} 
+{t \over (\ell_1+k_1)^2 (\ell_3+k_2)^2 (\ell_3-k_3)^2 (\ell_2-k_4)^2} \cr
& \hskip .3 cm 
%
+ {t \over (\ell_2+k_1)^2 (\ell_3+k_2)^2 (\ell_3-k_3)^2 (\ell_2-k_4)^2} \Bigr]
 \,. \cr}
\equn\label{TwistedN4ThreeCut}
$$
The complete $N=8$ result is given by simply inserting the $N=4$ results
(\ref{N4ThreeCut}) and (\ref{TwistedN4ThreeCut}) into
\eqn{ThreeCutSewingRelation}.

%
\begin{figure}[ht]
\centerline{
\epsfxsize 4.5 truein \epsfbox{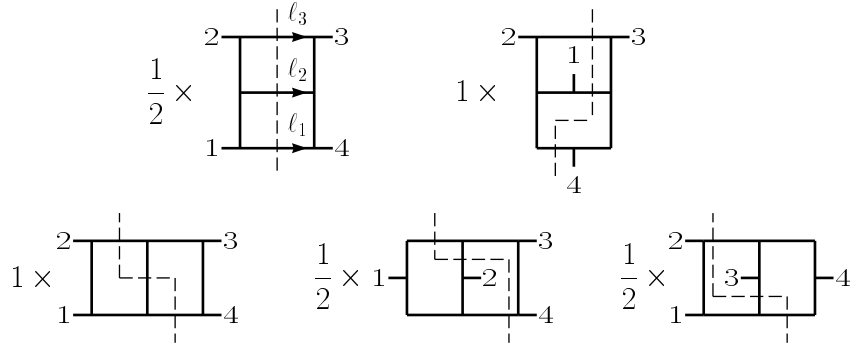} }
\caption[]{
\label{GravityTripleCutFigure}
\small
The three-particle $s$-channel cuts for gravity.  We must also sum
over $1\leftrightarrow 2$, $3\leftrightarrow 4$ and the $3!$
permutations of $\ell_1$, $\ell_2$ and $\ell_3$. We have included 
the appropriate combinatoric factors necessary for eliminating double
counts.}
\end{figure}

This may be compared with the $s$-channel
three-particle cuts of \eqn{TwoLoopGr}. Taking 
all $s$-channel three-particle cuts of the scalar integrals, as shown in 
\fig{GravityTripleCutFigure}, we obtain 
$$
\eqalign{
 s t u & M_4^{\rm tree}   \, 
\int {d^{D} \ell_1 \over (2\pi)^{D}} 
  {d^{D} \ell_2 \over (2\pi)^{D}} \; 
{1\over \ell_1^2\, \ell_2^2\, \ell_3^2} \cr
& \times
\biggl\{ \biggl[ \Bigl( 
{1\over 2}\, {t^2 \over (\ell_1+k_1)^2 (\ell_3+k_2)^2 (\ell_3-k_3)^2
  (\ell_1-k_4)^2}
+{t^2 \over (\ell_2+k_1)^2 (\ell_3+k_2)^2 (\ell_3-k_3)^2 (\ell_1-k_4)^2}
  \cr
& \hskip .5 cm 
+{s^2 \over (\ell_1+\ell_2)^2 (\ell_2+\ell_3)^2 (\ell_3+k_2)^2 (\ell_1-k_4)^2}
+{1\over 2} \, 
{s^2 \over (\ell_2+\ell_3)^2 (\ell_3+k_1)^2 (\ell_2+k_2)^2 
   (\ell_1-k_4)^2}   \cr
& \hskip .5 cm   
+{1\over 2}\, 
{s^2 \over (\ell_1+\ell_2)^2 (\ell_3+k_2)^2 (\ell_2-k_3)^2 (\ell_1-k_4)^2}   
\Bigr)
 + \hbox{perms}(\ell_1, \ell_2, \ell_3) \biggr] \cr
& \hskip 5 cm 
 + \{1 \leftrightarrow 2 \}
   + \{ 3 \leftrightarrow 4\} 
 + \{1 \leftrightarrow 2 ,\, 3 \leftrightarrow 4\} 
\biggr\} \biggr|_{\ell_i^2 = 0} \,,
\cr}
\equn\label{GravAnsatz}
$$
where we have included all kinematic factors associated with each
scalar diagram in \fig{GravityTripleCutFigure}.  The $1
\leftrightarrow 2$ and $3 \leftrightarrow 4$ permutations
act on all terms in the square brackets.  Note that the cut result
(\ref{ThreeCutSewingRelation}) also contains a sum over
$1\leftrightarrow 2$ and $3\leftrightarrow 4$ interchanges.

We have verified that the expressions in eqs.~(\ref{GravAnsatz}) and
(\ref{ThreeCutSewingRelation}) are indeed equal.  The $t$- and 
$u$-channel cuts are identical, up to relabelings.  Since all two- and
three-particle cuts are correctly given by \eqn{TwoLoopGr} in
arbitrary dimensions, it is the complete expression for the two-loop
four-point $N=8$ supergravity amplitude, expressed in terms of scalar 
loop integrals.


\section{Supersymmetric Cancellations in MHV Cuts to All Loop Orders}
\label{MultiParticleCutsSection}

In this section we illustrate four-dimensional supersymmetric
cancellations that take place at {\it any} loop order, in all cuts where the
amplitudes on both sides of the cut are maximally helicity violating
(MHV).  The MHV contributions are the most tractable, which is why we
focus on them.  Supersymmetry Ward identities only relate amplitudes
with the same total helicity; thus when considering supersymmetric
cancellations the MHV contributions do not mix with the non-MHV
contributions.  This section provides a concrete demonstration 
of the role that the SWI's play in reducing the degree of divergence of
supersymmetric $S$-matrix elements.  For MHV cuts at any loop order we
will obtain the same supersymmetric reduction in powers of loop
momentum as we obtained in the entirely two-particle constructible
contributions, providing additional evidence that the power-counting formulas
(\ref{N4Finiteness}) and (\ref{N8Finiteness}) hold.

As a warm-up, we will first consider the case of $N=4$ Yang-Mills theory.
Since the supersymmetry identities of $N=8$ supergravity are closely
related to those of $N=4$ Yang-Mills theory, it is relatively simple to
extend the $N=4$ analysis to $N=8$ supergravity.

\subsection{$N=4$ Yang-Mills Theory Warm-up}

Consider the case of $N=4$ supersymmetric Yang-Mills theory.  The
$m$-particle $s$-channel cut of an $L$-loop amplitude is,
$$
\eqalign{
&{\cal A}_{4}^{\Lloop}(1^-, 2^-, 3^+, 4^+)
\Bigr|_{m\rm \hbox{-}cut} \! \!
= \int\!{d^{D}\ell_1\over (2\pi)^{D}} \, 
{d^{D}\ell_2\over (2\pi)^{D}} \;
\cdots
{d^{D}\ell_{m-1}\over (2\pi)^{D}} \;
\hskip - .2 cm  \sum_{N=4 \ \rm states} \hskip - .2 cm 
{\cal A}_{m+2}^{r \mbox{-} \rm loop} 
  (1^-,2^-,\ell_m,\ell_{m-1},\ldots,\ell_1)\cr
& \hskip 5.3 cm \times
\,{i\over \ell_1^2} \, {i\over \ell_2^2 } \,\ldots \,{i\over \ell_m^2} 
\, {\cal A}_{m+2}^{(L-r-m+1) \mbox{-} \rm loop}
       (3^+,4^+,-\ell_1, -\ell_2, \ldots, -\ell_m) 
\biggr|_{\ell_i^2 = 0} \, ,\cr}
\equn
$$
where we have suppressed color and particle labelings of intermediate
states.  The complete sum over $N=4$ states consists of both MHV and
non-MHV contributions.  (For the case of two- and three-particle cuts,
all the contributions are MHV.)  We need only consider the $s$-channel
since the other channels are related via the supersymmetry identity
(\ref{PermIdentityYM}).  The number of distinct MHV particle
configurations (not counting the SO(4) multiplicities) that can cross an
$m$-particle $s$-channel cut is
$$
{(M+3)!\over 4! \, (M-1)!} + 1 \, , 
\equn
$$
where $M = m(m-1)/2$.  Although this is a rapidly increasing function
of the number of cut lines, $m$, $N=4$ supersymmetry will ensure that the
various configurations combine neatly.  

\begin{figure}[ht]
\centerline{ \epsfxsize 2.5 truein \epsfbox{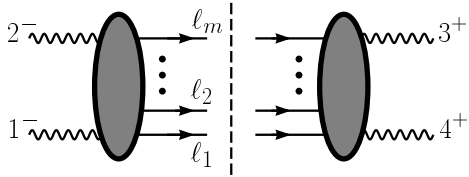} }
\vskip -.5 cm
\caption[]{
\label{NParticleCutFigure} 
An $m$-particle cut of an $L$-loop amplitude. }
\end{figure}

In order to perform the analysis, it is useful to divide the $s$-channel
MHV contributions into a `singlet' piece and `non-singlet' pieces,
analogous to the separation used in the two-loop three-particle cut analysis
of \app{ThreeParticleCutsAppendix}.  The singlet piece consists of the
single configuration where three identical-helicity gluons cross the cut;
this piece does not have to be combined with any other terms to obtain
supersymmetric cancellations.  The singlet and non-singlet pieces may
also be characterized in terms of the total helicity of the amplitudes
appearing on the left- and right-hand sides of the cut in
\fig{NParticleCutFigure}.  For the singlet, the total helicity of the
amplitude on the left of the cut is $m-2$ while the total helicity of
the amplitude on the right of the cut is $-(m-2)$.  For the
non-singlet contributions, the total helicities on the left- and right-hand
sides of the cut are reversed.  Since supersymmetry identities only
relate amplitudes with the same total helicity, supersymmetric
cancellations do not occur between the singlet and non-singlet
contributions. 
(For the case of $m=2$ there is only a singlet contribution.)

Using the $N=4$ supersymmetry identity (\ref{PermIdentityYM}), we may
relate the amplitude appearing on the left-hand side of the singlet cut
with an amplitude that appears as a non-singlet contribution and
exhibits no supersymmetric cancellation by itself,
$$
 {\cal A}_{m+2}(1^-, 2^-, \ell_1^+, \ldots \ell_m^+) = 
{\spa1.2^4 \over \spa{\ell_i}.{\ell_j}^4} \,
[ {\cal A}_{m+2}(1^-, 2^-, \ell_1^-,\ldots, \ell_{i-1}^-, \ell_i^+,
\ell_{i+1}^-,\ldots, \ell_{j-1}^-, 
\ell_j^+, \ell_{j+1}^-, \ldots, \ell_m^-) ]^\dagger \,,
\equn\label{SingletSWI}
$$
where the dagger means that we interchange 
$\spa{a}.{b} \leftrightarrow \spb{b}.{a}$.  
A similar identity holds for the amplitude
appearing on the right-hand side of the singlet cut.  This explicitly
shows that we may extract a factor of $\spa1.2^4 \spb3.4^4$ from the
singlet cut, as compared to a contribution with no supersymmetric
cancellations.  

More interestingly, the sum over non-singlet contributions exhibits 
a cancellation dictated by the supersymmetry Ward identities.
\Tab{NonSingletSWI} of \app{SusyIdentityAppendix} contains the
relative SWI factors between the
amplitudes that appear on the right-hand side of the cuts.  In
\tab{NonSingletSewingTable} we have collected the various types of
contributions to the non-singlet MHV cuts; all remaining such
contributions are given by relabelings.  Each entry in the first
column of \tab{NonSingletSewingTable} is obtained by multiplying one
of the 23 entries in \tab{NonSingletSWI} with the conjugate factor
from the amplitude on the left-hand side of the cut.  (The conjugate
amplitudes with two cut fermion lines have an additional minus sign, making 
up for the fermion loop sign.)

Collecting all MHV non-singlet contributions together yields an 
overall factor of 
$$
\Bigl( \sum_{i \not= j} a_{ij} \Bigr)^4 = s^4 
\equn\label{MHVcancel}
$$
in the amplitude, where 
$a_{ij} = (\ell_i + \ell_j)^2 = 2\ell_i\cdot\ell_j$.  
Equation~(\ref{MHVcancel}) is an
explicit all-loop demonstration of the reduction in powers of loop
momentum implied by supersymmetry.  The SWI (\ref{PermIdentityYM}) implies
that the same types of cancellations occur for the $t$- and
$u$-channel cuts, although different states cross the cuts.


\begin{table}[ht]
\begin{center}
\begin{tabular}{|c|c|}
\hline
$\vphantom{\Big|}$ {\bf Relative Factor} & {\bf Multiplicity} \\
\hline
$\vphantom{\Big|^A} a_{12}^4$ & 1 \\
\hline
$\vphantom{\Big|^A} a_{12}^3 a_{34}$\,,\,
$a_{12}^3 a_{23}$ & 4 \\
\hline
$\vphantom{\Big|^A} a_{12}^2 a_{34}^2$\,,\,
     $a_{12}^2 a_{23}^2$ & 6 \\
\hline
$\vphantom{|^{{B^A}^C}} a_{12}^2 a_{34} a_{56} $\,, 
 & \\
$a_{12}^2 a_{23} a_{45} $\,,\,
$a_{12}^2 a_{34} a_{45} $\,,\,
$ a_{12}^2 a_{23} a_{34} $ & 12 \\
$ a_{12}^2 a_{23} a_{14} $\,,\,
$ a_{12}^2 a_{23} a_{13} $\,,\,
$ a_{12}^2 a_{13} a_{14} $ & \\
\hline
$ \vphantom{|^{{B^A}^C}} a_{12} a_{34}
             a_{56} a_{78} $\,,\,
$a_{12}a_{23}
           a_{45} a_{67}$\,, & \\
$a_{12} a_{23}a_{34} a_{56}$\,,\,
$a_{12} a_{23} a_{45} a_{56}$\,,\,
$a_{12}a_{23} a_{13} a_{45}$ & \\
$a_{12} a_{23} a_{34}a_{45}$\,,\,
$a_{12} a_{23} a_{34} a_{41}$\,,\,
$a_{12} a_{13} a_{14} a_{56}$ & 24 \\
$a_{12} a_{13} a_{14}
         a_{45}$\,,\,
$a_{12} a_{13} a_{14} a_{42}$\,,\,
$a_{12} a_{13}a_{14} a_{15}$ & \\
\hline
\end{tabular}
\end{center}
\small
\caption[]{The relative numerator factors dictated by the
supersymmetry Ward identities, arising from the non-singlet sewing of
two MHV $N=4$ Yang-Mills amplitudes, where $a_{ij} = (\ell_i +
\ell_j)^2$.  The 23 different terms in the left column of the table
correspond to the 23 different particle configurations in
\tab{NonSingletSWI}, which contains the relative factor for the amplitude
on the right-hand side of the cut.
}
\label{NonSingletSewingTable}
\end{table}

\subsection{Supersymmetry Cancellations in $N=8$ Supergravity}

The analysis of the $m$-particle cuts at $L$ loops for
the case of $N=8$ supergravity is quite similar to that 
of $N=4$ Yang-Mills theory. 
For $N=8$ supergravity, the total number of distinct MHV particle
configurations that can cross the cut is,
$$
{(M+7)!\over 8! \, (M-1)!} + 1 \,,
\equn
$$
where $M = m(m-1)/2$.  These MHV contributions to the $m$-particle cuts
can again be divided into singlet and non-singlet contributions,
characterized by the total helicity of the amplitudes appearing on 
either side of the cuts.
For the singlet, the total helicity of the amplitude on the left side of
the cut is $2(m-2)$, while the total helicity of the amplitude on the
right of the cut is $-2(m-2)$.  For the non-singlet contributions,
the total helicities on the left and right sides of the cut are
reversed.  

As for $N=4$ Yang-Mills theory, the simplest case to consider is the singlet
where $m$ identical-helicity gravitons cross the cut.  For the
singlet, the SWI (\ref{PermIdentity}) implies that one can extract the
factor $\spa1.2^8 \spb3.4^8$ from the cut integrand as compared to a
factor of $\spa{\ell_i}.{\ell_j}^8 \spb{\ell_i}.{\ell_j}^8$ for a pure
graviton contribution to the non-singlet.  Note that these factors are
the squares of the corresponding $N=4$ Yang-Mills factors. 

For the non-singlet case to which all remaining MHV configurations
contribute, we need to combine the various terms in order to obtain
a reduction in the degree of divergence.  An easy way to determine
the relative factors between amplitudes appearing on either side of
the non-singlet cut is from the KLT relations on tree amplitudes.
For $(m+2)$-point gravity tree amplitudes these relations are of the form
$$
\eqalign{
M_{m+2}^{\rm tree}(1, 2, \ell_1, \ldots, \ell_{m} ) \sim &
\sum_{\{ i\} \{ j \} }
\biggl(\, \prod_{m-1\ \rm factors} (\ell_p + \ell_q)^2  \biggr)
A_{m+2}^{\rm tree}(1, \ell_{i_1}, \ldots,\ell_{i_n},2,\ell_{i_{n+1}},
\ldots,\ell_{i_{m}}) \cr
& \hskip 2cm \times
A_{m+2}^{\rm tree}(1,\ell_{j_1},\ldots,\ell_{j_k},2,\ell_{j_{k+1}},
\ldots,\ell_{j_{m}})\,, \cr }
\equn\label{KLTPowerCount}
$$
where $\{i_1,i_2,\ldots,i_{m}\}$ and $\{j_1,j_2,\ldots,j_{m}\}$ are
permutations of $\{3,4,\ldots,m+2\}$ which are determined from the KLT
expressions.  Since the relative factors for any MHV $N=4$ tree
amplitude are known from the SWI, the KLT relations (\ref{KLTPowerCount})
immediately imply that they are known for any MHV $N=8$ tree
amplitude.  Furthermore, since the supersymmetry Ward identities hold
to {\it all} orders of the perturbative expansion, the relative factors
determined in this way are the same ones that appear at loop level.  A
more direct analysis using the $N=8$ Ward identities yields the same
relative factors.

After including the appropriate multiplicity factors, following the
previous discussion for $N=4$ Yang-Mills theory, again we may combine all
relative factors in the non-singlet MHV $s$-channel cuts to obtain an
overall factor of
$$
\Bigl(\sum_{i\not = j} a_{ij} \Bigr)^8
= s^8 \,,
\equn
$$ 
which may be extracted from the cut integral.  
The symmetry under permutations of external legs implied by the 
SWI~(\ref{PermIdentity}) guarantees that one can also extract the same
factor from the $t$- and $u$-channel MHV cuts.

In summary, the above analysis illustrates that supersymmetry
cancellations in the MHV cuts occur to all loop orders, for the
cases of $N=4$ Yang-Mills theory and $N=8$ supergravity.  
In fact, these cancellations are closely related to those for the
two-particle sewing relations, when one negative gluon helicity appears
on each side of the cut.  (See eq. (5.6) of ref.~\cite{SusyFour}.) 
Since these sewing relations led to the power-counting 
eqs.~(\ref{N4Finiteness}) and (\ref{N8Finiteness}), 
the MHV results presented in this section suggest that these power 
counts hold more generally.



\section{Conclusions}
\label{ConclusionsSection}

In this paper we investigated the relationship between gravity and
gauge theory $S$-matrix elements.  Supergravity theories are
interesting because they have softer ultraviolet behavior than ordinary
gravity theories and also appear in the low-energy limit of
superstring theories.  We chose to investigate the case of
maximally supersymmetric theories (i.e. $N=4$ Yang-Mills theory and
$N=8$ supergravity); such theories are heavily constrained and
their amplitudes are therefore relatively simple.

At tree and one-loop level the four-point amplitudes for $N=4$
Yang-Mills theory and $N=8$ supergravity exhibit a `squaring'
relationship (summarized in
\figs{TreeRelationFigure}{OneLoopRelationFigure}).  One of the results
contained in this paper is a complete calculation of the $N=8$
two-loop four-point amplitude in terms of scalar integrals, similar to
the two-loop $N=4$ Yang-Mills calculation performed in
ref.~\cite{BRY}.  From these amplitudes we have obtained the
ultraviolet divergences in $D=7,9,11$, and have shown that the
squaring relationship persists to two loops (as depicted in
\fig{TwoLoopRelationFigure}).  Motivated by these results and by the
structure of string amplitudes, we postulate that a similar
relationship will extend to all loop orders.

Evidence for this conjecture is provided by cutting
techniques~\cite{SusyFour,SusyOne,Review,BRY}, which allow a complete
analytic reconstruction of an $L$-loop amplitude, if one computes all
possible cuts, up to and including the $(L+1)$-particle cuts.  For the
$N=8$ four-point amplitude we have iterated the one- and two-loop
two-particle cut analysis to {\it all} loop orders.  To do this we used
the known~\cite{KLT,BGK} relationship between gravity and gauge theory
tree amplitudes to recycle the previous calculation of the $N=4$
Yang-Mills cuts~\cite{BRY}.  Further checks of the conjecture could be
provided by cuts with more intermediate states, though this becomes
increasingly difficult.  Nevertheless, the cutting method is
overwhelmingly simpler than a corresponding Feynman diagram calculation.

Of supergravity theories, $N=8$ is believed to have the softest
ultraviolet behavior, so a complete proof that it actually does
diverge in four dimensions would provide strong evidence that all 
supergravity theories are divergent.  
Although superspace power-counting arguments~\cite{HoweStelle}
suggest the existence of a three-loop $D=4$ counterterm, its
presence has never been verified in an actual calculation.  
Our results indicate that for $N=4$ Yang-Mills theory and for $N=8$ 
supergravity these superspace power-counting arguments were too conservative
regarding the possible cancellation of infinities.  In particular, our
computation of the $N=8$ two-loop amplitudes prove that there is no
counterterm in $D=5$, contrary to expectations from superspace power
counting~\cite{HoweStelleTownsend,HoweStelle}.

Furthermore, in four dimensions our cut calculations indicate, but do not
yet prove, that there is no three-loop counterterm for $N=8$ supergravity,
contrary to expectations from superspace power-counting bounds.  On the
other hand, from the two-particle cuts we infer a counterterm at five
loops with non-vanishing coefficient.  In order to have a complete proof
that the divergence detected at five loops does not cancel, one would need
to investigate the cuts with up to six intermediate particles.
Alternatively, if one could prove that the numerators of all $N=8$
loop-momentum integrals are squares of the corresponding ones for $N=4$
Yang-Mills integrals (i.e. they always appear with the same sign), there
would be no need for a detailed investigation of the cuts; we have shown
that the iterated two-particle cuts have the required squaring property.
It would also be of interest to directly verify whether or not the 
coefficient of the three-loop counterterm vanishes for the case of 
$N=1$ supergravity.  It should be possible to address
these and other issues with the methods used in this paper.

Throughout the calculations presented in this paper we have found
simple relationships between $N=4$ Yang-Mills theory and $N=8$
supergravity.  Without fail we have been able to exploit the $N=4$
Yang-Mills results to streamline $N=8$ supergravity calculations.
It would be interesting to investigate whether such
relationships between gravity and gauge theories can be found for
cases with less than the maximal supersymmetry.

\vskip .3 cm 
{\bf Acknowledgements}

We thank K.S. Stelle for a number of important discussions and for
encouragement.  We also thank S. Deser, S. Ferrara, D.R.T. Jones,
R. Kallosh, D.A. Kosower, E. Marcus, and E.T. Tomboulis for
discussions.


\appendix

\section{Two-particle Cuts Via Helicity}
\label{TwoParticleCutsAppendix}

In this appendix we obtain the two-particle sewing equation
(\ref{BasicGravityCutting}) using a helicity basis.  Four-dimensional
helicity methods make explicit calculations quite a bit easier,
especially when the momenta also reside in four dimensions. However,
restricting the momenta in the cuts to four dimensions may in
principle drop terms necessary for continuation to other dimensions.
Nonetheless, the four-dimensional calculation makes the formal proof
of \eqn{BasicGravityCutting} transparent and gives us the opportunity
to introduce the notation which we use in the following appendix, where
we calculate the two-loop three-particle cuts using a helicity basis.

In four dimensions a massless external particle will have either
positive or negative helicity.  
The four-graviton tree amplitudes $M_4^{\rm tree}(1^+, 2^+,3^+, 4^+)$ 
and $M_4^{\rm tree} (1^-, 2^+, 3^+, 4^+)$ vanish
for all particle types.  This can be seen from \eqn{GravYMFour}, since
the corresponding gauge theory amplitudes vanish.  For a supergravity
theory, this vanishing is true not just at tree-level, but at each 
order in perturbation theory, due to supersymmetric Ward
identities~\cite{SWI}.  Thus, the only independent (under relabelings)
non-zero four-graviton amplitude is $M_4(1^-, 2^-,3^+,4^+)$.

Therefore, in the sewing equation (\ref{BasicGravityCutting})
we need only consider a single helicity configuration for the external
legs, which for convenience, we take to be $(1^-, 2^-, 3^+, 4^+)$. 
There are three possible sewings: the $s$-, $t$- and $u$-channels. 

For the $s$-channel, the  left-hand side of \eqn{BasicGravityCutting} is
$$
\sum_{\lambda, \lambda^\prime} 
 M_4^{\rm tree}(-\ell_1^{\lambda^\prime},  1^-, 2^-, \ell_2^\lambda) \times
  M_4^{\rm tree}(-\ell_2^{-\lambda}, 3^+, 4^+, \ell_1^{-\lambda^\prime} ) \,,
\equn
$$
where the sum runs over all helicities of the $N=8$ supermultiplet,
$\{-2, -\threehalf, -1, -\half, 0, \half, 1, \threehalf, 2\}$.  Again
using supersymmetric Ward identities~\cite{SWI} $M_4^{\rm
tree}(-\ell_1^\lambda, 1^-, 2^-, \ell_2^{\lambda^\prime})$ will vanish
unless the intermediate states are both gravitons of identical
helicity. The sum over $N=8$ states then reduces to the single term,
$$
M_4^{\rm tree}(-\ell_1^+,  1^-, 2^-, \ell_2^+) \times
  M_4^{\rm tree}(-\ell_2^-, 3^+, 4^+, \ell_1^- ) \, , 
\equn\label{sGravitySum}
$$
where all legs are gravitons.  From \eqn{TreeSquared} the two
non-vanishing tree amplitudes are
$$
\eqalign{
M_4^{\rm tree}(-\ell_1^+, 1^-, 2^-, \ell_2^+) &  = 
-i \biggl(s {\spa1.2 \over \spb1.2} 
{\spb{\ell_1}.{\ell_2} \over \spa{\ell_1}.{\ell_2}} \biggr)^2 
\biggl[ {1\over (\ell_1 - k_1)^2 } + {1\over (\ell_1 - k_2)^2} \biggr]\,, \cr
M_4^{\rm tree}(-\ell_2^-, 3^+, 4^+, \ell_1^-) & = 
-i \biggl(s {\spb3.4 \over \spa3.4} 
{\spa{\ell_1}.{\ell_2} \over \spb{\ell_1}.{\ell_2}} \biggr)^2 
\biggl[ {1\over (\ell_2 - k_3)^2 } + {1\over (\ell_2 - k_4)^2} \biggr]\,. \cr}
\equn
$$
Inserting these into~\eqn{sGravitySum} 
yields
$$
\eqalign{
M_4^{\rm tree}(-\ell_1^+, & 1^-, 2^-, \ell_2^+) \times
  M_4^{\rm tree}(-\ell_2^-, 3^+, 4^+, \ell_1^-) \cr \
& 
= -\biggl(  s^2 {\spa1.2 \over \spb1.2} {\spb3.4 \over \spa3.4} \biggr)^2
\biggl[ {1\over (\ell_1 - k_1)^2 } + {1\over (\ell_1 - k_2)^2} \biggr]
\biggl[ {1\over (\ell_2 - k_3)^2 } + {1\over (\ell_2 - k_4)^2} \biggr]
\cr
&
= i\,  s t u \, M_4^{\rm tree}(1^-, 2^-, 3^+, 4^+) 
\biggl[{1\over (\ell_1 - k_1)^2 } + {1\over (\ell_1 - k_2)^2} \biggr]
\biggl[{1\over (\ell_2 - k_3)^2 } + {1\over (\ell_2 - k_4)^2} \biggr] \,,\cr}
\equn\label{BasicGravityCuttingApp}
$$
which is the gravity sewing equation (\ref{BasicGravityCutting})
derived in the main body of the paper. 

From the supersymmetry Ward identity (\ref{PermIdentity}) given in
appendix~{\ref{SusyIdentityAppendix}, the $s$-channel is sufficient to
ensure the $t$- and $u$-channels are just given by relabelings
of~\eqn{BasicGravityCuttingApp}.  In these cases the cut receives
contributions from all the states in the $N=8$ multiplet.  When these
contributions are summed \eqn{BasicGravityCuttingApp} is reproduced,
up to relabelings.  The details may be found in ref.~\cite{DN}.


\section{Three-particle Cuts via Helicity}
\label{ThreeParticleCutsAppendix}

In this appendix we explicitly calculate the two-loop three-particle
cuts using four dimensional momenta and helicity in the cuts. 
This is useful because it is a concrete calculation, verifying
the more formal discussion in \sec{ThreeParticleSubSection}.
However, in general, this may drop pieces
containing $(-2\eps)$-dimensional momenta, which can prevent analytic
continuations away from $D=4$. (The discussion of
\secs{TwoParticleSubSection}{ThreeParticleSubSection} is valid in
any dimension, so for the two-loop $N=8$ case there are no such 
dropped pieces.) 

Consider the $s$-channel three-particle cut,
$$
\eqalign{
M_4^{\twoloop}& (1^-,  2^-,3^+,4^+) 
 \Bigr|_{s\mbox{-}{\rm channel}\ 3\rm \hbox{-}cut} 
\cr
& =
\int {d^{D} \ell_1 \over (2\pi)^{D}} 
 {d^{D} \ell_2 \over (2\pi)^{D}} 
\sum_{N=8\ {\rm states}}
M_5^{\rm tree} (1^-, 2^-, \ell_3, \ell_2, \ell_1)
{i\over \ell_1^2} 
{i\over  \ell_2^2} 
{i\over  \ell_3^2} 
M_5^{\rm tree} (3^+, 4^+, -\ell_1, -\ell_2, -\ell_3)
 \Bigr|_{\ell_i^2 = 0} \,, \cr}
\equn\label{ThreeCutHelicity}
$$
where $\ell_1$, $\ell_2$ and $\ell_3$ are the three momenta crossing
the cuts, as in \fig{ThreeParticleFigure}.  In performing the cut
calculation we must sum over all possible intermediate helicity and
particle configurations of the $N=8$ multiplet --- of which there are
46 distinct non-zero possibilities (not counting the $SO(8)$ multiplicities).

We characterize the contributions in terms of the total helicity (i.e.
the sum over helicity of the external states) of the amplitudes.
Since the only non-vanishing $N=4$ five-point Yang-Mills tree
amplitudes are those which have total helicity $+1$ or $-1$, the only
potential contributions to the sum over $N=8$ states are those where the
total helicity is either $+2$, $-2$ or $0$.  However, for the latter
case the amplitudes all vanish as required by an $N=8$ SWI.  This can
be directly verified using the explicit values of the $N=4$ Yang-Mills
tree amplitudes.  For example, for a one-scalar four-graviton helicity zero
amplitude we have,
$$
\eqalign{
M_5^{\rm tree}(\ell_1^s, 1^-, 2^-, \ell_3^+, \ell_2^+) 
& = i (\ell_1 + k_1)^2 (\ell_3 + k_2)^2 
A_5^{\rm tree}(\ell_1^+, 1^-, 2^-, \ell_3^+, \ell_2^+)\,
A_5^{\rm tree}(1^-, \ell_1^-, \ell_3^+, 2^-, \ell_2^+) \cr
& \hskip 3 cm 
+ \{ 1 \leftrightarrow 2 \}\cr
& = i (\ell_1 + k_1)^2 (\ell_3 + k_2)^2 
{\spa1.2^4 \over \spa{\ell_1}.1 \spa1.2 \spa2.{\ell_3} 
 \spa{\ell_3}.{\ell_2} \spa{\ell_2}.{\ell_1} }\,
{\spb{\ell_3}.{\ell_2}^4 \over \spb1.{\ell_1} \spb{\ell_1}.{\ell_3}
       \spb{\ell_3}.2 \spb2.{\ell_2} \spb{\ell_2}.1} \cr
& \hskip 3 cm 
+ \{ 1 \leftrightarrow 2 \} \cr
& = 0 \,. \cr}
\equn
$$
Similarly, all other $N=8$ amplitudes with total helicity $0$ vanish.
(This result can also be obtained using the $SO(4)$ global symmetry.)
The only non-vanishing contributions are the total helicity $+2$ and
$-2$ MHV contributions. 

The calculation thus breaks up into two distinct pieces: a `singlet' and a
`non-singlet' piece corresponding to $+2$ and $-2$ helicity amplitudes
on the left side of the cut in \fig{ThreeParticleFigure}.  
The singlet piece is composed of the single
helicity and particle configuration where three identical helicity
gravitons cross the cut. The `non-singlet' contribution is composed of
45 distinct helicity and particle configurations which must be
combined to get a simple result.  All supersymmetric cancellations
occur within each piece separately because the SWI's do not change the
total helicity.

In performing these calculations we will recycle the algebra that was
used to evaluate the three-particle cuts of the two-loop $N=4$
Yang-Mills amplitudes \cite{BRY}.  This is carried out
by expressing the five-point gravity tree amplitudes in terms of gauge
theory amplitudes using \eqn{GravYMFive}.

\subsection{The `Singlet' Contribution}

First consider the singlet case where three identical-helicity
gravitons cross the cut.  Since the two terms in
\eqn{ThreePartRightCutTree} are related by $1 \leftrightarrow 2$
symmetry, we can focus on the first term, obtaining the second by
symmetry.  Similarly, we need only focus on the first term in
\eqn{ThreePartLeftCutTree} because of $3 \leftrightarrow 4$ symmetry.

In the product of gravity tree amplitudes in \eqn{ThreeCutHelicity} we
have the following two products of Yang-Mills tree amplitudes
$$
\eqalign{
&\hbox{(a)} \hskip .5 cm
 A_5^{\rm tree} (1^-, 2^-, \ell_3^+, \ell_2^+, \ell_1^+) \times
 A_5^{\rm tree} (3^+, 4^+, -\ell_1^-, -\ell_2^-, -\ell_3^-)  \,, \cr
& \hbox{(b)} \hskip .5 cm
 A_5^{\rm tree}(1^-, \ell_1^+, \ell_3^+, 2^-, \ell_2^+) \times
 A_5^{\rm tree}(3^+, -\ell_3^-, -\ell_1^-, 4^+, -\ell_2^-) \,. \cr}
\equn
$$

These products have already been evaluated in the computation of the $N=4$
Yang-Mills amplitudes~\cite{BRY}.  To evaluate the product (a)
we use the explicit forms of the Yang-Mills amplitudes,
$$
\eqalign{
A_5^{\rm tree} (1^-, 2^-, \ell_3^+, \ell_2^+, \ell_1^+) & = 
i {\spa1.2^4 \over \spa1.2 \spa2.{\ell_3} \spa{\ell_3}.{\ell_2} 
      \spa{\ell_2}.{\ell_1} \spa{\ell_1}.1} \,, \cr
A_5^{\rm tree} (3^+, 4^+, -\ell_1^-, -\ell_2^-, -\ell_3^-) & = 
- i {\spb3.4^4 \over \spb3.4 \spb4.{-\! \ell_1} \spb{-\ell_1}.{-\!\ell_2} 
      \spb{-\ell_2}.{-\!\ell_3} \spb{-\ell_3}.3}\,, \cr}
\equn
$$
giving, 
$$
\eqalign{
A_5^{\rm tree} (1^-, 2^-, & \ell_3^+, \ell_2^+, \ell_1^+)  
\times 
A_5^{\rm tree} (3^+, 4^+, -\ell_1^-, -\ell_2^-, -\ell_3^-) \cr
& = - \spa1.2^2 \spb3.4^2 
 {\trplus [1 \ell_1 43 \ell_3 2] \over 
 (\ell_1 + \ell_2)^2 (\ell_2 + \ell_3)^2 (\ell_3 - k_3)^2 
(\ell_1 - k_4)^2 (\ell_3 + k_2)^2 (\ell_1 + k_1)^2} \,, \cr}
\equn
$$
where $\trplus [1 \ell_1 \cdots] = \tr [(1+\gamma_5) \ksl_1 \lsl_1 \cdots]/2$.
Similarly, for the product (b),
$$
\eqalign{
 & A_5^{\rm tree}(1^-, \ell_1^+, \ell_3^+, 2^-, \ell_2^+)  \times
   A_5^{\rm tree}(3^+, -\ell_3^-, -\ell_1^-, 4^+, -\ell_2^-) \cr
& \hskip0.1cm
 = -{\spa1.2^2 \spb3.4^2 \, \trplus[1 \ell_1 43 \ell_2 2] \,
     \trplus[ 1    \ell_2 4 3 \ell_3 2   ]
    \over (\ell_1 + \ell_3)^2
     (\ell_1 + k_1)^2 (\ell_3+k_2)^2 (\ell_2 + k_2)^2 (\ell_2 + k_1)^2 
     (\ell_3-k_3)^2 (\ell_1 - k_4)^2 (\ell_2 - k_4)^2 (\ell_2 - k_3)^2}
\,.\cr}
\equn
$$

Putting together the type (a) and type (b) contributions with the 
prefactors contained in eqs.~(\ref{ThreePartRightCutTree}) and 
(\ref{ThreePartLeftCutTree}) we therefore have for the singlet 
contributions to the cut (\ref{ThreeCutHelicity}), 
$$
\eqalign{
M_4^\twoloop(1^-,2^-,& 3^+,4^+)\Bigr|_{\rm singlet\ 3\hbox{-}cut} = 
s t u M^{\rm tree}_4(1^-,2^-,3^+,4^+) 
\int {d^{D} \ell_1 \over (2\pi)^{D}} 
  {d^{D} \ell_2 \over (2\pi)^{D}} \cr
&  \hskip .5 cm \times
\biggl[
{\trplus [1 \ell_1 43 \ell_3 2 1 \ell_1 43 \ell_2 2 1 \ell_2 4 3 \ell_3 2] 
\over\ell_1^2 \, \ell_2^2 \, \ell_3^2 \, (\ell_3 - k_3)^2 
    (\ell_1 - k_4)^2 (\ell_3 + k_2)^2 (\ell_1 + k_1)^2\; 
    (\ell_1 + \ell_2)^2 (\ell_2 + \ell_3)^2 } \cr
&  \hskip .5 cm \times
{1\over 
 (\ell_1 + \ell_3)^2 
     (\ell_2 + k_2)^2 (\ell_2 + k_1)^2
      (\ell_2 - k_4)^2 (\ell_2 - k_3)^2}  \cr
& \hskip 4 cm 
+  \{1 \leftrightarrow 2 \} + \{3 \leftrightarrow 4 \}
 + \{1 \leftrightarrow 2 ,\, 3 \leftrightarrow 4 \}
 \biggr]
 \biggr|_{\ell_i^2 = 0} \,, \cr}
\equn\label{TargetTripleGravity}
$$
where we used
$$
\eqalign{
\trplus [1 \ell_1 & 43 \ell_3 2] \trplus[1 \ell_1 43 \ell_2 2] 
 \trplus[ 1    \ell_2 4 3 \ell_3 2   ]       
= 
\trplus [1 \ell_1 43 \ell_3 2 1 \ell_1 43 \ell_2 2 1 \ell_2 4 3 \ell_3 2] \,.
\cr }
\equn\label{GravTrace}
$$
This completes the calculation of the singlet contribution.
(The identical particle phase-space factor of $1/3!$ is canceled by the
$3!$ permutations of the $\ell_i$.)  After combining this result with the 
non-singlet contribution, computed in the following subsection, we will 
compare the total with the three-particle cuts of the final form of the 
amplitude, \eqn{TwoLoopGr}.

\subsection{The Non-Singlet Contribution}
\label{NonSingletGravitySubsection}

We now include the sum over all remaining helicity and particle
configurations that can cross the cut.  As we shall see, there are a
total of 45 `non-singlet' contributions which combine to give a
contribution which is almost identical to the singlet case.
The first step in computing the cuts is to organize the set of 
tree amplitudes that will contribute.

A simple way to obtain these tree amplitudes is via the KLT relations
(\ref{ThreePartRightCutTree}) and (\ref{ThreePartLeftCutTree}).  
The factorizations of the required gravity
tree amplitudes into gauge theory partial amplitudes via
\eqn{ThreePartRightCutTree} are collected in the second and fourth
columns of \tab{FactorizationTable}.  The third and fifth columns
contain the relative factors between the various gauge theory amplitudes.
The pure gluon amplitude is 
$$
A_5^{\rm tree}(3^+, 4^+, \ell_1^-, \ell_2^-, \ell_3^+) = 
i {\spa{\ell_1}.{\ell_2}^4 \over \spa3.4 \spa4.{\ell_1} \spa{\ell_1}.{\ell_2}
                                 \spa{\ell_2}.{\ell_3} \spa{\ell_3}.3} \,.
\equn
$$
The entry in the first row and third column of \tab{FactorizationTable}
corresponds to the factor $\spa{\ell_1}.{\ell_2}^4$ appearing in the
numerator. 
As discussed in \app{SusyIdentityAppendix}, the relative factors between 
the various amplitudes are
easily obtained via supersymmetry Ward identities~\cite{SWI} and are
same ones used in the $N=4$ Yang-Mills two-loop calculation~\cite{BRY}.
For example, from supersymmetry Ward identities we have
$$
A_5^{\rm tree}(3^+, 4^+, \ell_1^-, \ell_{2\gT}^-, \ell_{3\gT}^+) = 
{\spa{\ell_1}.{\ell_3} \over \spa{\ell_1}.{\ell_2} } \, 
A_5^{\rm tree}(3^+, 4^+, \ell_1^-, \ell_2^-, \ell_3^+)
 = i {\spa{\ell_1}.{\ell_2}^3 \spa{\ell_1}.{\ell_3}
    \over \spa3.4 \spa4.{\ell_1} \spa{\ell_1}.{\ell_2}
           \spa{\ell_2}.{\ell_3} \spa{\ell_3}.3 } \,,
\equn
$$
whose numerator corresponds to the entry in the second row and third column of
\tab{FactorizationTable}.

\begin{table}[ht]
\hbox{
\def\tend{\cr \noalign{\hrule}}
\def\hs{\hskip .2 cm}
\vbox{\offinterlineskip
{
\hrule
\halign{
        &\vrule#
        &\strut\hs#\hfil\vrule
        &\hs\hfil\strut # \hfil \vrule
        &\hs\hfil\strut # \hfil \vrule
        &\hs\hfil\strut # \hfil \vrule
        &\hs\hfil\strut # \hfil
        \cr
height13pt  &{\bf Amplitude}  &{\bf Left YM} & {\bf Left Factor}
            &{\bf Right YM} & {\bf Right Factor} & \tend
height12pt & $h^+ h^+ h^- h^- h^+$
&  $g^+ g^+ g^- g^- g^+$  &  $\spa\ell_1.{\ell_2}^4$
&  $g^+ g^+ g^- g^- g^+$  &  $\spa\ell_1.{\ell_2}^4$ & \tend
height12pt & $h^+ h^+ h^- \tilde h^- \tilde h^+ $
&  $g^+ g^+ g^- \tilde g^- \tilde g^+$   
&  $\spa\ell_1.{\ell_2}^3\spa\ell_1.{\ell_3}$
&  $g^+ g^+ g^- g^- g^+ $   & $\spa\ell_1.{\ell_2}^4$ &
\tend
height12pt & $h^+ h^+ h^- v^- v^+$
& $g^+ g^+ g^- s^- s^+ $ & $\spa\ell_1.{\ell_2}^2\spa\ell_1.{\ell_3}^2$
& $g^+ g^+ g^- g^- g^+ $ & $\spa\ell_1.{\ell_2}^4$ & \tend
height12pt & $h^+ h^+  h^- \tilde v^- \tilde v^+$
& $g^+ g^+  g^- s^- s^+ $ & $\spa\ell_1.{\ell_2}^2\spa\ell_1.{\ell_3}^2$
& $g^+ g^+ g^- \tilde g^- \tilde g^+$ 
& $\spa\ell_1.{\ell_2}^3\spa\ell_1.{\ell_3}$ & \tend
height12pt & $h^+ h^+ h^- s^- s^+ $
& $g^+ g^+ g^- s^- s^+$   & $\spa\ell_1.{\ell_2}^2\spa\ell_1.{\ell_3}^2$ 
&  $g^+ g^+ g^- s^- s^+$  & $\spa\ell_1.{\ell_2}^2\spa\ell_1.{\ell_3}^2$
& \tend
%
height12pt & $h^+ h^+ \tilde h^- \tilde h^- v^+$
& $g^+ g^+ \tilde g^- \tilde g^- s^+$ 
& $i\spa\ell_1.{\ell_2}^2\spa\ell_1.{\ell_3} \spa\ell_2.{\ell_3}$ 
& $g^+ g^+  g^- g^- g^+$ & $\spa\ell_1.{\ell_2}^4$ &\tend
height12pt & $h^+ h^+ \tilde h^- v^- \tilde v^+$
& $g^+ g^+ g^- s^- s^+$ & $\spa\ell_1.{\ell_2}^2\spa\ell_1.{\ell_3}^2$ 
& $g^+ g^+ \tilde g^- g^- \tilde g^+$ &
$-\spa\ell_1.{\ell_2}^3\spa\ell_2.{\ell_3}$ 
&\tend
height12pt & $h^+ h^+ \tilde h^- \tilde v^- s^+$
& $g^+ g^+  g^- s^- s^+$ & $\spa\ell_1.{\ell_2}^2\spa\ell_1.{\ell_3}^2$
& $g^+ g^+ \tilde g^- \tilde g^- s^+$ 
& $i\spa\ell_1.{\ell_2}^2\spa\ell_1.{\ell_3}\spa\ell_2.{\ell_3}$ & \tend
height12pt & $h^+ h^+ v^-  v^- s^+$
& $g^+ g^+ \tilde g^-  \tilde g^- s^+$ 
& $i\spa\ell_1.{\ell_2}^2\spa\ell_1.{\ell_3}\spa\ell_2.{\ell_3}$
& $g^+ g^+ \tilde g^-  \tilde g^- s^+$ 
& $i\spa\ell_1.{\ell_2}^2\spa\ell_1.{\ell_3}\spa\ell_2.{\ell_3}$ &\tend
height12pt & $h^+ h^+ v^-  \tilde v^- \tilde v^-$
& $g^+ g^+ g^-  s^- s^-$ & $-\spa\ell_1.{\ell_2}^2\spa\ell_1.{\ell_3}^2$
& $g^+ g^+ s^-  \tilde g^- \tilde g^-$ 
& $i\spa\ell_1.{\ell_2}\spa\ell_1.{\ell_3} \spa\ell_2.{\ell_3}^2$ &\tend
}
}
}
}
\caption[]{
\label{FactorizationTable}
\small Factorization of $N=8$ supergravity amplitudes into sums of
products of $N=4$ Yang-Mills amplitudes.  The factors in the third and
fifth columns are the relative numerator factors appearing in the
gauge theory amplitudes.  The corresponding factors in the
supergravity cases are products of the `left' and `right' gauge theory
amplitudes.}
\end{table}

One nice property of using the KLT relations to obtain the gravity
tree amplitudes is that we do not need to use the $N=8$ Lagrangian or Feynman
rules.  As a check, we have independently
verified the results for the gravity amplitudes contained in 
\tab{FactorizationTable} using the $N=8$ supersymmetry algebra; 
the product of each `left' and `right' factor contained in
\tab{FactorizationTable} corresponds to the supersymmetry Ward identity
factor contained in \tab{NEightTable} in \app{SusyIdentityAppendix}.

We comment that there is some ambiguity as to how to
perform the factorization. For example, left and right can be
interchanged.  As a less trivial example, the last entry in
\tab{FactorizationTable} can be factorized as
$$
h^+h^+v^- \vT^- \vT^- \rightarrow  g^+ g^+ \gT^- \gT^- s^- \; \times \; 
                        g^+ g^+ \gT^-  s^- \gT^-
\, .
\equn
$$
However, in each case where an alternate factorization is available, it
leads to the same overall factor as the factorization
in~\tab{FactorizationTable}, after the left and right factors are
multiplied together, and ignoring an irrelevant phase.  (This ambiguity in
factorizing arises because a string is composed of different `sectors',
such as the Ramond or Neveu-Schwarz sectors.)

When calculating the non-singlet contributions to the cut
in \eqn{ThreeCutHelicity}, we first evaluate the case of three
gravitons crossing the cut, and later use \tab{FactorizationTable} to
obtain the remaining non-singlet contributions.
Just as in the singlet case, we may recycle a Yang-Mills calculation.  
Here we have to evaluate the following products of tree amplitudes,
$$
\eqalign{
&\hbox{(a)} \hskip .5 cm
 A_5^{\rm tree} (1^-, 2^-, \ell_3^-, \ell_2^+, \ell_1^+) \times
 A_5^{\rm tree} (3^+, 4^+, -\ell_1^-, -\ell_2^-, -\ell_3^+)  \,, \cr
& \hbox{(b)} \hskip .5 cm
 A_5^{\rm tree}(1^-, \ell_1^+, \ell_3^-, 2^-, \ell_2^+) \times
 A_5^{\rm tree}(3^+, -\ell_3^+, -\ell_1^-, 4^+, -\ell_2^-)\,; \cr}
\equn
$$
again these products have already been evaluated in the analogous 
cut-calculation for $N=4$ Yang-Mills theory~\cite{BRY}.

The net result is that for $h^- h^- h^+$ crossing the cut we obtain
$$
\eqalign{
M_4^{\twoloop}(1^-,2^-,& 3^+,4^+)
  \Bigr|_{\rm non\hbox{-}singlet\ 3\hbox{-}cut}  = 
s t u M^{\rm tree}_4(1^-,2^-,3^+,4^+) 
\int {d^{D} \ell_1 \over (2\pi)^{D}} 
  {d^{D} \ell_2 \over (2\pi)^{D}} \,
{(\ell_1 + \ell_2)^{16} \over s^8}\,  \cr
& \hskip .5 cm \times
\biggl[
{\trminus [1 \ell_1 43 \ell_3 2 1 \ell_1 43 \ell_2 2 1 \ell_2 4 3 \ell_3 2]
\over\ell_1^2 \, \ell_2^2 \, \ell_3^2 \, (\ell_3 - k_3)^2 
    (\ell_1 - k_4)^2 (\ell_3 + k_2)^2 (\ell_1 + k_1)^2\; 
    (\ell_1 + \ell_2)^2 (\ell_2 + \ell_3)^2 } \cr
& \hskip .5 cm \times
{1\over 
 (\ell_1 + \ell_3)^2 
     (\ell_2 + k_2)^2 (\ell_2 + k_1)^2
      (\ell_2 - k_4)^2 (\ell_2 - k_3)^2} \cr
& \hskip 4 cm 
+ \{1 \leftrightarrow 2 \} + \{3 \leftrightarrow 4 \}
 + \{1 \leftrightarrow 2 , \, 3 \leftrightarrow 4 \}
 \biggr]_{\ell_i^2 = 0} \ .
 \cr}
\equn\label{hhhmmpCut}
$$ 
This result is exactly the one for the singlet except that 
$\trplus \rightarrow \trminus$ and there is a prefactor of 
$((\ell_1 + \ell_2)^2/s)^8$. 

We use~\tab{FactorizationTable} to find the contribution of any 
other particle and helicity configuration from \eqn{hhhmmpCut}.  
(In all cases, the helicity assignments are for particles that cross
the cut from right to left.)
One simply multiplies each `left' and
`right' relative factor in the table together and then by their
complex conjugate to obtain the total relative factors in the cuts.  In
\tab{GravityCutTable} we have collected the relative factors between
the various contributions to the cuts.  All other non-vanishing
contributions may be obtained by simple relabelings of the entries in
the table.

\begin{table}
\vskip .5 cm
\hskip 3 truecm
\hbox{
\def\tend{\cr \noalign{\hrule}}
\def\t#1{\tilde{#1}}
\def\tw{\theta_W}
\vbox{\offinterlineskip
{
\hrule
\halign{
        &\vrule#
        &\quad\hfil\strut#\hfil\quad\vrule
        &\quad\hfil\strut#\hfil\quad
        \cr
height13pt  &{\bf Intermediate States}  &{\bf Relative Cut Factor}  &\tend
height12pt & $ h^- h^- h^+$
& $a^{8}$ &\tend
height12pt & $ h^- \tilde h^- \tilde h^+$
& $8\, a^7 c$  &\tend
height12pt & $ h^- v^- v^+ $
& $28\, a^6 c^2$ &\tend
height12pt & $ h^- \tilde v^- \tilde v^+ $
& $56\, a^5 c^3$  &\tend
height12pt & $ h^- s^- s^+ $
& $70\, a^4 c^4 $ &\tend
%
height12pt & $\tilde h^- \tilde h^- g^+$
& $56 \, a^6bc $  & \tend
height12pt & $ \tilde h^- v^- \tilde v^+$
& $168\, a^5 b c^2 $ &\tend
height12pt & $ \tilde h^- \tilde v^- s^+$
& $280\, a^4 b c^3$  &\tend
height12pt & $ v^-  v^- s^+$
& $420\, a^4 b^2 c^2 $  &\tend
height12pt & $ v^-  \tilde v^- \tilde v^-$
& $560 \,a^3 b^2 c^3$ & \tend
}
}
}
}
\vskip .2 cm
\nobreak
\caption{\label{GravityCutTable} \small The relative contributions to
the non-singlet three-particle cuts obtained from taking the absolute
value squared of the product of the `left' and `right' factors
in \tab{FactorizationTable}. The numbers appearing in the second column
are the multiplicities for each type of cut.  In the table $a = (\ell_1
+ \ell_2)^2$, $b = (\ell_2 + \ell_3)^2$ and $c = (\ell_1 + \ell_3)^2$.}
\end{table}

Summing over all contributions that cross the cut, with the $N=8$ 
multiplicities, we have
$$
\eqalign{
& a^8 
+ 8\, a^7c 
+ 28\, a^6 c^2
+ 56\, a^5 c^3
+ 70\, a^4 c^4
+ 56 \, a^6bc 
+ 168\, a^5 b c^2 
+ 280\, a^4 b c^3
+ 420\, a^4 b^2 c^2
+ 560 \,a^3 b^2 c^3 \cr
& \hskip 3 cm 
+ \hbox{relabelings} \, , \cr}
\equn
$$
where $a = (\ell_1
+ \ell_2)^2$, $b = (\ell_2 + \ell_3)^2$ and $c = (\ell_1 + \ell_3)^2$.
This may be recombined to give
$$
(a+b+c)^8 = s^8 \,.
\equn
$$
The factor of $s^8$ exactly cancels the $1/s^8$ appearing in
\eqn{hhhmmpCut}.  Thus, after summing over all `non-singlet' states
crossing the cut, we obtain exactly the same result as for the singlet
given in \eqn{hhhmmpCut}, except that $\trplus$ is replaced with
$\trminus$.  Of course, since the $\gamma_5$ terms vanish for
four-point kinematics the results are actually completely identical,
after loop-momentum integration.
The net effect is simply to double the singlet result.

\subsection{Comparison to Ansatz}

In order to verify that the cut results agree with the expression in
\eqn{GravAnsatz}, we must compare the integrands, eqs.~(\ref{GravAnsatz})
and (\ref{TargetTripleGravity}), after accounting for the factor of two
arising from combining the singlet and non-singlet contributions.  We have
verified that the results are indeed equal. (When evaluating the traces
analytically one must account for the fact that $\det(k_i \cdot k_j)$
vanishes in four dimensions for five or more independent momenta. The
analysis of \sec{ThreeParticleSubSection}, which uses $D$-dimensional
expressions for the $N=4$ Yang-Mills cuts as input, does not require the
use of this four-dimensional constraint.)

The $t$- and $u$-channel three-particle cuts may be evaluated similarly.
However, it is much simpler to appeal to the supersymmetry
Ward identity (\ref{PermIdentity}) which guarantees that the $t$- and
$u$-channel cuts are identical to the $s$-channel cut, up to
relabelings.  Combining these results with those of 
\app{TwoParticleCutsAppendix} shows that \eqn{TwoLoopGr} is consistent 
with all cuts, but with $D$-dimensional momenta replaced with 
four-dimensional ones.

\subsection{Exactness of Cut Calculation}
\label{CutExactnessSubsection}

We now argue that although the previous cut calculations
use spinor helicity with four-dimensional cut momenta, no terms were
dropped.  In principle, we may have dropped pieces
containing $(-2\eps)$-dimensional momenta of the form,
$$
\int {d^4 p \over (2\pi)^4}  {d^{-2\eps} \mu_p \over (2\pi)^{-2\eps}} \;
     {d^4 q \over (2\pi)^4}  {d^{-2\eps} \mu_q \over (2\pi)^{-2\eps}} \;
      {f(p,q, k_i) \times \bigl\{ \mu_p^2, \, \mu_q^2, \, \mu_p\cdot \mu_q,
            \ldots \bigr\} \over
      (p^2 - \mu_p^2) (q^2 - \mu_q^2) \cdots } \,,
\equn\label{ErrorTerms}
$$
where we have explicitly separated the loop momenta into four- and
$(-2\eps)$-dimensional parts. The $(-2\eps)$-dimensional components
of the two loop momenta are $\mu_p$ and $\mu_q$. (A discussion of the
$(-2\eps)$-dimensional parts of loop momenta can be found, for
example, in refs.~\cite{Mahlon,Massive}.)  Terms containing
such factors are necessarily suppressed by a power of $\eps$. However,
this suppression may be cancelled if the integrals contain poles in
$\eps$.

To avoid the need for an explicit calculation of contributions
of the form~(\ref{ErrorTerms}),
we make use of the KLT tree-level relations which relate the tree
amplitudes on either side of the three-particle cuts to the ones
appearing in the $N=4$ cut calculation.  In ref.~\cite{BRY} it was
argued that the light-cone superspace power-counting of
Mandelstam~\cite{Mandelstam} ruled out the appearance of terms of the
form (\ref{ErrorTerms}) in the two-loop $N=4$ three-particle cut
sewing algebra.  Since this argument is based on superspace
cancellations it applies to the integrands before integration over
loop momenta.

Our observation here is rather simple: Since the $N=4$ Yang-Mills sewing
algebra using four-dimensional momenta in the cuts is exact, then the
$N=8$ sewing algebra must also be exact since it is composed of sums of
products of the $N=4$ algebra.  Thus there can be no terms of the form
(\ref{ErrorTerms}) in the $N=8$ supergravity sewing algebra, given that
there are no such terms in the $N=4$ algebra.  Moreover, this conclusion 
holds at the level of the integrand.

In summary, the above calculation of the three-particle cuts shows
that the expression for the two-loop amplitude (\ref{TwoLoopGr}) is
exact.  This agrees with the result obtained in
\sec{ThreeParticleSubSection} via a more formal $D$-dimensional
calculation.


\section{Extraction of Ultraviolet Infinities from Two-Loop Integrals}
\label{UVExtractionAppendix}

In this appendix, we shall evaluate the ultraviolet divergences of the 
dimensionally-regulated planar and non-planar double-box 
integrals~(\ref{TwoLoopScalarInts}). These integrals are ultraviolet finite in 
$D \leq 6$;
since there is already a one-loop divergence for $D=8$ and $D=10$, the 
cases of $D=7$, 9, and 11 are of more interest.  
(If one uses a different regulator, for example a proper time cutoff, 
there may also be one-loop divergences in $D=9$ and 11, but these linear and 
higher order divergences are absent in dimensional regularization.)

A straightforward Feynman parameterization of the 
integrals~(\ref{TwoLoopScalarInts}) gives
$$
 \I_4^{\twoloop,\, X}(s,t) = {\Gamma(7-D) \over(4\pi)^D}
 \int_0^1 d^7a \ \delta\Bigl( 1 - \sum_{i=1}^7 a_i \Bigr) \, 
   \bigl(-Q_X(s,t,a_i)\bigr)^{D-7} \bigl(\Delta_X(a_i)\bigr)^{7-3D/2}\,,
\hskip.8cm X=\P,\NP,
\equn\label{feynparam}
$$
where
$$
\eqalign{
  \Delta_\P(a_i)  &= (a_1+a_2+a_3)(a_4+a_5+a_6) + a_7(1-a_7)\,, \cr
  \Delta_\NP(a_i) &= (a_1+a_2)(a_3+a_4) + (a_1+a_2+a_3+a_4)(a_5+a_6+a_7) \,, 
\cr}\equn\label{Deltadenom}
$$
and
$$
\eqalign{
  Q_\P(a_i)  &= s \, \bigl( a_1 a_3 (a_4+a_5+a_6) + a_4 a_6 (a_1+a_2+a_3) 
                       + a_7 (a_1+a_4)(a_3+a_6) \bigr) 
              + t \, a_2 a_5 a_7, \cr 
  Q_\NP(a_i)  &= s \, \bigl( a_1 a_3 a_5 + a_2 a_4 a_7 
                       + a_5 a_7 (a_1+a_2+a_3+a_4) \bigr)
              + t \, a_2 a_3 a_6 + u \, a_1 a_4 a_6 \,. 
\cr}\equn\label{Qnumer}
$$      

Fig.~\ref{DoubleBoxParamFigure} depicts the two integrals with the
labels on the propagators specifying the Feynman parametrization used 
in~\eqn{feynparam}.

%
\begin{figure}[ht]
\centerline{\epsfxsize 3 truein \epsfbox{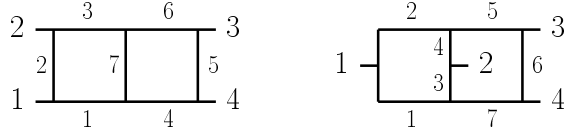}}
\vskip -.2 cm
\caption[]{
\label{DoubleBoxParamFigure}
\small
The planar and non-planar double box integrals, labeled with Feynman
parameters.}
\end{figure}

For $D=7$, 9, and 11, the absence of one-loop sub-divergences implies that the
two-loop divergence should have only a single $1/\e$ pole, which is
contributed by the prefactor $\Gamma(7-D)$.  In extracting the coefficient
of this pole, one might therefore hope to be able to set $\e=0$
immediately in the remaining parameter integral.  This is possible for 
$D=7$, but it is not directly possible for $D=9$ or 11, because the 
remaining parameter integrals do not converge for $\e=0$, and have to be 
defined through analytic continuation.  

Note that the factors $\Delta_X(a_i)$ depend only on special combinations
of Feynman parameters, namely the sums corresponding to the three lines of
the vacuum diagram obtained by deleting the four external legs.  
In contrast, the $Q_X(s,t,a_i)$ depend on all the parameters,
making it difficult to perform the parameter integrals if $Q_X$ 
should appear raised to a non-integer power.  Fortunately, in extracting
the $1/\e$ pole term for $D=n-2\e$ it is legitimate to replace 
$\I_4^{\twoloop,\, X}(s,t)$ with
$$
 \I_4^{\twoloop,\, X} (s,t)\Bigr|_{\rm pole}  = {\Gamma(7-D) \over(4\pi)^D}
 \int_0^1 d^7a \, \delta\Bigl( 1 - \sum_{i=1}^7 a_i \Bigr) \, 
   \bigl(-Q_X(s,t,a_i)\bigr)^{n-7} \bigl(\Delta_X(a_i)\bigr)^{7-n-D/2},
\hskip.5cm X=\P,\NP \,.
\equn\label{newfeynparam}
$$
This can be shown by integrating by parts on the uniform scaling
parameter.  (Alternatively, one can use the fact that the leading 
divergence has a known, uniform dimension~\cite{MarcusSagnottiIntegrals}.)
Now the factor containing $Q_X$ is a polynomial in the Feynman 
parameters, and it is straightforward to integrate all but two of the
six remaining Feynman parameter integrals.  

For the case $D=7-2\e$, where we can set $\e=0$ from the beginning 
(except in the $\Gamma$-function prefactor), we get:
$$
\eqalign{
 \I_4^{\twoloop,\, \P,\, D=7-2\e}\Bigr|_{\rm pole} 
 &= {1\over2\e\ (4\pi)^7} {1\over4} \int_0^1 dy\ y^2(1-y)^2 
   \int_0^1 dx \, { x^{3/2} \over [1-x(1-y(1-y))]^{7/2} } \cr
 &= {1\over2\e\ (4\pi)^7} {\pi\over10} \,.
\cr}\equn\label{planarseven}
$$

For $D=9-2\e$, setting $\e=0$ would lead to 
$$
\eqalign{
 \I_4^{\twoloop,\,  \P,\,  D=9-2\e}\Bigr|_{\rm pole} 
 &= {1\over4\e\ (4\pi)^9} {1\over498960} \int_0^1 
   {dy \over [y(1-y)]^{3/2} }
   \biggl[ 3 \, s^2\ \Bigl( 16 y^2(1-y)^2 - 77 y(1-y) + 132 \Bigr) \cr
&\hskip4cm
         + 8 \, st\ y(1-y) \Bigl( 2 y(1-y) + 11 \Bigr) 
         + 80\, t^2\ y^2(1-y)^2 \biggr] \,.
\cr}\equn\label{planarninea}
$$
The terms in \eqn{planarninea} that are proportional to $st$ and to
$t^2$ have convergent, elementary $y$-integrals.  However, the $s^2$
term is divergent, so we must retain the full $\e$-dependence
from \eqn{newfeynparam} in evaluating it.  
Doing this, we find that 
the $(x,y)$-integral for the $s^2$ term can be conveniently written 
as the sum of two terms,
$$
\eqalign{
 \I_4^{\twoloop,\,  \P,\,  D=9-2\e}\Bigr|_{{\rm pole},\ s^2\ {\rm term}} 
 &= {s^2\over4\e\ (4\pi)^9} \int_0^1 dy \int_0^1 dx 
 \Bigl[ C_1(x,y) + C_2(x,y) \Bigr] \,,
\cr}\equn\label{planarnineb}
$$
where 
$$
\eqalign{
 C_1(x,y) &= {[y(1-y)]^4\over480} { x^{5/2+\e} 
   \Bigl( -x^2 y(1-y) + (1-x)(2-3x) \Bigr)
   \over [1-x(1-y(1-y))]^{13/2-\e} } \,, \cr
 C_2(x,y) &= {[y(1-y)]^2\over360} 
   { x^{5/2+\e} \over [1-x(1-y(1-y))]^{9/2-\e} } \,.
\cr}\equn\label{planarninec}
$$
The integral of $C_1$ converges for $\e=0$, and 
$\int_0^1 dx \, dy\ C_1(x,y) = -5\pi/11088$.

The integral over $C_2$ requires analytic continuation in $\e$,
which we handle with identities (3.197.3) and (9.131.2) of 
Gradshteyn and Ryzhik~\cite{GradRyzh}.  Somewhat more generally, we need
$$
\eqalign{
I(p,q,\alpha) &\equiv \int_0^1 dy\ [y(1-y)]^p \int_0^1 dx \, 
{ x^{\alpha-q-2+\e} \, (1-x)^q \over [1-x(1-y(1-y))]^{\alpha-\e} } \cr
& = { \Gamma(\alpha-q-1+\e)\Gamma(q+1) \over \Gamma(\alpha+\e) }
  \int_0^1 dy\ [y(1-y)]^p 
    ~_2F_1\bigl( \alpha-\e, \alpha-q-1+\e; \alpha+\e ; 1-y(1-y) \bigr)
  \cr
& = { \Gamma(\alpha-q-1+\e)\Gamma(q+1) \over \Gamma(\alpha+\e) }
  \int_0^1 dy\ [y(1-y)]^p \cr
&\hskip0.5cm \times \Biggl\{
    { \Gamma(\alpha+\e)\Gamma(-\alpha+q+1+\e) 
      \over \Gamma(2\e) \Gamma(q+1) }
    ~_2F_1\bigl( \alpha-\e, \alpha-q-1+\e; \alpha-q-\e ; y(1-y) \bigr) \cr
&\hskip0.5cm 
 +  [y(1-y)]^{-\alpha+q+1+\e}
    { \Gamma(\alpha+\e)\Gamma(\alpha-q-1-\e) 
      \over \Gamma(\alpha-\e) \Gamma(\alpha-q-1+\e) }
    ~_2F_1\bigl( 2\e, q+1; -\alpha+q+2+\e ; y(1-y) \bigr) \Biggr\} \,,
\cr}\equn\label{auxinta}
$$
where $p$ and $q$ are positive integers and $\alpha$ is a positive
half-integer.  In the limit $\e\to0$, the factor of 
$1/\Gamma(2\e)$ in \eqn{auxinta} causes the term containing it to
vanish, and the surviving hypergeometric function can be set to 1.
Performing the remaining $y$-integral gives
$$
\eqalign{
I(p,q,\alpha) & = 
 { \Gamma(\alpha-q-1-\e)\Gamma(q+1) \over \Gamma(\alpha-\e) }
  \int_0^1 dy\ [y(1-y)]^{-\alpha+p+q+1+\e} \cr
& = 
 { \Gamma(\alpha-q-1)\Gamma(q+1) \Gamma^2(-\alpha+p+q+2)
   \over \Gamma(\alpha) \Gamma\bigl(2(-\alpha+p+q+2)\bigr) } \,.
\cr}\equn\label{auxintb}
$$
Thus $I(p,q,\alpha) = 0$ (after analytic continuation in $\e$) 
unless $\alpha < p+q+2$.
In the present case, $p=2$, $q=0$, and $\alpha=9/2$, 
so the integral of $C_2$ vanishes.

The case of $D=11-2\e$ is similar, although the leading $s^4$ term in 
$D=11-2\e$ requires a bit more separation,
along the lines of \eqn{planarnineb}, into convergent terms, and terms
to which \eqn{auxinta} can be applied.  
The final result for the planar double-box pole at $D=9-2\e$ 
and $D=11-2\e$ is then
$$
\eqalignno{
 \I_4^{\twoloop,\,  \P,\, D=9-2\e}\Bigr|_{\rm pole} 
 &= {1\over4\e\ (4\pi)^9} {\pi\over99792} 
 ( -45 s^2 + 18 st + 2 t^2 ) \,,
&\equnno\label{planarnine} \cr
 \I_4^{\twoloop,\,  \P,\,  D=11-2\e}\Bigr|_{\rm pole} 
 &= {1\over48\e\ (4\pi)^{11}} {\pi\over196911000} 
 ( 2100 s^4 - 880 s^3 t + 215 s^2 t^2 + 30 s t^3 + 12 t^4 ) \,. 
\hskip 1.5 truecm \null  &
 \equnno\label{planareleven} \cr}
$$

The non-planar double-box integrals are handled analogously, 
with the results:
$$
\eqalignno{
 \I_4^{\twoloop,\, \NP,\, D=7-2\e}\Bigr|_{\rm pole} 
 &= {1\over2\e\ (4\pi)^7} {\pi\over15} \,,
& \equnno\label{nonplanarseven}\cr
 \I_4^{\twoloop,\, \NP,\,  D=9-2\e}\Bigr|_{\rm pole} 
 &= {1\over4\e\ (4\pi)^9} {-\pi\over83160} ( 75 s^2 + 2 tu ) \,, &
\equnno\label{nonplanarnine} \cr
 \I_4^{\twoloop, \, \NP,\,  D=11-2\e}\Bigr|_{\rm pole} 
 &= {1\over48\e\ (4\pi)^{11}} {\pi\over1654052400} 
  ( 40383 s^4 - 1138 s^2 tu + 144 t^2 u^2 ) \,. &
\equnno\label{nonplanareleven} \cr}
$$

As mentioned in \sec{TwoLoopGravityAmplSubSection}, for $N=8$
supergravity we are also interested in the divergences at $D=10-2\e$,
which require subtractions of one-loop sub-divergences.  Here we
followed the approach of ref.~\cite{MarcusSagnottiIntegrals},
differentiating the subtracted momentum integrals with respect to the
external momenta until they are only logarithmically divergent.  For
completeness, we also quote the results for $D=8-2\e$, which were
obtained in the same way.  In each case a minimal ($1/\e$) subtraction
of the subdivergence was used:
$$
\eqalignno{
 \I_4^{\twoloop,\, \P,\, D=8-2\e}\Bigr|_{\rm subtracted,\ pole} 
 &= {1\over2\ (4\pi)^{8}} \biggl[ 
  - { 1 \over 72 } { s \over \e^2 } 
  + { 1 \over 864 }{ 5 s + 2 t \over \e } \biggr] \,,
& \equnno\label{planareight}\cr
 \I_4^{\twoloop,\, \P,\,  D=10-2\e}\Bigr|_{\rm subtracted,\ pole} 
 &= {1\over12\ (4\pi)^{10}} \biggl[
  - { 1 \over 25200 } { s^2 ( 4 s + t ) \over \e^2 } \cr
& \hskip1cm
  + { 1 \over 21168000 } 
        { - 704 s^3 + 55 s^2 t + 252 s t^2 + 63 t^3 \over \e } \biggr] \,, &
\equnno\label{planarten} \cr}
$$
$$
\eqalignno{
 \I_4^{\twoloop,\, \NP,\, D=8-2\e}\Bigr|_{\rm subtracted,\ pole} 
 &= {1\over2\ (4\pi)^{8}} \biggl[ 
  - { 1 \over 144 } { s \over \e^2 } 
  - { 1 \over 864 }{ s \over \e } \biggr] \,,
& \equnno\label{nonplanareight}\cr
 \I_4^{\twoloop,\, \NP,\,  D=10-2\e}\Bigr|_{\rm subtracted,\ pole} 
 &= {1\over12\ (4\pi)^{10}} \biggl[
   { 1 \over 7200 } { s^3 \over \e^2 }
 - { 1 \over 4536000 } { s ( 301 s^2 + 10 t u ) \over \e } \biggr]\,. 
& \equnno\label{nonplanarten} \cr}
$$

\section{Rearrangements of Color Factors for Comparison to 
Marcus and Sagnotti Calculation}
\label{YMColorAppendix}

In order to compare our results for the UV divergence of the
four-point two-loop $N=4$ Yang-Mills amplitudes to the result obtained by
Marcus and Sagnotti via explicit Feynman diagram calculation
(eqs.~(4.5) and~(4.6) of ref.~\cite{MarcusSagnotti}), we need to
convert the group theory factors that we use in \Eqn{SUSYdiv} into the
basis of non-planar double-box, single-box and tree group theory
factors used by them, using their non-standard normalizations.

In the graphical notation~\cite{Cvitanovic} for the Yang-Mills
group theory factors that was used in ref.~\cite{MarcusSagnotti}
(and converting our $\tilde{f}^{abc}$ to $f^{abc}$), 
our \eqn{SUSYdiv} reads:
$$
\eqalign{
T_7 &  = - {8\,g^6 \, \pi \, s \over (4\pi)^7 \, 2\e}
\Biggl[ {1 \over 10} \biggl( \mbox{\psfig{file=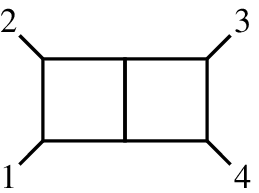,width=60pt,%
bbllx=-5pt,bblly=20pt,bburx=85pt,bbury=85pt}}
\ +\  \mbox{\psfig{file=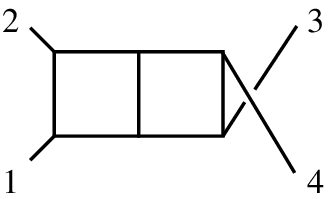,width=55pt,%
bbllx=5pt,bblly=25pt,bburx=85pt,bbury=85pt}}\ \ \biggl)
\ +\ {2 \over 15} \mbox{\psfig{file=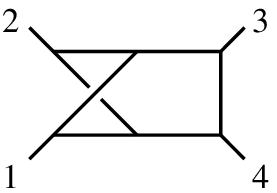,width=60pt,%
bbllx=-5pt,bblly=25pt,bburx=85pt,bbury=90pt}} \Biggr]\ +\ \hbox{cyclic} ,\cr
T_9 &  = - {8\,g^6 \, \pi \, s \over (4\pi)^9 \, 4\e}
\Biggl[  {1\over99792} (-45s^2+18st+2t^2) 
 \mbox{\psfig{file=pb.eps,width=60pt,%
bbllx=-5pt,bblly=20pt,bburx=85pt,bbury=85pt}}
\ +\  {1\over99792} (-45s^2+18su+2u^2) 
\mbox{\psfig{file=cpb.eps,width=55pt,%
bbllx=5pt,bblly=25pt,bburx=85pt,bbury=85pt}}\ \ 
 \cr & \hskip 2 cm  -\ {2\over83160} (75s^2+2tu) 
\mbox{\psfig{file=npb.eps,width=60pt,%
bbllx=-5pt,bblly=25pt,bburx=85pt,bbury=90pt}} \Biggr]\ +\ \hbox{cyclic} ,\cr
}
\equn\label{GraphSUSYdiv}
$$
where `$+$~cyclic' instructs one to add the two cyclic permutations
of (2,3,4). 

We then make use of the graphical identities described in Appendix A
of ref.~\cite{MarcusSagnotti}. We collect the relevant identities in
fig.~\ref{GroupTheoryFigure}, in which a 3-vertex represents a Lie
algebra structure constant $f^{abc}$ and an internal line indicates an
index contraction, $\delta^{ab}$.  Fig.~\ref{GroupTheoryFigure}a
defines the normalization of the structure constants (consistent with
non-standard normalizations of ref.~\cite{MarcusSagnotti}).
Fig.~\ref{GroupTheoryFigure}b graphically depicts the Jacobi identity.
Figs.~\ref{GroupTheoryFigure}c,d,e and f can be derived from
fig.~\ref{GroupTheoryFigure}a and b.  Fig.~\ref{GroupTheoryFigure}e
allows us to remove all planar double-box group theory factors in
favor of non-planar double-boxes and single boxes.
Fig.~\ref{GroupTheoryFigure}d allows the legs of the single boxes to
be uncrossed.  Finally, fig.~\ref{GroupTheoryFigure}b lets us remove
unwanted tree factors.

Applying these identities, we can rewrite \Eqn{GraphSUSYdiv} as
$$
\eqalign{
T_7\ &=\ {3\over2} (-4) {g^6 \, \pi \over (4\pi)^7 \, 2\e} \Biggl\{
 s \Biggl[ {1\over90} \biggl( \mbox{\psfig{file=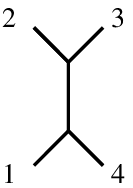,width=40pt,%
bbllx=-15pt,bblly=20pt,bburx=50pt,bbury=50pt}}
\ +\  \mbox{\psfig{file=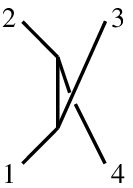,width=40pt,%
bbllx=-15pt,bblly=20pt,bburx=50pt,bbury=50pt}} \biggl)
\ +\ {4\over9} \mbox{\psfig{file=npb.eps,width=60pt,%
bbllx=-5pt,bblly=25pt,bburx=85pt,bbury=90pt}} \Biggr]\ +\ \hbox{cyclic} 
 \Biggr\} \,, \cr 
T_9\ &=\ (-4) {g^6 \, \pi \over (4\pi)^9 \, 2\e} \Biggl\{
 {5\over3024} stu \Biggl[ \mbox{\psfig{file=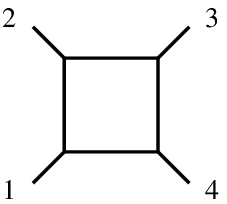,width=40pt,%
bbllx=-15pt,bblly=20pt,bburx=50pt,bbury=50pt}}
\ +\ {1\over6} \biggl( \mbox{\psfig{file=vt.eps,width=40pt,%
bbllx=-15pt,bblly=20pt,bburx=50pt,bbury=50pt}}
\ +\ \mbox{\psfig{file=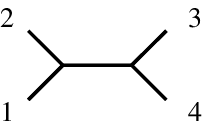,width=50pt,%
bbllx=5pt,bblly=15pt,bburx=70pt,bbury=65pt}} \biggr) \Biggr] \cr
&\hskip2cm
 - \Biggl[ s^3 \Biggl( {5\over133056} \biggl( 
  \mbox{\psfig{file=vt.eps,width=40pt,%
bbllx=-15pt,bblly=20pt,bburx=50pt,bbury=50pt}}
\ +\  \mbox{\psfig{file=ct.eps,width=40pt,%
bbllx=-15pt,bblly=20pt,bburx=50pt,bbury=50pt}} \biggl)
\ +\ {13\over4536} \mbox{\psfig{file=npb.eps,width=60pt,%
bbllx=-5pt,bblly=25pt,bburx=85pt,bbury=90pt}} \Biggr)\ +\ \hbox{cyclic} 
 \Biggr]  \Biggr\} \,.  
\cr}\equn\label{SevenNineResults}
$$
This form can be compared directly to eqs.~(4.5) and~(4.6) of
ref.~\cite{MarcusSagnotti}.  Our $2\eps$ corresponds to their $\eps$,
and the $(-4)$ prefactor accounts for a different normalization of
the operator $t_8 F^4$, as deduced from the one-loop case.

%
\begin{figure}[ht]
\centerline{\epsfxsize 3.5 truein \epsfbox{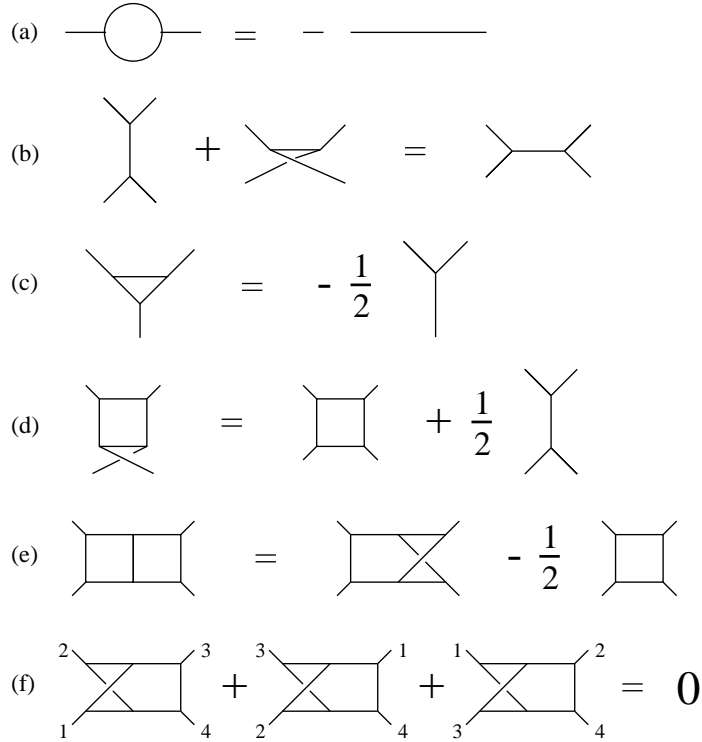}}
\vskip -.2 cm
\caption[]{
\label{GroupTheoryFigure}
\small
Graphical representation of group theory relations.
(a) Normalization. (b) Jacobi identity.
(c), (d), (e) and (f) are used as described in the text.}
\end{figure}


\section{Extended Supersymmetry Ward Identities}
\label{SusyIdentityAppendix}

In this appendix, we review extended supersymmetry Ward identities
which relate the various amplitudes to all perturbative orders.  Using
these relations one can, for example, explicitly verify the
decomposition in \tab{FactorizationTable} of $N=8$ gravity tree
amplitudes in terms of $N=4$ Yang-Mills tree amplitudes.  These identities
are also used in \sec{MultiParticleCutsSection} to demonstrate
supersymmetry cancellations in the MHV cuts.

\subsection{$N=1$ Identities}

First recall the $N=1$ supersymmetry identities~\cite{SWI},
$$
[Q(p), g^\pm(k)] = \mp \Gamma^\pm(k,p)\, \gT^\pm(k) \,, \hskip 2 cm 
[Q(p), \gT^\pm(k)] = \mp \Gamma^\mp(k,p)\, g^\pm(p) \,,
\equn\label{Commutator}
$$
where $Q(p)$ is the supercharge and
$$
\Gamma^+ (k,p) = \bar \theta \spa{p}.{k} \,, \hskip 2 cm 
\Gamma^- (k,p) = \theta \spa{p}.{k} \,.
$$
Because $\Gamma^-$ is proportional to a Grassmann variable $\theta$
and $\Gamma^+$ is proportional to $\bar \theta$, the coefficients of
$\Gamma^+$ and $\Gamma^-$ in a supersymmetric Ward identity are independent.

Since the supercharge $Q(p)$ annihilates the vacuum,
one can construct a typical $N=1$ supersymmetry
Ward identity in the following way:
$$
\eqalign{
0 = \langle 0| [Q,g^-g^-\gT^+g^+g^+] |0 \rangle = & 
\Gamma^-(k_1,p)A_5(1_\gT^-,2_g^-,3_\gT^+,4_g^+,5_g^+) 
+ \Gamma^-(k_2,p)A_5(1_g^-,2_\gT^-,3_\gT^+,4_g^+,5_g^+) \cr
& - \Gamma^-(k_3,p)A_5(1_g^-,2_g^-,3_g^+,4_g^+,5_g^+) 
- \Gamma^+(k_4,p)A_5(1_g^-,2_g^-,3_\gT^+,4_\gT^+, 5_g^+) \cr
& - \Gamma^+(k_5,p)A_5(1_g^-,2_g^-,3_\gT^+,4_g^+,5_\gT^+) \,. }
\equn
$$
Since the coefficients of $\Gamma^+$ and $\Gamma^-$ 
are independent, the sum of the terms with $\Gamma^-$
prefactors must vanish. Thus by choosing $p=k_1$ we obtain 
the following identity:
$$
A_5(1_g^-,2_\gT^-,3_\gT^+,4_g^+,5_g^+) = {\spa1.3 \over \spa1.2}
A_5(1_g^-,2_g^-,3_g^+,4_g^+,5_g^+) \,.
\equn
$$
These relationships are valid to all orders in perturbation
theory.

\subsection{Extended Supersymmetric Ward Identities for 
$N=4$ Yang-Mills Theory}
\label{extendedYM}

In order to obtain the complete set of supersymmetry
relations between amplitudes, one must, of course, use the complete 
supersymmetry algebra of a particular theory.
The $N=2$ commutation relations are~\cite{SWI}
$$
\eqalign{
[Q_a, g^\pm(k)] & = \mp \Gamma^\pm(k,p) \gT_a^\pm\,,\cr
[Q_a,  \gT^\pm_b(k)] & = \mp \Gamma^\mp(k,p) g^\pm \delta_{ab} 
        \mp i\, \Gamma^\pm(k,p) s_{ab}^\pm \eps_{ab}\,, \cr
[Q_a, s_{ab}^\pm(k)] & 
  = \pm i \, \Gamma^\mp(k,p) \eps_{ab} \gT_b^\pm\,,\cr}
\equn
$$
where the subscripts label the different particle states in the $N=2$
super-multiplet. (Although the scalars do not carry helicity, for 
convenience we keep the helicity label since it helps keep track of 
the relationship of the scalars to the other particles in the multiplet.)
The $N=2$ extended supersymmetry Ward identities can be used to calculate 
$N=4$ Yang-Mills amplitudes.  Consider, for example, the following equation,
$$
\eqalign{
0 = \langle 0| [Q_1,g^+g^+g^-\gT_2^-s_{12}^+] |0 \rangle = & 
\Gamma^-(k_3,p)A_5(1_g^+,2_g^+,3_{\gT_1}^-,4_{\gT_2}^-, 5_{s_{12}}^+) 
+i \Gamma^-(k_4,p)A_5(1_g^+,2_g^+,3_g^-,4_{s_{12}}^-, 5_{s_{12}}^+) \cr
& +i \Gamma^-(k_5,p)A_5(1_g^+,2_g^+,3_g^-,4_{\gT_2}^-, 5_{\gT_2}^+) \,,
}
\equn
$$
where we have dropped terms proportional to $\Gamma^+$.  If we now
choose $p=k_4$, we can solve for the amplitude with an external scalar
and two external fermions,
$$
A_5(1_g^+,2_g^+,3_{\gT_1}^-,4_{\gT_2}^-,5_{s_{12}}^+) 
 = -i {\spa4.5 \over \spa4.3}
A_5(1_g^+,2_g^+,3_g^-,4_{\gT_2}^-,5_{\gT_2}^+) 
 =  +i {\spa3.5 \spa4.5 \over \spa3.4^2}
A_5(1_g^+,2_g^+,3_g^-,4_g^-,5_g^+) \,.
\equn
\label{FermionFermionScalar}
$$

\begin{table}[ht]
\begin{center}
\begin{tabular}{|c|c||c|c|}
\hline
$\vphantom{\Big|}$ {\bf Amplitude} & {\bf Relative Factor} & {\bf Amplitude} 
   & {\bf Relative Factor} \\
\hline
$ \vphantom{\Big|}
g^-g^-g^+g^+$ & $\spa1.2^4$ &
$\gT_1^-\gT_1^-\gT_1^+\gT_1^+$ & $-\spa1.2^3\spa3.4$ \\
\hline
$ \vphantom{\Big|}
g^-\gT_1^-\gT_1^+g^+$ & $\spa1.2^3\spa1.3$ &
$\gT_1^-s_{23}^-s_{23}^+\gT_1^+$ & $\spa1.2\spa1.3^2\spa2.4$ \\
\hline
$ \vphantom{\Big|}
g^-s_{12}^-s_{12}^+g^+$ & $\spa1.2^2\spa1.3^2$ &
$\gT_1^-s_{23}^-s_{12}^+\gT_3^+$ & $\spa1.2\spa1.3\spa1.4\spa2.3$ \\
\hline
$ \vphantom{\Big|}
\gT_1^-\gT_2^-s_{12}^+g^+$ & $i\spa1.2^2\spa1.3\spa2.3$ &
$s_{12}^-s_{23}^-s_{12}^+s_{23}^+$ & $-\spa1.2\spa1.4\spa2.3\spa3.4$ \\
\hline
$ \vphantom{\Big|}
g^-s_{12}^-\gT_1^+\gT_2^+$ & $-i\spa1.2^2\spa1.3\spa1.4$ &
$s_{12}^-s_{12}^-s_{12}^+s_{12}^+$ & $\spa1.2^2\spa3.4^2$ \\
\hline
$ \vphantom{\Big|}
\gT_1^-\gT_2^-\gT_2^+\gT_1^+$ & $-\spa1.2^2\spa1.3\spa2.4$ &
& \\
\hline
\end{tabular}
\end{center}
\caption[]{
\label{GluonTable}
\small The extended supersymmetric Ward identities give the relative
factors necessary to obtain all four-point $N=4$ Yang-Mills amplitudes.
The first entry $g^-g^-g^+g^+$ corresponds to $A_4(1^-, 2^-, 3^+, 4^+)$.}
\end{table}

Extended supersymmetry Ward identities relate all the possible external 
particle configurations for four-point Yang-Mills amplitudes.
(See table~\ref{GluonTable}.)  Thus, for the four-point amplitude in an
$N=4$ Yang-Mills theory one only needs to calculate the amplitude with
external gluons.  A similar result holds for the four-point amplitudes
of $N=8$ supergravity.

Extended supersymmetry Ward identities also lead to the following
relationship for $N=4$ MHV amplitudes with $n$ external
gluons~\cite{DimShift},
$$
{1\over\spa{i}.{j}^4}\,
 {\cal A}_n(1^+, 2^+, \ldots, i^-, \ldots, j^-, \ldots n^+) = 
{1 \over \spa{a}.{b}^4} \,
{\cal A}_n(1^+, 2^+, \ldots, a^-, \ldots, b^-, \ldots n^+) \,,
\equn\label{PermIdentityYM}
$$
where $i$ and $j$ are the only negative helicity legs on the
left-hand side and $a$ and $b$ are the only negative helicities on the
right-hand side.  This identity means that these amplitudes are 
symmetric under the interchange of any two legs, up to a trivial prefactor.
Similar equations relate MHV amplitudes for other external states.
We have collected the relative prefactors in \tab{NonSingletSWI};
these results are used in \sec{MultiParticleCutsSection} to demonstrate the
supersymmetric cancellations in the MHV cuts at any loop order. 
The significance of \eqn{PermIdentityYM} at four points
is that the $t$- and $u$-channel cuts must be identical
to the $s$-channel cut, up to relabeling and a simple prefactor.

\begin{table}[ht]
\hskip -1. cm 
\begin{small}
\begin{center}
\begin{tabular}{|l|c||l|c|}
\hline
\vp
{\bf Cut Legs} & {\bf SWI Factor} & {\bf Cut Legs} & {\bf SWI Factor}\\
\hline
$\vp g^-g^-g^+g^+\cdots$ & $\spa{\ell_1}.{\ell_2}^4$ &
$s_{12}^-s_{23}^-s_{12}^+s_{23}^+\cdots$ & $-\spa{\ell_1}.{\ell_2}\spa{\ell_1}.{\ell_4}\spa{\ell_2}.{\ell_3}\spa{\ell_3}.{\ell_4}$
\\
\hline
$\vp g^-\gT_1^-\gT_1^+g^+\cdots$ & 
$\spa{\ell_1}.{\ell_2}^3\spa{\ell_1}.{\ell_3}$ &
$s_{12}^-s_{12}^-s_{12}^+s_{12}^+\cdots$ & 
$\spa{\ell_1}.{\ell_2}^2\spa{\ell_3}.{\ell_4}^2$ \\
\hline
$\vp g^-s_{12}^-s_{12}^+g^+\cdots$ & 
$\spa{\ell_1}.{\ell_2}^2\spa{\ell_1}.{\ell_3}^2$ &
$s_{12}^-s_{12}^-s_{12}^+\gT_1^+\gT_2^+\cdots$ &
$i\spa{\ell_1}.{\ell_2}^2\spa{\ell_3}.{\ell_4}\spa{\ell_3}.{\ell_5}$ \\
\hline
$\vp \gT_1^-\gT_2^-s_{12}^+g^+\cdots$ & 
$i\spa{\ell_1}.{\ell_2}^2\spa{\ell_1}.{\ell_3}\spa{\ell_2}.{\ell_3}$ &
$s_{12}^-s_{23}^-s_{23}^+\gT_1^+\gT_2^+\cdots$ &
$i\spa{\ell_1}.{\ell_2}\spa{\ell_1}.{\ell_3}\spa{\ell_2}.{\ell_4}
                    \spa{\ell_3}.{\ell_5}$ \\
\hline
$\vp g^-s_{12}^-\gT_1^+\gT_2^+\cdots$ & 
$-i\spa{\ell_1}.{\ell_2}^2\spa{\ell_1}.{\ell_3}\spa{\ell_1}.{\ell_4}$ &
$s_{12}^-s_{23}^-s_{13}^+\gT_2^+\gT_2^+\cdots$ &
$-i\spa{\ell_1}.{\ell_2}\spa{\ell_1}.{\ell_3}\spa{\ell_2}.{\ell_3}
                    \spa{\ell_4}.{\ell_5}$
\\
\hline
$\vp g^-\gT_1^+\gT_2^+\gT_3^+\gT_4^+\cdots$ &
$i\spa{\ell_1}.{\ell_2}\spa{\ell_1}.{\ell_3}\spa{\ell_1}.{\ell_4}
                   \spa{\ell_1}.{\ell_5}$ &
$s_{12}^-s_{12}^-\gT_1^+\gT_1^+\gT_2^+\gT_2^+\cdots$ &
$\spa{\ell_1}.{\ell_2}^2\spa{\ell_3}.{\ell_4}\spa{\ell_5}.{\ell_6}$ \\
\hline
$\vp \gT_1^-\gT_2^-\gT_2^+\gT_1^+\cdots$ & 
$-\spa{\ell_1}.{\ell_2}^2\spa{\ell_1}.{\ell_3}\spa{\ell_2}.{\ell_4}$ &
$s_{12}^-s_{23}^-\gT_1^+\gT_2^+\gT_2^+\gT_3^+\cdots$ &
$\spa{\ell_1}.{\ell_2}\spa{\ell_1}.{\ell_5}\spa{\ell_2}.{\ell_6}
\spa{\ell_3}.{\ell_4}$ \\
\hline
$\vp \gT_1^-\gT_1^-\gT_1^+\gT_1^+\cdots$ & 
$-\spa{\ell_1}.{\ell_2}^3\spa{\ell_3}.{\ell_4}$ &
$s_{12}^-s_{34}^-\gT_1^+\gT_2^+\gT_3^+\gT_4^+\cdots$ &
$-\spa{\ell_1}.{\ell_5}\spa{\ell_1}.{\ell_6}\spa{\ell_2}.{\ell_3}
\spa{\ell_2}.{\ell_4}$ \\
\hline
$\vp \gT_1^-s_{23}^-s_{23}^+\gT_1^+\cdots$ & 
$\spa{\ell_1}.{\ell_2}\spa{\ell_1}.{\ell_3}^2\spa{\ell_2}.{\ell_4}$ &
$\gT_1^-\gT_1^+\gT_1^+\gT_2^+\gT_3^+\gT_4^+\cdots$ &
$i\spa{\ell_1}.{\ell_4}\spa{\ell_1}.{\ell_5}\spa{\ell_1}.{\ell_6}
\spa{\ell_2}.{\ell_3}$ \\
\hline
$\vp \gT_1^-s_{23}^-s_{12}^+\gT_3^+\cdots$ &
$\spa{\ell_1}.{\ell_2}\spa{\ell_1}.{\ell_3}\spa{\ell_1}.{\ell_4}
                  \spa{\ell_2}.{\ell_3}$ &
$s_{12}^-\gT_1^+\gT_2^+\gT_1^+\gT_2^+\gT_3^+\gT_4^+\cdots$ &
$-\spa{\ell_1}.{\ell_6}\spa{\ell_1}.{\ell_7}
\spa{\ell_2}.{\ell_4}\spa{\ell_3}.{\ell_5}$ \\
\hline
$\vp s_{12}^-\gT_3^-\gT_3^+\gT_2^+\gT_1^+\cdots$ &
$i\spa{\ell_1}.{\ell_2}\spa{\ell_1}.{\ell_3}\spa{\ell_2}.{\ell_4}
                 \spa{\ell_2}.{\ell_5}$
& $\gT_1^+\gT_2^+\gT_3^+\gT_4^+\gT_1^+\gT_2^+\gT_3^+\gT_4^+\cdots$ & 
$-\spa{\ell_1}.{\ell_5}\spa{\ell_2}.{\ell_6}\spa{\ell_3}.{\ell_7}
                \spa{\ell_4}.{\ell_8}$ \\
\hline
$\vp s_{12}^-\gT_2^-\gT_2^+\gT_2^+\gT_1^+\cdots$ &
$-i\spa{\ell_1}.{\ell_2}^2\spa{\ell_2}.{\ell_5}\spa{\ell_3}.{\ell_4}$ && 
\\
\hline
\end{tabular}
\hskip 1. cm $\null$
\end{center}
\end{small}
\caption[]{\label{NonSingletSWI}
\small The relative factors between the MHV amplitudes 
obtained from the $N=4$
supersymmetric Ward identities.  The first entry 
$g^-\, g^-\, g^+\, g^+ \cdots$
corresponds to the amplitude $A_{m+2}(3^+, 4^+, \ell_1^-, \ell_2^-, 
\ell_3^+, \ldots, \ell_m^+)$ which appears on the right-hand side 
of the $m$-particle cuts discussed in \sec{MultiParticleCutsSection}.
The dots in the `cut legs' column
indicate that the remaining states crossing the cuts are all 
plus-helicity gluons. }
\end{table}

\subsection{Extended Supersymmetric Ward Identities for $N=8$ supergravity}

Since the $N=4$ extended SWI's are a subset of the ones for $N=8$, 
we can use the $N=4$ SWI's to relate many of the $N=8$ supergravity 
amplitudes to each other.  However, if one wants to obtain identities
relating amplitudes with external particle states consisting of the
full $N=8$ supergravity multiplet $(h,\hT,v, \vT,s)$, one needs to
make use of the complete $N=8$ algebra.  These
identities are governed by the following commutation relations
$$
\eqalign{
[Q_a, h^\pm] & = \pm \Gamma^\pm(k,p) \, \tilde h_a^\pm\,,  \cr
[Q_a,  \tilde h^\pm_b] & = \pm \Gamma^\mp(k,p) \, \delta^{ab} h^\pm    
                    \pm i\, \Gamma^\pm(k,p) \eps^{ab} v_{ab}^\pm \,, \cr
[Q_a, v_{bc}^\pm] & = \mp i \, \Gamma^\mp(k,p) \delta^{ab} \tilde h^\pm_c 
             \mp i \, \Gamma^\mp(k,p) \delta^{ac} \tilde h^\pm_b
             \mp \Gamma^{\pm}(k,p) \eps^{abc}  \tilde v_{abc}^\pm\,,\cr
[Q_a, \tilde v_{bcd}^\pm] & = \mp \Gamma^\mp(k,p) \delta^{ab} v_{cd}^\pm 
                            \mp \Gamma^\mp(k,p) \delta^{ac} v_{bd}^\pm 
                            \mp \Gamma^\mp(k,p) \delta^{ad} v_{bc}^\pm 
                            \mp i\, \Gamma^\pm(k,p) \eps^{abcd} s_{abcd}^\pm 
				\,, \cr
[Q_a, s_{bcde}^\pm] & = \pm i \Gamma^\mp(k,p) \delta^{ab} \tilde v^\pm_{cde} 
                   \pm i \Gamma^\mp(k,p) \delta^{ac} \tilde v^\pm_{bde} 
                   \pm i \Gamma^\mp(k,p) \delta^{ad}\tilde v^\pm_{bce}
                   \pm i \Gamma^\mp(k,p) \delta^{ae}\tilde v^\pm_{bcd} \cr
& \hskip 9.5cm \pm \Gamma^\pm(k,p)\, \eps^{abcde}\,\tilde v^\mp_{abcde}\,.\cr}
\equn
\label{NEightSusy}
$$
Since all $SO(8)$ group indices are totally antisymmetric we can
turn $n$ indices into $8-n$ indices via the totally antisymmetric
tensor.  (As a point of notation, the repeated indices should not be
summed over.)

As an example of an $N=8$ supergravity Ward identity consider
$$
\eqalign{
0 =  & \langle 0 \,| \, [Q_1, h^+ h^+ h^- \vT_{234}^- s_{1234}^+]
|\, 0\rangle = 
-  \Gamma^-(k_3, p) M_5(1_h^+, 2_h^+, 3_{\hT_1}^-, 
         4_{\vT_{234}}^-, 5_{s_{1234}}^+) \cr
&\hskip 1.5cm 
+ i \Gamma^-(k_4, p) M_5(1_h^+, 2_h^+, 3_h^-, 4_{s_{1234}}^-, 5_{s_{1234}}^+) 
+ i \Gamma^-(k_5, p) M_5(1_h^+, 2_h^+, 3_h^-, 4_{\vT_{234}}^-,
              5_{\vT_{234}}^+) \,,}
\equn
$$
which after choosing $p=k_5$ yields the following supergravity relationships,
$$
M_5(1_h^+, 2_h^+, 3_\hT^-, 4_\vT^-, 5_s^+) = 
i \, {\spa5.4\over \spa5.3}\,
  M_5(1_h^+, 2_h^+, 3_h^-, 4_s^-, 5_s^+)
= i \, {\spa4.5 \spa3.5^3 \over \spa3.4^4} \, 
  M_5(1_h^+, 2_h^+, 3_h^-, 4_h^-, 5_h^+) \,.
\equn
\label{SusyExampleOne}
$$
\Tab{NEightTable} summarizes the relations between the five-point 
trees necessary for the non-singlet three-particle cut, as deduced 
from the $N=8$ SWI's.
The same results can be obtained from \tab{FactorizationTable},
via the KLT tree relations.

\begin{table}[ht]
\begin{center}
\begin{tabular}{|c|c||c|c|}
\hline
$\vphantom{\Big|}$ {\bf Amplitude} & {\bf Relative Factor} & {\bf Amplitude} & 
{\bf Relative Factor} \\
\hline
$  \vphantom{\Big|}
h^+h^+h^-h^-h^+$ & $\spa3.4^8$ &
$h^+h^+h^-\hT^-\hT^+$ & $\spa3.4^7\spa3.5$ \\
\hline
$ \vphantom{\Big|} 
h^+h^+h^-v^-v^+$ & $\spa3.4^6\spa3.5^2$ &
$h^+h^+h^-\vT^-\vT^+$ & $\spa3.4^5\spa3.5^3$ \\
\hline
$ \vphantom{\Big|} 
h^+h^+h^-s^-s^+$ & $\spa3.4^4\spa3.5^4$ &
$h^+h^+\hT^-\hT^-v^+$ & $i\spa3.4^6\spa3.5\spa4.5$ \\
\hline
$ \vphantom{\Big|}
h^+h^+\hT^-v^-\vT^+$ & $\spa3.4^5\spa4.5\spa3.5^2$ &
$h^+h^+\hT^-\vT^-s^+$ & $i\spa3.4^4\spa4.5\spa3.5^3$ \\
\hline
$ \vphantom{\Big|}
h^+h^+v^-v^-s^+$ & $-\spa3.4^4\spa4.5^2\spa3.5^2$ &
$h^+h^+v^-\vT^-\vT^-$ & $-i\spa3.4^3\spa4.5^2\spa3.5^3$ \\
\hline
\end{tabular}
\end{center}
\caption[]{
\label{NEightTable}
\small 
The $N=8$ extended supersymmetry Ward identity relations between all
five-point supergravity amplitudes with two positive helicity 
external gravitons.}
\end{table}

Extended supersymmetry Ward identities imply that 
for $N=8$ pure graviton amplitudes,
$$
{1\over\spa{i}.{j}^8}\,
 {\cal M}_n(1^+, 2^+, \ldots, i^-, \ldots, j^-, \ldots n^+) = 
{1 \over \spa{a}.{b}^8} \,
{\cal M}_n(1^+, 2^+, \ldots, a^-, \ldots, b^-, \ldots n^+) \,,
\equn\label{PermIdentity}
$$
where $i$ and $j$ are the only negative helicity legs on the
left-hand side and $a$ and $b$ are the only negative helicities on the
right-hand side.  This identity is similar to \eqn{PermIdentityYM}
for $N=4$ Yang-Mills theory, and may be proven in the same way.


\end{document}